\documentclass[aps,rmp,reprint,amsmath,amssymb,graphicx,longbibliography]{revtex4-1} 

\usepackage{graphicx}
\usepackage{bmpsize}
\usepackage{titlesec}
\usepackage{subfig}
\usepackage{epsfig}
\usepackage{slashed}
\usepackage{soul}

\titleformat{\paragraph}
{\normalfont\normalsize\bfseries}{\theparagraph}{1em}{}
\titlespacing*{\paragraph}
{0pt}{3.25ex plus 1ex minus .2ex}{1.5ex plus .2ex}

\newcommand{\la}{\langle}
\newcommand{\ra}{\rangle}
\newcommand{\bfr}{{\bf r}}
\newcommand{\bfR}{{\bf R}}

\newcommand{\ben}{\begin{displaymath}}
\newcommand{\een}{\end{displaymath}}
\newcommand{\be}{\begin{equation}}
\newcommand{\ee}{\end{equation}}
\newcommand{\bea}{\begin{eqnarray}}
\newcommand{\eea}{\end{eqnarray}}

\newcommand{\eq}[1]{Eq.~(\ref{#1})}

\newcommand{\bfp}{{\bf p}}
\newcommand{\bfP}{{\bf P}}
\newcommand{\bfq}{{\bf q}}
\newcommand{\bfk}{{\bf k}}                                                           
\newcommand{\bfK}{{\bf K}} 

\newcommand{\xB}{\ifmmode x_B \else $x_B$ \fi}

\newcommand{\etal}{{\it et al.}}
\usepackage{bm}

\newcommand{\bfkappa}{\mbox{\boldmath$\kappa$}}

\def\p{\phi}

\def\g{\gamma}
\def\e{\epsilon}

\def\a{\alpha}
 \def\d{\delta}

\def\m{\mu}
\def\n{\nu}
\def\e{\epsilon}
\def\L{\Lambda}

\def\b{\beta}

\def\k{\kappa}

\begin{document}

\title{Nucleon-Nucleon Correlations, Short-lived Excitations, and the Quarks Within }

\author{Or Hen}
\affiliation{Massachusetts Institute of Technology, Cambridge, MA 02139}
\author{Gerald A. Miller} 
\affiliation{Department of Physics, University of Washington, Seattle, WA 98195}
\author{Eli Piasetzky}
\affiliation{School of Physics and Astronomy, Tel Aviv University, Tel Aviv 69978, Israel}
\author{Lawrence B. Weinstein}
\affiliation{Department of Physics, Old Dominion University, Norfolk,
  VA 23529}

\date{\today{}}

\begin{abstract}
  This article reviews our current understanding  of how the internal quark
  structure of a   nucleon     bound in nuclei differs from that of a free nucleon.
  We focus on the interpretation of measurements of the EMC effect for
  valence quarks, a reduction in the Deep Inelastic Scattering (DIS)
  cross-section ratios for nuclei relative to deuterium, and its
  possible connection to nucleon-nucleon Short-Range Correlations
  (SRC) in nuclei. Our review and new analysis (involving the
  amplitudes of non-nucleonic configurations in the nucleus) of the
  available experimental and theoretical evidence shows that there is
  a phenomenological relation between the EMC effect and the
  effects of SRC that is not an accident.  The influence of strongly
  correlated neutron-proton pairs involving highly virtual nucleons is
  responsible for both effects. These correlated pairs are temporary high-density fluctuations in the nucleus in which the internal structure of the nucleons is briefly modified. This conclusion needs to be
  solidified by the future experiments and improved theoretical
  analyses that are discussed herein.

\end{abstract}


\maketitle

\tableofcontents{}

\section{Introduction - Short Range Correlations (SRC) and Nuclear  Dynamics} 
\label{Intro}

Nuclear physics is one of the oldest fields in modern physics.  Its history~\cite{Wong:1998ex}
separate from atomic physics, can be said to start with the discovery
of radioactivity in 1896 by Henri Becquerel.  Fifteen years 
later Rutherford used backward  scattering 
of alpha particles to discover that the nucleus
is a tiny object at the heart of the atom. In 1932 Chadwick
discovered a neutral particle of about the same mass as the proton that he
called the neutron. This discovery allowed scientists to understand
that the binding energy accounted for less than one percent of the 
nuclear mass.  Thus it is natural to say that the nucleus is made of
neutrons and protons.  In 1935 Yukawa suggested a theory of the strong
force to explain how the nucleus holds together. In the Yukawa
interaction a virtual particle, later called a meson, mediated a force
between nucleons. This force explained why nuclei did not fall apart
due to proton repulsion, and it also explained why the
attractive strong force had a shorter range than the
electromagnetic proton repulsion.  Thus we may think of the
stable nucleus as a tight ball of neutrons and protons (collectively
called nucleons), held together
by the strong nuclear force.

This basic picture has been studied for many years. Early models
treated  heavy nuclei, which could contain hundreds of nucleons,
as classical liquid drops.  The liquid-drop model can reproduce
many features of nuclei, including the general trend of binding energy
with respect to mass number, as well as nuclear
fission.
 
The liquid drop idea cannot explain more detailed properties of
nuclei. Quantum-mechanical effects (which can be described using the
nuclear shell model developed initially by Mayer~\cite{PhysRev.78.16}
and Jensen~\cite{PhysRev.75.1766.2}) explained that nuclei with certain
numbers of neutrons and protons (the magic numbers 2, 8, 20, 28, 50,
82, 126, ...) are particularly stable because their shells are
filled.  
Many studies were devoted to understanding how the liquid
drop model, with its collective features, could be consistent with the
shell model.
 
Detailed studies of nucleon-nucleon scattering indicated that their
interaction contains something like a hard core, making the origin of
the shell model even more mysterious than its coexistence with the
liquid drop model.  Brueckner and other early workers (see the references in ~\cite{Gomes:1957zz}) showed that in the nuclear medium, the
large, short-ranged effects of the strong nucleon-nucleon potential
could be summed and treated in terms of a smoother object, defined as
a $G$ matrix. 
This idea allowed much of nuclear
phenomena to be understood (at least qualitatively) in terms of the
fundamental nucleon-nucleon interaction. The nucleus was made of
nucleons, with the occasional evanescent meson existing as it
propagated from nucleon to nucleon. 

After the single-particle shell model, the natural next step in
describing nuclei is including the effects of two-nucleon
correlations.  The strong short-ranged nucleon-nucleon force that is averaged to make the mean-field $G$-matrix also causes a
significant nucleon-nucleon correlation function (see the Appendix for
definitions).  However, definitive experimental evidence for
correlations had to await two kinds of high-energy reactions \cite{Frankfurt81}. These are the inclusive  $(e,e')$  scattering at values of Bjorken $\xB>1$~\cite{egiyan03,egiyan06,fomin12}
 and exclusive reactions that could isolate the effects of ground-state
correlations from the various two-body currents and final state interactions
that occur in nuclear reactions~\cite{tang03, Piasetzky:2006ai, shneor07, Subedi:2008zz,bagh10,korover14,hen14,Makek:2016utb,Monaghan:2013ah}.

Meanwhile, deep inelastic scattering on nucleons led to the
discovery that the nucleons are made of quarks.
However due to the small ($\le1\%$) nuclear binding energy and the
idea of quark-gluon confinement, it was thought that quarks had no explicit role in
the nucleus and that therefore nuclei could still be described in terms of
nucleons and mesons. 
The simple and compelling nucleon/meson picture of the nucleus was
shaken to its core by the 1982 discovery by the European Muon Collaboration~\cite{aubert83}, of 
the non-trivial dependence of the per-nucleon lepton deep inelastic
scattering cross section on the specific
nuclear target.  The EMC initially reported incorrect results for $\xB<0.15$.  As a result many refer to the EMC effect as  
the  reduction of the cross section per nucleon in the region $0.2<\xB<0.7$. This reduction has been observed many times and 
we use the term, `the   EMC effect'  to refer  to this region. 
The observation of this reduction, caused by the nucleus,  showed that the quarks have a small but definite role in
the nucleus. We need to understand this. 

There are a number of fundamental unanswered questions about nuclear physics. 
\begin{enumerate}
\item Is the  nucleus really made of nucleons and mesons only?

\item How does the nucleus emerge from QCD, a theory of quarks and gluons?

\item How does the partonic 
 content of the nucleus differ from that of $N$ free neutrons plus $Z$
 free protons?
\end{enumerate}
No one asked such questions before the discovery of the EMC effect. 

At first glance there appears to be little relation 
between  nucleon-nucleon correlations and the EMC effect. However,
there is a strong phenomenological connection between
them~\cite{weinstein11} that occurs for the valence quarks that carry large momentum  and that connection is the
subject of this review.  Indeed, the fundamental challenge for current explanations of the EMC effect is to explain also the inclusive and exclusive high momentum transfer reactions dominated by short ranged correlations which take up about 20\% of the wave function. The  data suggest
that the non-nucleonic admixture in these  correlations is at most  about 10\%, leading to a 2\% non-nucleonic contribution. However, the EMC effect is about 15\%, so that one needs to find an enhancement mechanism.  

We now summarize our most important conclusions for the benefit of the reader:
\begin{itemize}
\item there is much indirect and direct evidence for the existence of
  nucleon-nucleon short-ranged correlations (SRC),

\begin{itemize}
\item high energy $(e,e'pN)$ and $(p,2pN)$ reactions show that
  two-nucleon correlations exist in nuclei, dominate the high-momentum ($k\ge k_F$) tail of the nuclear momentum distribution, and are dominated, at certain nucleon
  momenta, by $np$ pairs, and

\item high energy $(e,e')$ reactions at large values of $\xB$ (the
  Bjorken scaling variable) show that all nuclei have similar momentum
  distributions at large momentum, consistent with the direct observation that strongly-correlated
  two-nucleon clusters exist in the nuclear ground state,
 
\item a consequence of the $np$-SRC dominance is  the possible inversion
  of the kinetic energy sharing in nuclei with $N>Z$ (i.e., that
  protons might have more kinetic energy than neutrons in neutron-rich
  nuclei).

\item this leads to a dynamic model of nuclei where SRC pairs are temporary large fluctuations in the local nuclear density.
\end{itemize}

\item conventional (non-quark) nuclear physics cannot account for the EMC effect,

\item models need to include nucleon modification to account for the
  EMC effect.  These models can modify the structure of either:
\begin{itemize}
\item predominantly mean-field nucleons, which are modified by
  momentum-independent interactions, or
\item predominantly nucleons belonging to SRC pairs, or
\item both mean-field and SRC nucleons,
\end{itemize}

\item there is a phenomenological connection between the strength of
  the EMC effect and the probability that a nucleon belongs to a
  two-nucleon SRC pair ($a_2(A)$).  This connection has also been derived using two
  completely different theories, so that it is no accident,

\item in contrast to previous static models of the EMC effect, the association with SRC implies that nucleons are temporarily modified only
  when they briefly fluctuate into an SRC pair,

\item the influence of SRC pairs can account for the EMC-SRC
  correlation  because both effects are driven by high virtuality
  nucleons ($p^2\ne M^2$),

\item high-virtuality nucleons have an enhanced but still small {\it amplitude} for non-nucleonic configurations. Interference effects between nucleonic and non-nucleonic components (linear in the amplitudes)  are responsible for the EMC effect,

\item modified nucleons, by definition, must contain a small fraction of baryons that
  are not nucleons. Amplitudes for such baryons, with effects enhanced in a coherent manner, exist in the short-ranged correlations, and
  are the source of the EMC effect.

\end{itemize}

 We aim   to critically discuss the reasons for these conclusions and provide enough details for the reader to appreciate the progress that has been made in recent years.
The remainder of this article describes the experimental and theoretical   evidence for  the existence of two-nucleon short range correlations and the properties thereof;  the theoretical and experimental facts regarding  deep inelastic scattering, nucleon structure functions and the EMC effect; and, the need for nucleon modification to explain the EMC effect.  It will then present the unexpected correlation between the strength of the EMC effect in a given nucleus and the probability that a nucleon in that nucleus belongs to an SRC pair.  The ensuing  discussion  presents theoretical ideas connecting SRC and EMC physics, and explores the idea that the SRC-EMC correlation can be used to determine the structure function of a free neutron. The final sections are concerned with other evidence that the nuclear medium modifies the structure of bound nucleons, and future directions for experimental and theoretical research. The Appendix  presents formal  definitions of the terms we use, and also  explains some equations used in the main text. Specific locations of the various subjects are listed in the Table of Contents.

\subsection{The challenge of describing nuclei } 
Nucleons bound in nuclei move under the influence of the strong
interaction as effected by short-ranged two and three body
potentials. Solving even the non-relativistic $A$-body Schroedinger equation 
was initially an impossibly daunting challenge, so that understanding
the vast array of relevant experimental data required the use of
models. 
 
The nuclear shell
model was one of the earliest and perhaps most powerful models. In
this model, each nucleon moves independently in the average field
produced by the other nucleons.  This shell model provides a
reasonable description of many nuclear properties and is the
fundamental starting point for all efforts to provide a theory of
nuclei. Its explanation of the nuclear magic numbers is a major
accomplishment in the history of physics.  Despite this, early
research involving collective degrees of freedom established that the
single particle picture of nuclei could not be complete. More
generally, corrections to the shell model can be classified broadly in
terms of the relevant distances needed to describe the various
phenomena. There are both long-ranged ($\sim$ the size of the nucleus)
and short-ranged  ($\sim$ the size of the nucleon) phenomena.

The strong 
nucleon-nucleon force  is known to bind medium and heavy nuclei, all   with about the same average central density of $\rho_A = 0.16$  nucleons/fm$^3$. 
Thus, the average distance between nucleons is  about $1/\rho_A^{1/3}=1.8 $ fm.  
The radius of a nucleon is about $r=$ 0.86 fm, so that most (but not all) of the time it does not overlap in space with other nucleons. The nucleon  has a volume of $V=\frac43 \pi r^3=2.5$ fm$^3$ and a corresponding density of $\rho_N = 0.4$ fm$^{-3}$.  Thus $\rho_N/\rho_A = 2.5$ and the {\it maximum} nuclear density, even without nucleons overlapping, is 2.5 times the {\it average} nuclear density. 

The fact that a   nucleon has  about $2.5$ times larger density than the nuclear central density and  that nucleons move in the  nucleus with about a quarter of the velocity of light opens up the possibility of large local density fluctuations.  These also lead to large local momentum fluctuations via the uncertainty principle. The strong short range repulsive force between nucleons restrains the size of these fluctuations, but since its range is smaller than a fermi, the density and momentum fluctuations  in nuclei can still be quite large.


The diverse features described above indicate that understanding the broad range of nuclear phenomena requires the
 use of many experimental tools. Since electromagnetic interactions
 are well-understood and presumably simple, electron scattering has
 long been used as a tool to investigate different aspects of nuclear
 structure.  We examine the use of electron scattering to probe the
 validity of the single-particle shell model in the next sub-section.



\subsection{The need for short range correlations/Beyond the nuclear shell model \label{NeedForSrc}}
\subsubsection{Spectroscopic factors} 
Data from electro-induced proton knockout reactions on nuclei, $A(e,e'p)$, provided early evidence for the validity of the shell model \cite{frullani84}. These studies complemented the use of low-energy nuclear reactions, such as $(d,p)$ and $(p,pp)$. Later on, more detailed studies using  higher energy electron beams explored the limits of the validity of the shell model.  We next explain how this happened.

In the $(e,e'p)$ reaction the electron knocks out a nucleon so that an initial nuclear state $|i\ra$ of $A$ nucleons is converted to a final nuclear state $|f\ra$ of $A-1$ nucleons. The reaction can be analyzed in terms of spectroscopic factors~\cite{RevModPhys.32.567}, which are probabilities that all but one of the nucleons willl find themselves in the final state. More formally, if one considers a single-particle state of quantum numbers $\a$, the spectroscopic factor $S_\a$ is given by the square of the overlap: $S_\a=\left|\la f |b_\a|i\ra \right|^2$, where $b_\a$ destroys a nucleon. If the independent particle model were exact, then $S_\a$ would be unity for each occupied state $\a$. Thus measuring $S_\a$ is a useful way to study the nuclear wave functions and the limitations of the independent particle model.

\begin{figure}
\includegraphics[width=\columnwidth]{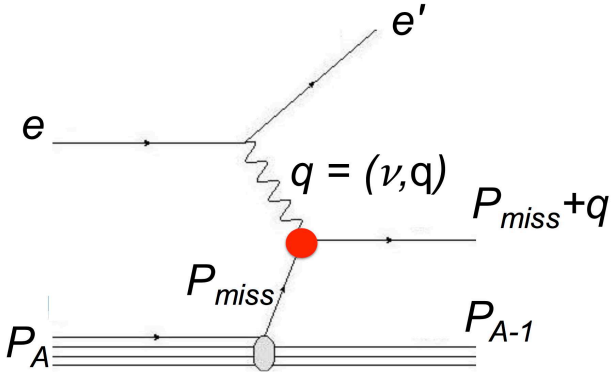}
\caption{\label{fig:pwiafirst} The $A(e,e'p)$ reaction in the Plane Wave Impulse Approximation.  A nucleus of four-momentum $P_A$ emits a nucleon of four-momentum $P_{\rm miss}$ that absorbs a virtual photon of four-momentum $q$ to make a nucleon of four momentum $P_{\rm miss}+q$, with $(P_{\rm miss}+q)^2=M^2$, where $M$ is the nucleon mass. The blob represents the in-medium electromagnetic form factors. }
\end{figure}

In the Plane Wave Impulse Approximation (PWIA), an electron transfers a single virtual photon with momentum ${\bf q}$ and energy $\nu$ (sometimes written $\omega$) to a single proton, then leaves the nucleus without reinteracting  and can thus be described by a plane wave (see Fig. \ref{fig:pwiafirst}).

In PWIA the cross section factorizes in the form \cite{kelly96}
\bea
\frac{d\sigma}{d\nu d\Omega_e d  E_{\rm miss} d\Omega_p} = K \sigma_{ep} S( E_{\rm miss},{\bf p} _{\rm miss}) \label{one}
\eea
where $K = E_p p_p / (2\pi)^3$ is a kinematical factor, $E_p$ and $p_p$ are the energy and momentum of the outgoing proton, $\sigma_{ep}$ is the electron  cross section \cite{deforest83} for scattering by a bound proton, and $S$ is the spectral function, the probability of finding a nucleon in the nucleus with momentum ${\bf p} _{\rm miss}$ and separation energy $ E_{\rm miss}$.  The missing momentum and missing energy are given by: 
\bea {\bf p} _{\rm miss}&=&{\bf q} - {\bf p}_p \cr  E_{\rm miss} &=& \nu - T_p - T_{A-1}   
\eea
where $T_p$ and $T_{A-1}$ are the kinetic energies of the detected proton and residual (undetected) $A-1$ nucleus.  

However, the knocked-out proton then interacts with other nucleons as it leaves the nucleus; these final state interaction (FSI) effects  have been typically calculated using either  an optical model at low momenta \cite{kelly96} or using the eikonal or Glauber approximations at higher momenta \cite{sargsian05,Ryckebusch:2003fc}. Calculations where the wave function of the knocked-out proton are distorted by FSI are referred to as distorted wave impulse approximation calculation (DWIA).  [Note that FSI effects mean that $p_{miss}$ is no longer equal to the initial momentum of the struck nucleon.]
In DWIA, the $(e,e'p)$  cross section does not exactly factorize as in the PWIA. However,  factorization is a good approximation at $Q^2 >> p_{\rm miss}^2$ and the cross section is approximately proportional to a distorted spectral function $S^D$ \cite{kelly96}.  Neither PWIA nor DWIA calculations
conserve current because the initial and final wave functions of the model calculations are not orthogonal and because the effective $NN$ interactions used in the initial and final states are different.  (Some models force current conservation by arbitrarily modifying kinematic variables such as $q^\mu$ \cite{DeForest:1983ahx}.)  Relativisitic DWIA models were developed by Van Orden and collaborators \cite{pickle85,pickle89} and later elaborated by \cite{udias93,udias95,udias99} and \cite{kelly99,Kelly:1994xh}.

\begin{figure}
\includegraphics[width=3.in]{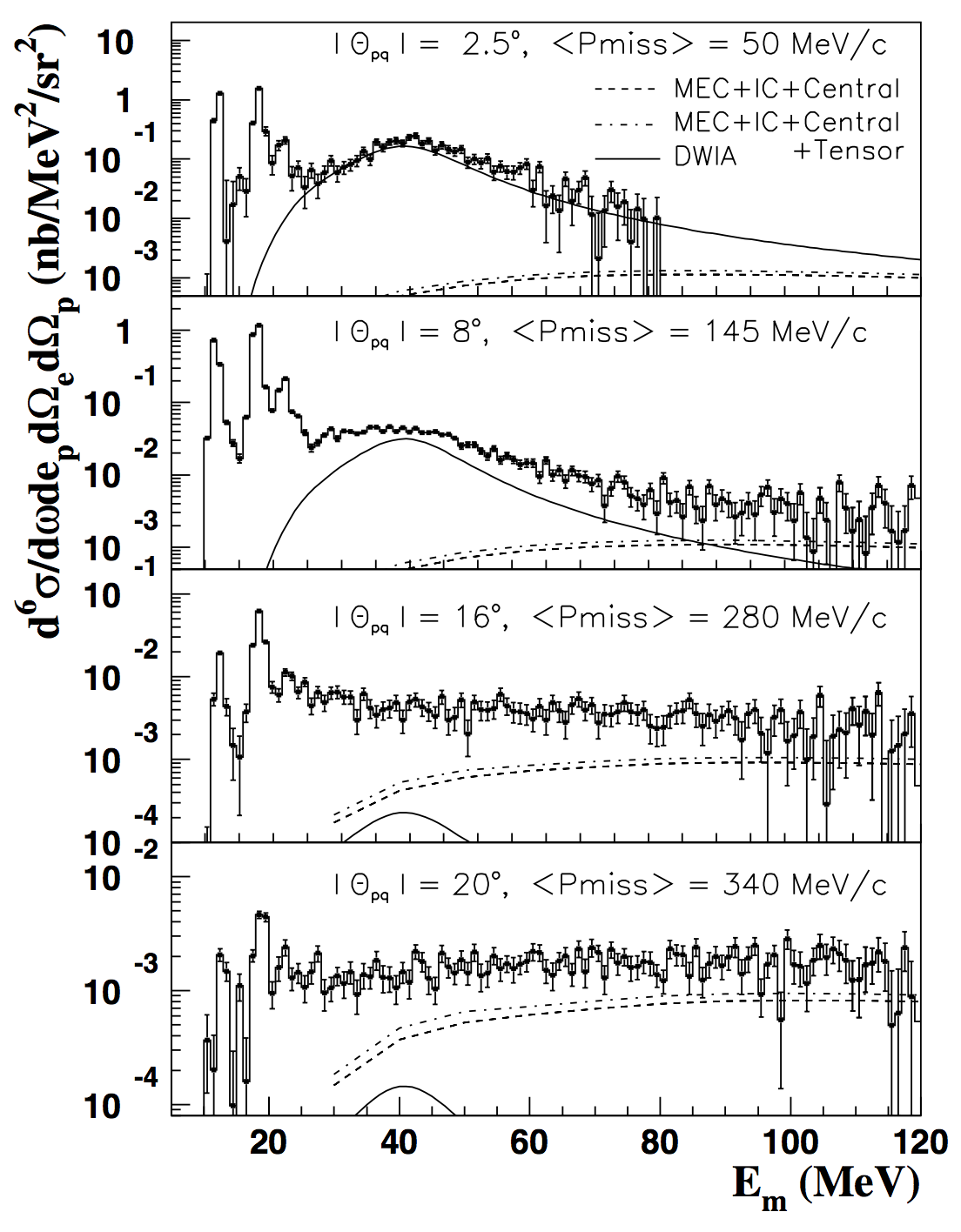}
\includegraphics[width=3.in]{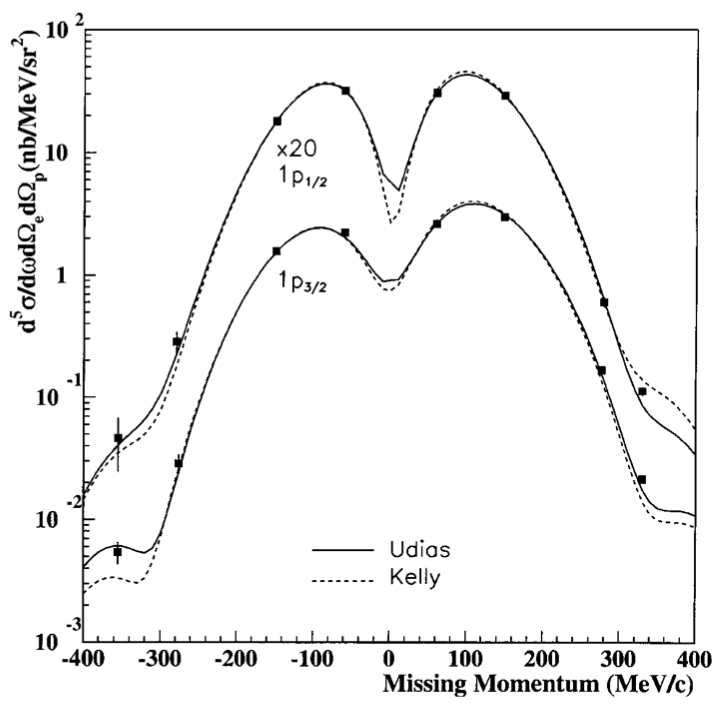}
\caption{\label{fig:Oeep} (upper)  The O$(e,e'p)$ cross section plotted versus missing energy at $Q^2=0.8$ GeV$^2$ and $\nu=0.439$ GeV for different angles, $\theta_{pq}$, between the proton spectrometer and {\bf q}.  The curve labelled DWIA is a distorted wave impulse approximation calculation of $s$-shell knockout; the other curves are calculations of two-nucleon knockout including meson exchange currents (MEC), delta production (IC), and central and/or tensor correlations. Figure from \cite{liyanage01}. (lower) The cross section plotted versus missing momentum for the $1p_{1/2}$ and $1p_{3/2}$ 
states.  Figure  from \cite{gao00}.  The curves show DWIA calculations.  See \cite{liyanage01,gao00,fissum04} for details.}
\end{figure}

Thus, $(e,e'p)$ measurements should be sensitive to the spectral function, i.e., to the momentum and energy distributions of nucleons in the nucleus.  Fig. \ref{fig:Oeep} shows the $^{16}$O$(e,e'p)$ cross section at $Q^2=0.8$ GeV$^2$ and $\nu=0.439$ GeV plotted versus missing energy at several different missing momenta and plotted versus missing momentum for the two $p$-shell states.  There are sharp peaks at $ E_{\rm miss}=12$ and 18 MeV, corresponding to proton knockout from the $1p_{1/2}$ and $1p_{3/2}$ shells, a broad peak at $ E_{\rm miss}\approx40$ MeV corresponding to proton knockout from the $1s$ shell (and other processes), and a long tail extending to large $ E_{\rm miss}$, especially at the largest missing momenta.  The momentum distribution calculations shown in Fig. \ref{fig:Oeep}(lower) use an optical potential, a modern bound state wave function, and an off-shell cross section $\sigma_{ep}$ and fit only the magnitude (see Ref. \cite{gao00} for details).  The calculations describe the data well, except for the fact that the ratio of data to theory (the spectroscopic factor) is approximately 0.7.  This means that the experiment only measured 70\% of the expected number of $p$-shell protons.

\begin{figure}
\includegraphics[width=3.in]{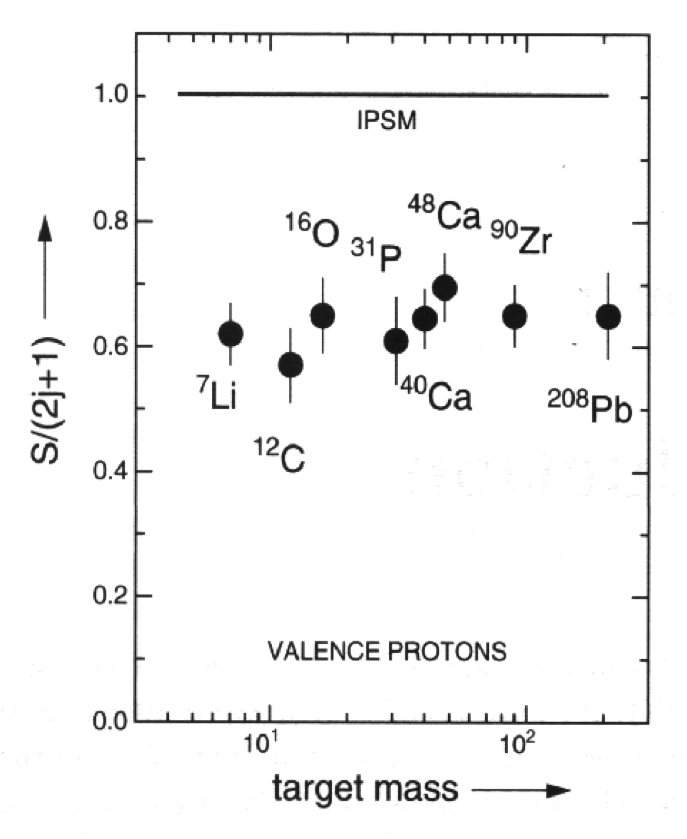}
\caption{\label{fig:specfac} The fractional spectroscopic factors (the ratio of measured cross sections to those calculated with the Independent Particle Shell Model) for valence nucleon knockout $(e,e'p)$.  Reproduced based on \cite{lapikas97}.}
\end{figure}

This depletion of the spectroscopic factor was observed over a wide range of the periodic table at relatively low momentum transfer (see Fig. \ref{fig:specfac}) for both valence nucleon knockout using the $(e,e'p)$ reaction \cite{lapikas97} and stripping using the $(d,^3{\rm He})$ reaction \cite{kramer01}.  Only about 60--70\% of the expected valence nucleon strength was observed.  The missing strength implies the existence of collective effects  (long range correlations)  and short range correlations in nuclei.  The spectroscopic factors and the size of the collective effects
depend on momentum transfer~\cite{Lapikas:1999ss,Frankfurt:2000ty}.  In addition, the spectroscopic strength for valence proton knockout (e.g., $1p_{3/2}$ proton knockout from C) is distributed over many states and not all of these states are included when measuring the spectroscopic factor. The results in Fig. \ref{fig:specfac}) cannot be directly related to the probability of short range correlations in nuclei due to the effects of momentum transfer-dependence, state-splitting, and collective effects.
Our focus will be on the short range correlations as observed using high-momentum transfer probes.

In the DWIA independent particle shell model we would expect that the spectroscopic factors are unity and that there is little cross section at large $ E_{\rm miss}$.  The fact that spectroscopic factors are significantly less than unity for all nuclei, and that there is significant cross section at large missing energy indicates that this  simple model picture omits important physics. This is not surprising, since the short-ranged nature of the strong nuclear forces implies that nucleons must be influenced by nearby nucleons. There is no fundamental one-body potential in the nucleus, unlike the central one-body Coulomb potential that binds electrons to form the structure of the atom.
 
Indeed, since the  $NN$ forces are short ranged, the fact that the shell model approximation has any relevance is somewhat surprising.  In the early days of nuclear physics, the fundamental question of nuclear physics was: how does the very successful shell model of the nucleus emerge in spite of the strong short-ranged interactions between nucleons?  

We next  answer this fundamental question, then examine  the consequences of the answer. 

  \subsubsection{From the $NN$ Interaction to the Shell Model and Beyond} 
How can the mean-field shell model arise from a system 
made of nucleons interacting by strong short-ranged forces?
An answer to this question 
was provided early on by Brueckner \& Goldstone, see the
review by Bethe~\cite{Bethe:1971xm}. The strong two-nucleon
interactions encoded by the potential $V$, constructed to reproduce
experimentally measured $NN$ scattering observables and believed to include
strong repulsion at short distance and attraction at longer ranges,
are summed to form the $T$ matrix of scattering theory and the
$G$-matrix for bound states. The operator $G$ is obtained from $T$ by
modifying the propagator of the Lippmann-Schwinger equation to include
the effects of the Pauli principle and to use the appropriate
self-consistent (single) nucleon energies. The $G$ matrix is
considerably weaker than $V$. For example,  even if the potential is infinitely strong, the
product $V\Psi$ of the potential with the  wave function obtained from the chosen Hamiltonian  would be
finite and well behaved. Schematically, one has $G\Phi=V\Psi$, where
$\Phi$ is the shell-model two-nucleon wave function.  Calculations
show that $G$ is reasonably smooth  and can be used as input in higher-order calculations. 

The theory proceeds by forming the nuclear mean field $U$ through the
Hartree-Fock method employing the $G$-matrix, and the first
approximation to the wave function is the anti-symmetrized product of
single particle wave functions engendered by $U$. However, the
complete nuclear wave function is obtained in a perturbative hole-line
expansion that includes two-particle -- two-hole excitations and other
excitations which incorporate correlations. The presence of such
correlations is demanded by the theory.  

Later work formulated a
relativistic version of Brueckner theory in which the Dirac equation
replaces the Schroedinger
equation~\cite{Anastasio:1984gy,Brockmann:1984qg}. There is also a
light front version~\cite{Miller:1999ap,Miller:2000kv}.

The Brueckner theory approach described above presumes that the
two-nucleon potential contains strong short-distance repulsion.  Early
attempts to construct soft potentials (i.e., lacking the strong repulsion)
that also reproduce scattering data did not succeed in obtaining
interactions that could be used perturbatively to calculate nuclear bound
states~\cite{Bethe:1971xm}.  This failure is  now known to be caused in large measure by  the 
omission of three-body forces. Relativistic $G$-matrix calculations include important three-body forces~\cite{Anastasio:1984gy,Brockmann:1984qg,Miller:1999ap,Miller:2000kv}. There are
also  fundamental three-nucleon  forces, such as those involving an intermediate $\Delta$ resonance. In addition to true three-body forces, 
induced multi-nucleon forces occur as a  result of  using unitary transformations to produce soft, two-nucleon interactions~\cite{Bogner:2009bt}.

Much more has been learned since Bethe's 1971 review.
 (1) Our  understanding of the connection through symmetries
between the $NN$ interaction  and the underlying theory of QCD is much improved. (2) Our ability to make fundamental first-principles calculations of nuclear  energies is also much improved. (3) However, it is possible that  improved treatments of nuclear
energy levels decrease our ability to understand  the nuclear high-momentum transfer interactions of interest in this review. (4) We now know that 2$^{nd}$ order interactions of the $NN$ potential have a major effect on the density distribution and the correlation function in all existing approaches.

(1) Chiral effective field theory provides a low-energy version of QCD, guided by chiral symmetry, in which one obtains the potential as an expansion in powers of $(Q/\Lambda_\chi)$ where $Q$ is a generic external momentum or the pion mass,  and $\Lambda_\chi $ is the chiral symmetry breaking scale of about 1 GeV.
Such approaches have the advantage of being systematically improvable for low-energy observables. 
 See for example the review~\cite{Bedaque:2002mn}. In such theories the short distance interaction can be treated as a contact interaction, modified by the inclusion of a cut-off, and the longer ranged interactions are accounted for by one and two pion (or more) exchange interactions ~\cite{Machleidt20111}.  The advantage gained is that different parts of the potential are divided between more easily understood long ranged contributions and presumably unknown short-ranged contributions.

(2) Modern first-principles calculations of nuclear spectra have been applied to an ever increasing mass range.  One of the main tools is the use of soft potentials, which do not connect low-relative momentum states to those of high relative momentum.  This greatly simplifies the calculations by increasing the validity of perturbation theory and other approximation techniques.

The softness (involving low momentum) or hardness
(involving higher momentum) of the potential is determined by the
value of the cutoff see {\it e.g.}  \cite{Machleidt20111,Epelbaum:2008ga}. Such  potentials introduce a cutoff in momentum space at fairly low values of momenta.
Typically, the momentum-space potential obtained from Feynman diagrams,  $V(\bfp,\bfp')$, is replaced:
\bea V(\bfp,\bfp')\rightarrow V(\bfp,\bfp')   e^{-({p'\over \L})^n}e^{-({p\over \L})^n}\label{cut} 
\eea
with $p=|\bfp|,\,p'=|\bfp'|,$  
 $\L$ ranges between 400 and 500 MeV and $n$ ranges from 2 to 4. These are very strong cutoffs in momentum that introduce significant non-locality to the nucleon-nucleon interaction. This causes difficulties in maintaining conservation of the electromagnetic currents~\cite{Gross:1987bu}.
 
Another approach uses renormalization group methods to generate a
soft $NN$ potential from a hard interaction either by integrating out
high momentum components (in the case of $V_{low-K}$), or by using the
similarity renormalization group~\cite{Bogner:2009bt}.  This potential is perturbative in the sense that the Born
series for scattering converges. Furthermore, many-body perturbation theory starting from a Hartree-Fock bound state can be applied
to the nuclear bound state problem.


(3) But there is another more general issue that arises in trying to understand high momentum transfer nuclear reactions.  The ability to originate and predict the  results of experiments that probe short-ranged 
correlations (as was done in ~\cite{Frankfurt81,Frankfurt88}) depends on the idea that the simple  impulse approximation is the best  way to think about the relevant kinematics and reaction physics. This simplicity 
may be lost if one uses  dynamics generated by the different intent of simplifying nuclear spectroscopy. We explain.
Let us suppose that the renormalization group successfully  eliminates matrix elements of
the nucleon-nucleon (or inter-nucleon)  potential connecting low and high relative
momentum states, leading to an accurate reproduction of 
nuclear binding energies and spectra. This procedure  would also lead to wave
functions without high-momentum components and truly short
ranged-correlations.  However, it would be necessary to consistently
transform all other operators~\cite{Anderson:2010aq,Neff:2015xda} in order to calculate observables. For high momentum transfer reactions,
the renormalization group changes a known simple probe, described by a
single-nucleon operator, into a more complicated probe describable by
unknown (in practice) $A$-nucleon operators. This could prevent the efficient 
analysis of any high momentum transfer experiment. 
The same remark holds for  chiral potentials.  The use of a cutoff, as in \eq{cut}, 
 leads to the violation of current conservation in electromagnetic interactions unless the currents are modified
 substantially. For example, one could use minimal substitution, which would introduce terms involving several powers of the electromagnetic potential $A^\m$.  This means that the simplicity  of using electromagnetic probes  would be lost because of the need to use very complicated operators to analyze experiments. Again we reach the same conclusion: the use of  potentials with strong 
 momentum-dependence is not  optimum for the purpose of using  high momentum transfer electromagnetic processes  to understand the short-range structure of nuclei.

It is worthwhile to put comments (1)-(3) into a broader perspective. The goal of EFT is to obtain results that are independent of the chosen cutoff. In principle, this can be done. In practice, one chooses a given scale to simplify the problem at hand.  The use of low momentum scales  simplifies nuclear structure calculations, but complicates the currents needed to understand high-momentum transfer reactions. The use of one-body currents of the impulse approximation simplifies the understanding of high-momentum transfer nuclear reactions, but involves NN potentials that do not have low-momentum cutoffs.  Bjorken scaling, \cite{Bjorken:1966jh} obtained via the use of the simple currents of the non-interacting quark model (impulse approximation)  offers a useful historical example. If Bjorken had been overly concerned with issues of QCD evolution, Bjorken scaling and the existence of quarks might never have been discovered. Therefore,  we take the experiment-based, discovery-based   view that we are using an implicit  momentum scale at which the impulse approximation offers  a reasonable first approximation to the physics at hand throughout this review. 
 
(4)  Second-order effects of the tensor term of the one-pion exchange potential are common to 
all of these approaches, since the beginning  \cite{Brown,Bethe:1971xm,Machleidt:1989tm,Bogner:2005sn,Holt:2013fwa} and through to the current days of effective field theory.
 These effects are large enough to cause convergence difficulties in the application 
of Brueckner theory~\cite{Vary:1973zz}, and also cause challenges in defining the  power counting which defines any effective field theory~\cite{Bedaque:2002mn}.

The effect of this on the relative $s$-wave function of two nucleons in nuclei can
be characterized by the effective potential
\bea
V_{00} = V_T {1\over E-H_0}QV_T,\label{V00}
\eea
where $V_T$ is the tensor potential,  the subscript $00$ indicates an s-wave to s-wave matrix element, $H_0$ is the Hamiltonian in the
absence of $V_T$, and $Q$ is a projection operator taking the Pauli
principle into account. The operator $V_{00}$ has a major effect on the density distribution and correlation function 
(as discussed in the Appendix). These  effects occur in {\it all} existing approaches. A major purpose of this review is to show that the influence of the  correlations induced by the tensor force is manifest in high momentum transfer reactions.

To summarize, nuclear theorists have made tremendous progress in understanding the connections between $NN$ potentials and QCD, as well as in calculating nuclear energies and states.
High momentum transfer experiments are easier to analyze using well-defined current operators, rather than using transformed $A$-nucleon operators with a renormalization-group-transformed potential. These well-defined current operators can be used if the effects of correlations are maintained in the nuclear wave function instead of being hidden in the current operators through the use of the renormalization group or very soft $NN$ potentials. However, regardless of approach, the influence of the correlations induced by the tensor force is manifest in all theoretical approaches to date, and, as we shall see, is manifest in high momentum transfer reactions.

\subsubsection{Short-ranged two-nucleon clusters  \label{phen}}



As discussed in previous Sections,  in the nucleus,  nucleons behave  approximately  as independent particles in a mean field created by their average interaction with the other nucleons. But occasionally ($20-25\%$ in medium/heavy nuclei) two nucleons get close enough to each other so that temporarily their singular short range interaction cannot be well described by a mean field approximation. {\bf These are the two nucleon short-ranged correlations ($2N$-SRC), defined operationally in experiments as having
  small center of mass momentum and large relative momentum.}  These pairs are predominantly 
 neutron-proton pairs. Colle et al. \cite{colle15} show that it is predominantly nucleon-nucleon pairs in a nodeless relative-$S$ state of the mean-field that create these 2N-SRC. The force between the nucleons in the pair is predominantly a tensor force which creates a pair with the quantum numbers of the deuteron ($S=1, T=0$), a neutron-proton system~\cite{Vanhalst:2011es}.

 The two nucleons in $2N$-SRC have a typical distance of about $1$ fm which means that their local density is a few times higher  than the average nuclear density. The relative momentum of the two nucleons in the pair can be a few times the Fermi momentum, $k_F$, which is large. 
  SRC of more than two nucleons probably also exist in nuclei, and might have  higher density than that of the  $2N$-SRC.  However their probability is expected to be significantly smaller than the probability of $2N$-SRC~\cite{Bethe:1971xm}. 

 The $2N$-SRC are isospin-dependent fluctuations.  For example, the deuteron is the only bound two-nucleon system.  We know now that density fluctuations involving one neutron and one proton occur more often than those involving like-nucleons, see  Sect. IIC.  Therefore we examine the deuteron  first.


\begin{figure}
\includegraphics[width=8cm, height=8cm]{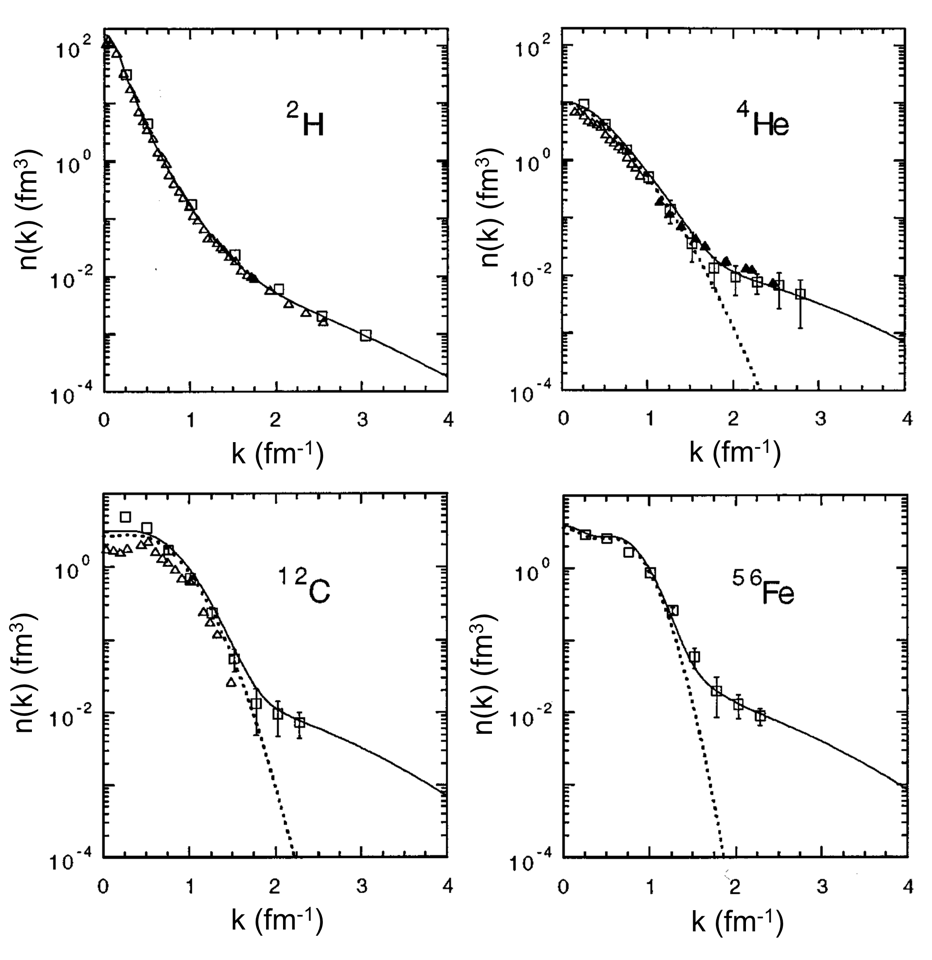}
\caption{\label{fig:nk_ciofi} The nucleon momentum distributions $n_0(k)$ (dashed line) and $n(k)$ (solid line) plotted versus momentum in fm$^{-1}$ for the deuteron, $^4$He, $^{12}$C and $^{56}$Fe.  Adapted from \cite{cda96}.}
\end{figure}

The simplest nucleus, the deuteron,  has spin $S=1,$ isospin $T=0,$ and $J^\pi=1^+$. The relevant quantity for electron scattering is $n(k)$ which is the probability of finding a nucleon of momentum between $k$ and $k+dk$. This function is the sum of  two terms, one arising from the $l=0$ ($s$-wave), and the other from the $l=2$ ($d$-wave).  At momenta of interest for short range correlated pairs (i.e., $p$ significantly greater than $p_{F}\approx 250$ MeV/c, where $p_{F}$ is the typical Fermi momentum for medium and heavy nuclei), the otherwise-small $d$-wave becomes very important. This is especially true at   $p\approx 400$ MeV/c where there is a minimum in the $s$-wave.  In the Argonne V18 potential \cite{wiringa14} the $d$-wave component is due to the  tensor force.  The combination of $d$- and $s$-waves leads to a ``broad shoulder'' in the deuteron momentum distribution, which extends from about 300 to 1400 MeV/c in the AV18 potential. See Sect.~\ref{Appendix} for an explanation. This broad shoulder is also a  dominant feature in the tail of  the single-nucleon momentum distributions computed with realistic internucleon interactions,  see Fig.~\ref{fig:nk_ciofi}, in particular 
 with the AV18 potential  for  $A\le 12$ \cite{wiringa14} and more effective approaches for heavier systems~\cite{CiofidegliAtti:1995qe,ryckebusch15}.

\begin{figure}[t]
  \centering
    \includegraphics[width=8cm, height=6cm]{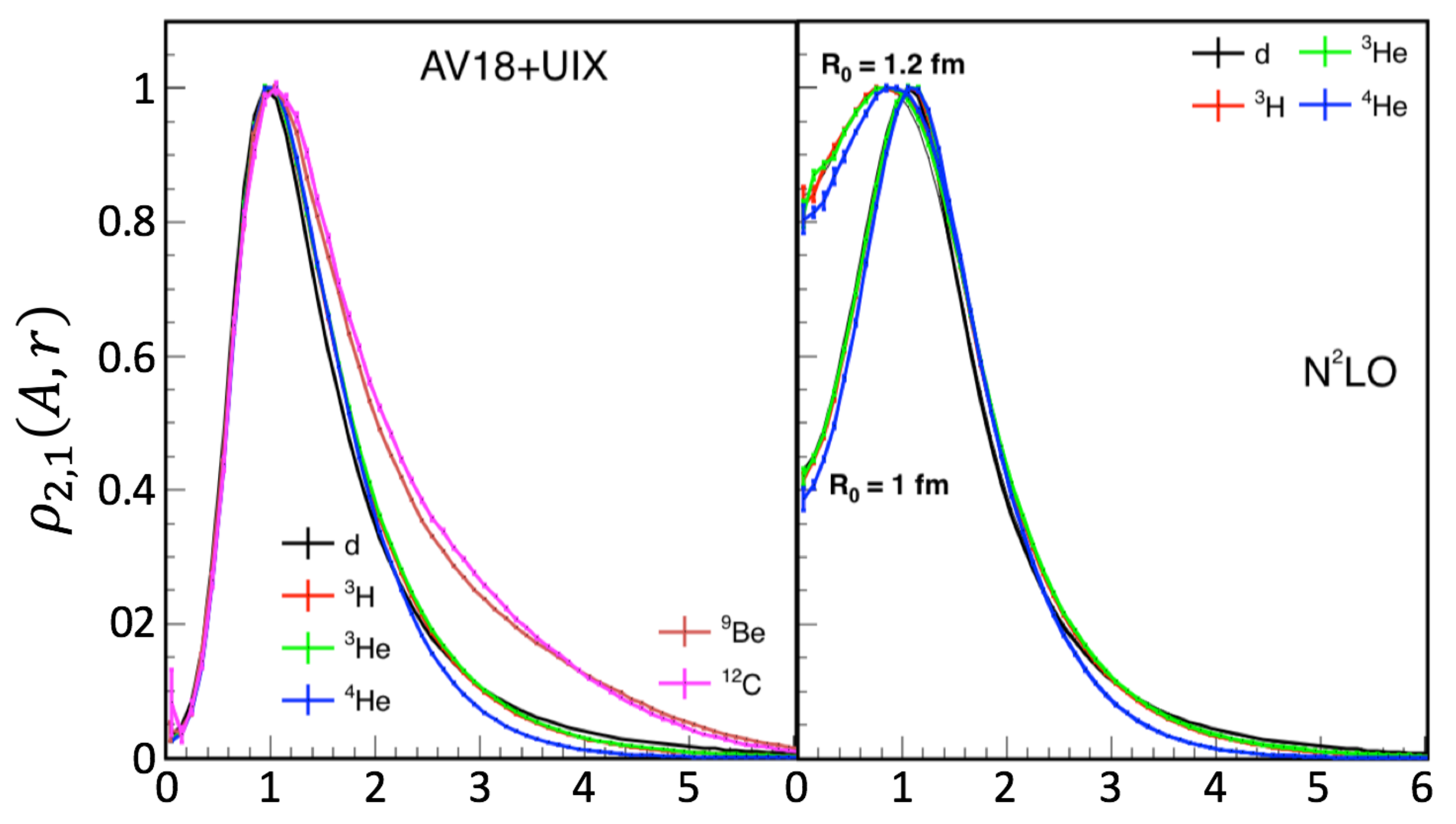} 
    \caption{ Scaled two-body distribution function $\rho_{2,1}^A(r)/A$  (see \eq{r21}) for nuclei with $A=2,3,4$.  A correlation hole is seen for all of these nuclei.  The two sets of curves are obtained with the AV18+UIX (left) and N$^2$LO (right) potentials.  
   Figure reproduced based on~\cite{Chen:2016bde}. The meaning of $R_0$ is discussed in the text. }
     \label{fig:CorrelationHole} 
\end{figure}

We can also consider the spatial wave function of the nucleus.
The short range part of the $NN$ interaction gives a correlation hole at small $NN$ relative distances, see Fig.~\ref{fig:CorrelationHole}.  Precise definitions are given in Sect.~\ref{Appendix}. Calculations with various bare realistic interactions show that, apart from a   normalization factor depending upon the different number of pairs in different nuclei, the relative two-nucleon density $\rho_{rel}(r)$ and its spin-isospin components $\rho_{ST}^{N_1N_2}(r)$ at $r\le1.5$ fm exhibit similar correlation holes, generated by the interplay of the short-range repulsion and the intermediate-range tensor attraction of the $NN$ interaction, with the tensor force governing the overshooting at $r \simeq 1.0$ fm. The correlation hole is universal, in that it is almost independent of the mass $A$ of the nucleus \cite{ciofi15}.  The depth of the correlation hole depends on the short-distance behavior of the potential. The value of $R_0$ shown in  Fig.~\ref{fig:CorrelationHole} refers to  the cutoff  on the short distance N$^2$LO nucleon-nucleon potential, as defined in~\cite{Gezerlis:2014zia}. A correlation hole is seen to occur for $R_0=1$ fm, but is much less deep for $R_0=1.2 $ fm. The use of such a soft potential is  not suitable in the present experiment-based high-scale context. Furthermore, this soft potential predicts erroneous nucleon-nucleon phase shifts for the $^3D_1$ partial wave, and also for lab energies greater than 250 MeV.  



In momentum space, the existence of this universal correlation hole translates into nucleon momentum distributions $n_A(p)$ that are significant at large momentum ($p\ge p_F$) and that are similar for all nuclei, $n_A(p) \propto n_d(p)$, at these large momenta \cite{Frankfurt81,Frankfurt88,cda96,alvioli13}.  Frankfurt and Strikman realized that these could be measured with hard probes (see Section \ref{HardScattering}).

Ciofi degli Atti and Simula \cite{cda91,cda96} used this similarity to model  the nucleon spectral function $P(\bfp,E)$ (the joint probability to find a nucleon in a nucleus with momentum $\bfp$ and removal energy $E$) for all nuclei 
  \bea  P(\bfp,E)=\la \Psi| b^\dagger(\bfp)\delta(E-H)b(\bfp)\Psi\ra,\eea
  where $|\Psi\ra$ represents the nuclear wave function and spin, isospin and nuclear ($A$) labels are suppressed for simplicity.
  The momentum density $n(\bfp)$ is given by
  \bea n(\bfp)=\int dE  P(\bfp,E).\eea
   These authors write 
  \bea P(\bfp,E)=P_0(\bfp,E) +P_1(\bfp,E)\label{eq:01}\eea
  where the subscript zero refers to  values of $E$ corresponding to low-lying intermediate excited states and the subscript one refers to high-lying continuum states that are caused by the short-ranged correlations. Therefore one also has $n(\bfp)=n_0(\bfp)+n_1(\bfp)$, where $n_1(\bfp)$ is associated with   the high momentum 
caused by short-ranged correlations.
 $n_0(p) $ is typically dominant for $p< 250$  MeV/c or so and $n_1(p)$ becomes dominant for larger values.  Furthermore,  $n_1(p)$ is almost independent of $A$ at $p>400$ MeV/c; they attribute this to $NN$ correlations.  See Fig. \ref{fig:nk_ciofi}. 
It should be noted that   for $^3$He Ref.~\cite{cda91}   showed that the proposed model spectral function agrees with the one obtained by direct calculation. 

SRC pairs are conventionally defined in momentum space as a pair of nucleons with high relative momentum and low center of mass (c.m.) momentum, where high and low are relative to the Fermi momentum of medium and heavy nuclei. Thus the most prominent effect of SRC will be to populate high-momentum states in the nuclear momentum distribution. As conventional mean-field theories predict only a very small high-momentum tail, the effect of SRCs there should be substantial.
Formally, one needs 
the two-nucleon momentum density, $n(\bfp_1,\bfp_2)$  (see Section \ref{Appendix}),  where $\bfp_{tot}=\bfp_1 + \bfp_2$ and $\bfp_{rel}=\frac12(\bfp_1 - \bfp_2)$ are the center of mass and relative momenta of the two nucleons.  Studies of spectral functions show that  at large values of $p_{rel}$, the two-nucleon momentum density factorizes:
\begin{equation}
n(\bfp_{tot},\bfp_{rel}) = n(p_{tot})n(p_{rel}).
\end{equation}
A justification of this factorization is presented in Sect.~\ref{Appendix}.

The coordinate-space  correlation holes (Fig.~\ref{fig:CorrelationHole}) give similar $NN$ relative ($p_{rel}$) momentum distributions (at large $p_{rel}$ in all nuclei.  Exact calculations with the AV18 potential for $^4$He, show that, at small $p_{tot}$ there is a minimum in $p_{rel}$ for $pp$ pairs at $p_{rel}=400$ MeV/c.  This is because, at small  $p_{tot}$, the $pp$ pair must be in a relative $s$-state which has a minimum at $p_{rel}=400$ MeV/c, just like in deuterium.  For $np$ pairs, this minimum is filled in by the $d$-wave caused by the short range pion-exchange tensor force \cite{wiringa14}.

Thus, the combination of the minimum in the $s$-wave momentum distribution at $p\approx 400$ MeV/c and the filling in of this minimum by the $d$-wave pion-exchange tensor force, leads to the expected dominance of $np$ correlated pairs over $nn$ and $pp$ pairs at $300\le p\le 500$ MeV/c.  This ratio of $np$ to $pp$ pairs should decrease at relative momentum significantly greater than 400 MeV/c, the $s$-wave minimum (as we will discuss  Section \ref{HardScattering}).
   

Short-range correlations in light nuclei have been examined recently theoretically from several points of view~\cite{Feldmeier:2011qy,wiringa2014,Rios:2013zqa,Atti:2015eda,Weiss:2015mba,ryckebusch15,vanhalst12,Vanhalst:2011es}. One consistent finding of such work is the dominance of $np$ deuteron-like pairs ($ST=10$) over other pairs at high momentum.

These facts described in this sub-section  lead to an effective description of nuclei in momentum space as having two important regions: (1) a mean-field region ($k\le p_F$), which accounts for about $80\%$ of the nucleons, where the many-body dynamics result in single nucleons moving under the influence of an effective potential created by the residual $A-1$ system and (2) a high-momentum region ($p\ge p_F$), which accounts for about $20\%$ of the nucleons (but 70\% of the kinetic energy ~\cite{Benhar:1989aw,Polls:1994zz}), where nucleons are predominantly in the form of $pn$-SRC pairs, having a very weak interaction with the residual $A-2$ system.  As noted above, it is possible to use unitary transformations to derive a low-momentum effective interaction that weaken the strong short-ranged correlations  present in the original interactions. However, the one and two body  density operators also need to be transformed. It is necessary to include three or more body effects  to obtain accurate results with these soft interactions~\cite{Feldmeier:2011qy}. This approach  complicates the analyses of experiments.


To summarize, the high momentum nucleons in nuclei are mainly due to 2N-SRC and are therefore associated with high density fluctuations in the nucleus. In what follows (see Section \ref{SRCvsMF})
 we will examine the hypothesis that these temporary high density/large momentum `hot spots' are the sites where the nucleon internal structure is modified and the EMC effect is created.  First, we will present the experimental evidence for short range correlations.




\section{Hard scattering and Short-Range Correlations \label{HardScattering}}

\subsection{Hard Reactions \label{HardReactions}}
In optics the resolving power is the minimum distance at which an imaging device can separate two closely spaced objects. This is normally proportional to the wavelength of the light. The smaller the wavelength, the better the resolution.

We often scatter particles to try to resolve the internal  structure of a complex target. The sizes of the target and its constituents define the required resolving power. For example, to observe the nucleus of an atom one needs a spacial resolution of about 10 fm, to observe nucleons in nuclei one needs a resolution of about $1$ fm, and to observe the partonic structure of a nucleon one needs sub-fermi resolution.

The spacial resolution of a scattering experiment is determined by the de Broglie wave length ($\lambda$) of the probe (scattering particle) and the momentum transfer of the reaction ($q$). We define as 'hard' a process that fulfills the following conditions: $\lambda \ll R$ and $q  R \gg 1$, where $R$ is the size of the target or the structure to be studied. In practice, we shall see that the results of  measurements  can be interpreted as observing a hard reaction even though these kinematic conditions are not always rigorously met.  

Another important lepton-scattering length scale is the coherence length, or Ioffe length~\cite{Ioffe:1969kf,Gribov:1965hf}: $l_I={2\over M\xB}\approx{0.4\,{\rm fm}\over \xB},$ where $\xB={Q^2\over 2M \nu}$. Here $M$ is the nucleon mass, $Q^2$ is the negative of the square of the virtual exchanged photon four-momentum, and $\n$ is its energy. This length is the typical distance between the absorption and re-emission of the virtual photon. This length must be short enough to resolve the relevant inter-nucleon distance scales of the order of a fermi. Thus, we will focus on the region $\xB>0.3$ where valence quarks are dominant and the sea is almost invisible.
 
In this paper we are dealing with two reactions and the connection between them. Deep inelastic scattering (DIS) attempts to resolve the partonic structure of nucleons and quasielastic scattering  (QE) attempts to resolve the nucleonic structure of nuclei. These reactions have different required resolutions and hence different kinematical conditions  to achieve them. 

For $(e,e')$ DIS reactions, which are typically measured as a function of  $\xB=Q^2/2M\nu$ for $\xB<1$, there are two important parameters, 
the $4$-momentum transfer squared of the virtual photon, $Q^2$, and the invariant mass of the virtual photon plus struck nucleon, $W = \sqrt{M^2 + 2M\nu - Q^2}$. Since $\xB, Q^2$, and $W$ are all functions of the same two variables, only two   are independent. For the inelastic scattering to be considered deep (the ``D'' in DIS), experiments typically require $W\ge 2$ GeV.  This allows the experiments to be sensitive to the internal structure of a proton or neutron and avoid   the influence of  individual nucleon resonances, which  cause the cross section to fluctuate rapidly with $W$. 

Early studies at the high energy facilities (SLAC and CERN) measured DIS for $5 \le Q^2 \le 50$ GeV/c$^2$ and found  that the ratios of DIS cross sections for $0.3\le \xB \le 0.7$ are largely independent of $Q^2$ \cite{Norton03}. The newer JLab experiments used lower lepton energies (typically $4-5$ GeV) and therefore lower $Q^2$, $4\le Q^2 \le 6$ GeV$^2$ \cite{Seely09}.  The higher-energy SLAC and CERN measurements required $W\ge 2$ GeV. However the lower-energy JLab data required only $W\ge 1.4$ GeV.  

For inclusive $(e,e')$ QE scattering, there are again only two independent kinematical variables,  normally chosen to be $Q^2$ and $x_B$. However, in addition to making sure that the resolving power is sufficient, we also need to optimize the kinematics to select scattering from high-momentum nucleons in the nucleus and to reduce the effects of non-single-nucleon currents.  In order to resolve nucleons in SRC pairs, measurements are typically made at $Q^2>1.5$ (GeV/c)$^2$. Large ($p>p_F$)  minimum initial momentum of the struck nucleon (assuming no final state interactions) can be selected at $Q^2 >1.5$ (GeV/c)$^2$ by choosing either $\xB \ge 1.5$  or  $\xB \le 0.6$ (see Section \ref{Inclusive}).   $\xB \geq 1$ is preferred, so that the energy transfer is smaller, inelastic processes (resonance production, meson exchange currents [MEC] and isobar configurations [IC]) are suppressed, and the reaction is more sensitive to the nuclear momentum distribution. Increasing $Q^2$ further suppresses MEC contributions. The inclusive QE scattering data discussed in Section \ref{Inclusive} were measured at $\xB\ge 1.5$. 

In exclusive and semi-exclusive reactions, $(e,e'p)$ and $(e,e'pN)$, large initial nucleon momenta can be selected directly and the \xB{} restrictions can be relaxed (see Section \ref{Exclusive}).

\subsection{Exclusive Scattering \label{Exclusive}}
The study of SRCs using exclusive reactions has a long history that extends beyond the scope of this review. Here we focus only on exclusive measurements performed with high energy probes and large momentum transfer (hard reactions). See~\cite{kelly96} and references therein for a review of the older measurements. We   use the term exclusive to refer to measurements in which, in addition to the scattered probe particle,  two knocked-out nucleons are measured in the final state.

In the context of SRC studies, exclusive reactions are hard processes in which a probe scatters from one nucleon in an  SRC pair and all particles emitted in the final state (e.g., the scattered probe and both nucleons of the pair) are detected. The energy of the probe and the momentum transfer must be  large enough so that the probe interacts with a single, high-momentum ($p_i>p_F$) nucleon in the pair. If the pair was at rest ($p_{cm}=0$) and neither nucleon rescattered as it left the nucleus, then the struck nucleon's correlated partner would recoil with momentum $\bfp_2=-\bfp_i$.  This back-to-back angular correlation between the initial momentum of the knocked out nucleon and the momentum of the recoil nucleon is a clear experimental signature for exactly two nucleons being involved in the interaction. We   note that these reactions can be analyzed in terms of the decay function introduced by \cite{Frankfurt88}.

However, other reaction mechanisms can also involve two nucleons, leaving the residual $A-2$ nucleus almost at rest.  The probe can scatter from one nucleon, which can rescatter from a second (FSI), the probe can scatter from a meson being exchanged between two nucleons (MEC), or the probe can excite the first nucleon which can then de-excite via interaction with a second nucleon (IC).  Disentangling these competing and interfering effects can be difficult.  It is important to realize that  the effects of MEC and IC are  dramatically decreased by choosing kinematics with $\xB>1$ and with larger values of $Q^2$.  The effects of FSI can also be dramatically decreased by (a) choosing kinematics where the relative momentum of the two final-state nucleons is large and (b) avoiding kinematics where the opening angle between the two outgoing nucleons is $70-90^\circ$.  (Non-relativistically, when one billiard ball scatters from a second billiard ball at rest, the opening angle in the final state is $90^\circ$.)  


The detection of the outgoing nucleons in exclusive reactions provides complementary information to the inclusive reactions discussed \st{above} {\bf below}.  By detecting the struck nucleon at large $p_{miss}$ and looking for the recoil partner nucleons, exclusive measurements can measure the fraction of high-momentum nucleons belonging to SRC pairs. They can also extract information on the SRC pair isospin structure and $p_{cm}$ distribution, as well as their $A$ and momentum dependence.

This additional information however comes at the price of increased sensitivity to FSI. FSI can be generally split into two main contributions: re-scattering between the nucleons of the pair, and re-scattering between the nucleons of the pair and the residual $A-2$ system. Rescattering between the nucleons of the pair will alter the measured relative momentum but leave $p_{cm}$ unchanged.  Rescattering between the nucleons of the pair and the residual $A-2$ system  will change the momentum of the outgoing nucleons and  ``attenuate'' them.  The attenuation of the nucleons as they traverse the nucleus is usually referred to as the 'nuclear transparency' and limits the spatial region probed in the experiment to the outer part of the nucleus. It can be calculated in the Glauber approximation (for large enough nucleon momentum).  The momentum changes also affect the measured kinematical distributions. Here the use of high momentum transfer, as required for hard reactions, also allows using the Glauber approximation to calculate to the effects of FSI and to select kinematics to minimize their effects, either in the measured cross sections or the kinematical distributions. 

Specifically, at $Q^2\ge 1.5-2$ (GeV/c)$^2$ and $x_B\ge1$ (or proton scattering experiments at $|t|, |u|, |s| \ge 2$ GeV/c$^2$) Glauber calculations show that the outgoing nucleons predominantly rescatter from each other and not from the residual $A-2$ system \cite{Frankfurt81,Arrington:2011xs,frankfurt93,Frankfurt88,cda96}. This implies that certain quantities such as the total pair momentum, $p_{cm}$, and pair isospin structure are insensitive to rescattering while other quantities like the pair relative momentum, $p_{rel}$, are very sensitive to rescattering and thus cannot be reliably extracted from the experimental data, see~\cite{shneor07,Frankfurt:1996xx} for details. The contribution of Meson Exchange Currents (MEC) and Isobar Currents (IC)  are also minimized at high $Q^2$ and $x_B\ge1$. 

The first exclusive hard two nucleon knockout experiments, measuring the $^{12}$C$(p,2pn)$ and $^{12}$C$(e,e'pN)$ reactions, were done at BNL and JLab, respectively~\cite{tang03, Piasetzky:2006ai, shneor07, Subedi:2008zz}. These experiments scattered $5$ - $9$ GeV/c protons (BNL) and electrons (JLab) off high initial momentum ($300\le p_i \le 600$ MeV/c) protons in $^{12}$C and looked for a correlated recoil nucleon emitted in the direction of the missing momentum. The JLab experiment measured both proton and neutron recoils, whereas the BNL experiment only measured recoiling neutrons. Both experiments measured at large momentum transfer ($Q^2\approx2$ (GeV/c)$^{2}$), which suppressed competing reaction mechanisms and largely confined FSI to be between the nucleons of the pair.

\begin{figure}[tbp]
\includegraphics[width=7cm]{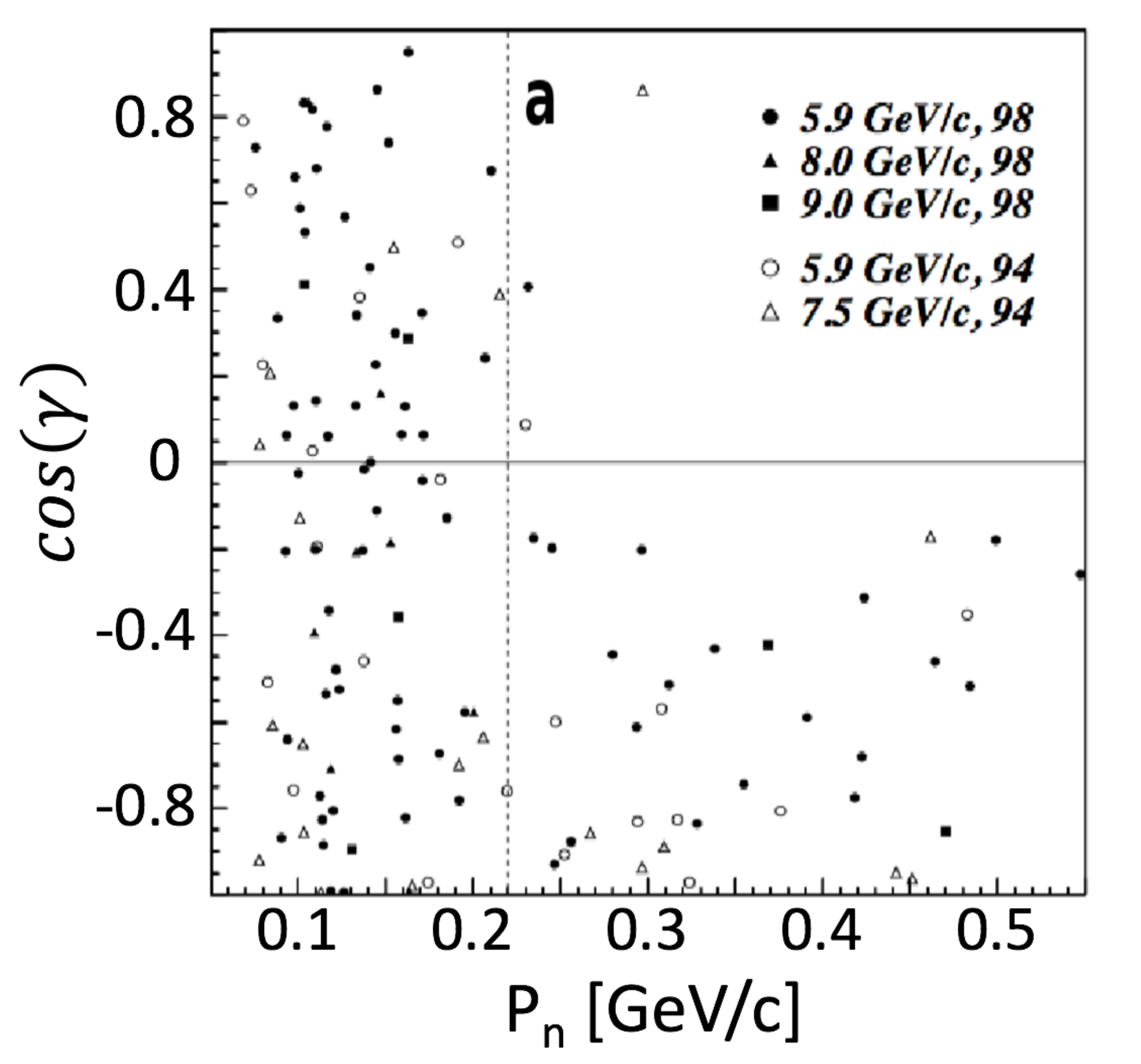}
\includegraphics[width=4.5cm]{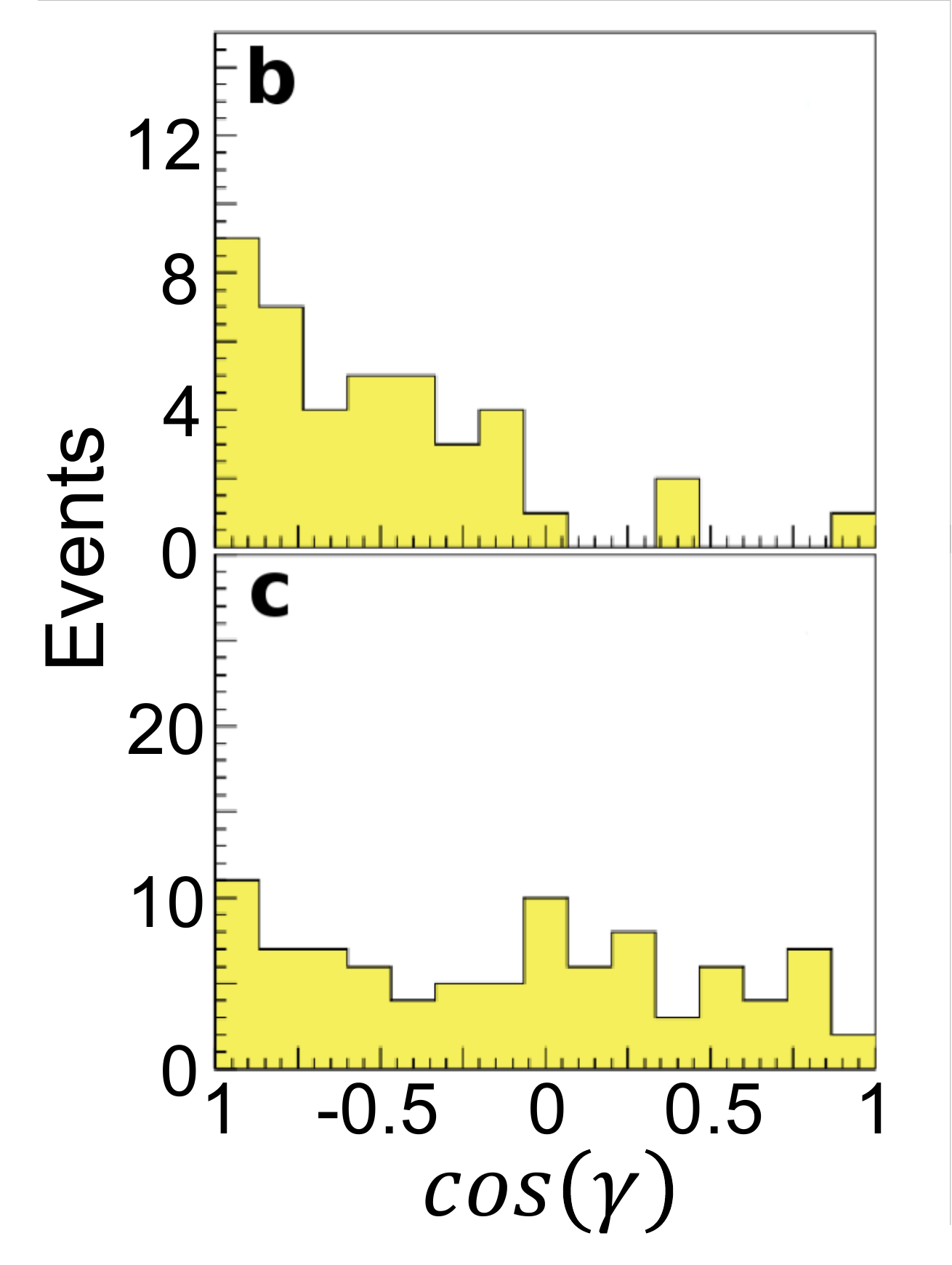}
\caption {\label{fig:OpeningAngle1} Distributions of the relative angle ($\gamma$) between the reconstructed initial momentum of the knockout proton and the recoil neutron.  Results for $^{12}$C$(p,2pn)$ events from BNL, shown as a function of the momentum of the recoil neutron (a) and for events with recoiling neutron momentum greater than (b) and less than (c) $k_F=225$ MeV/c.  Note the transition from an isotropic distribution to a correlated one at about $k_F=225$ MeV/c.  Figures adapted from \cite{tang03, Piasetzky:2006ai}.}
\end{figure}

\begin{figure}[tbp]
\includegraphics[width=3in]{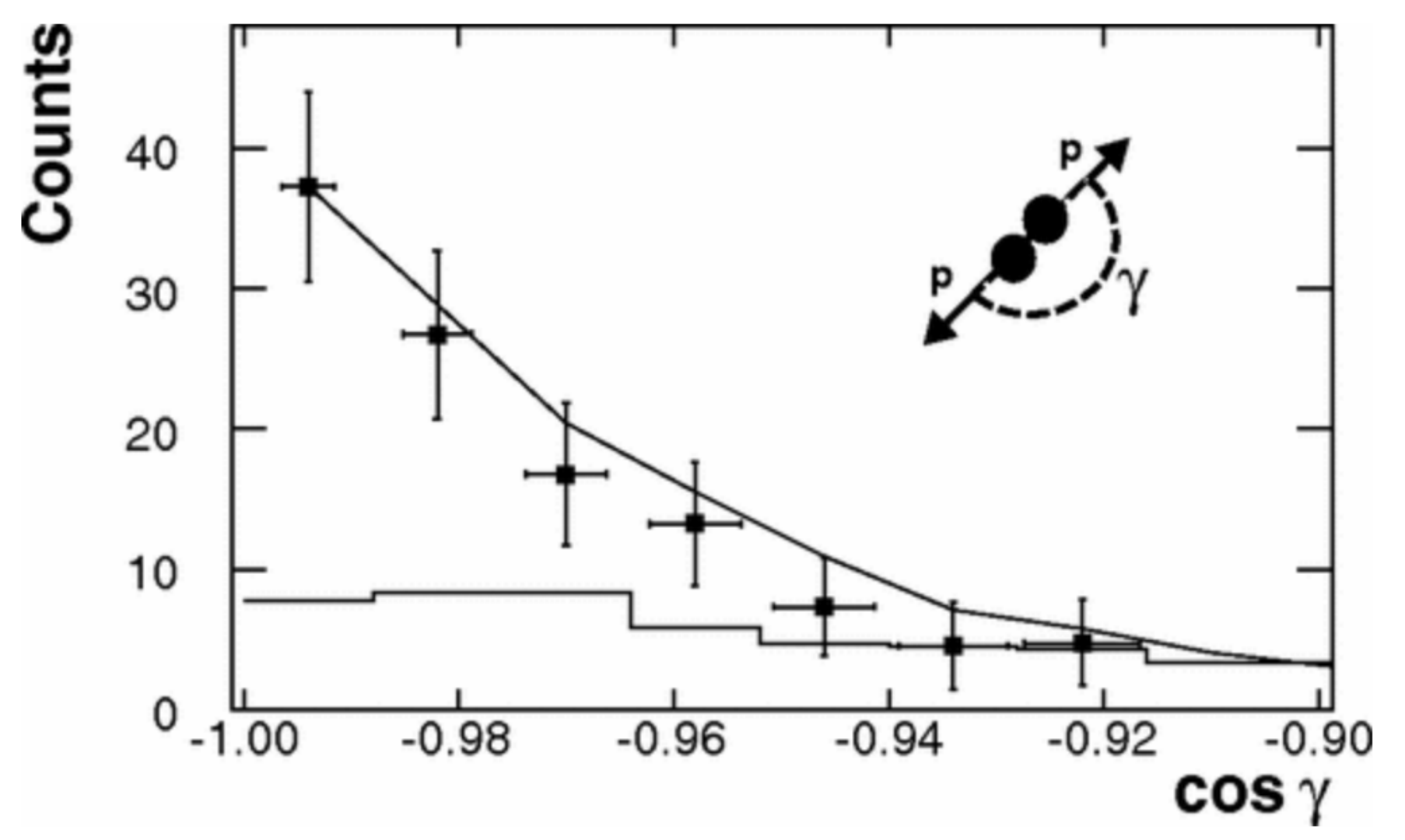} 
\caption {\label{fig:OpeningAngle2} Distributions of the relative angle ($\gamma$) between the reconstructed initial momentum of the knockout proton and the recoil nucleon.  Results for $^{12}$C$(e,e'pp)$ events from JLab at kinematics corresponding to scattering off $\sim500$ MeV/c initial momentum protons. Figure from \cite{shneor07}. }
\end{figure}

\begin{figure}[tbp] 
\centering 
\includegraphics[width=8.5cm]{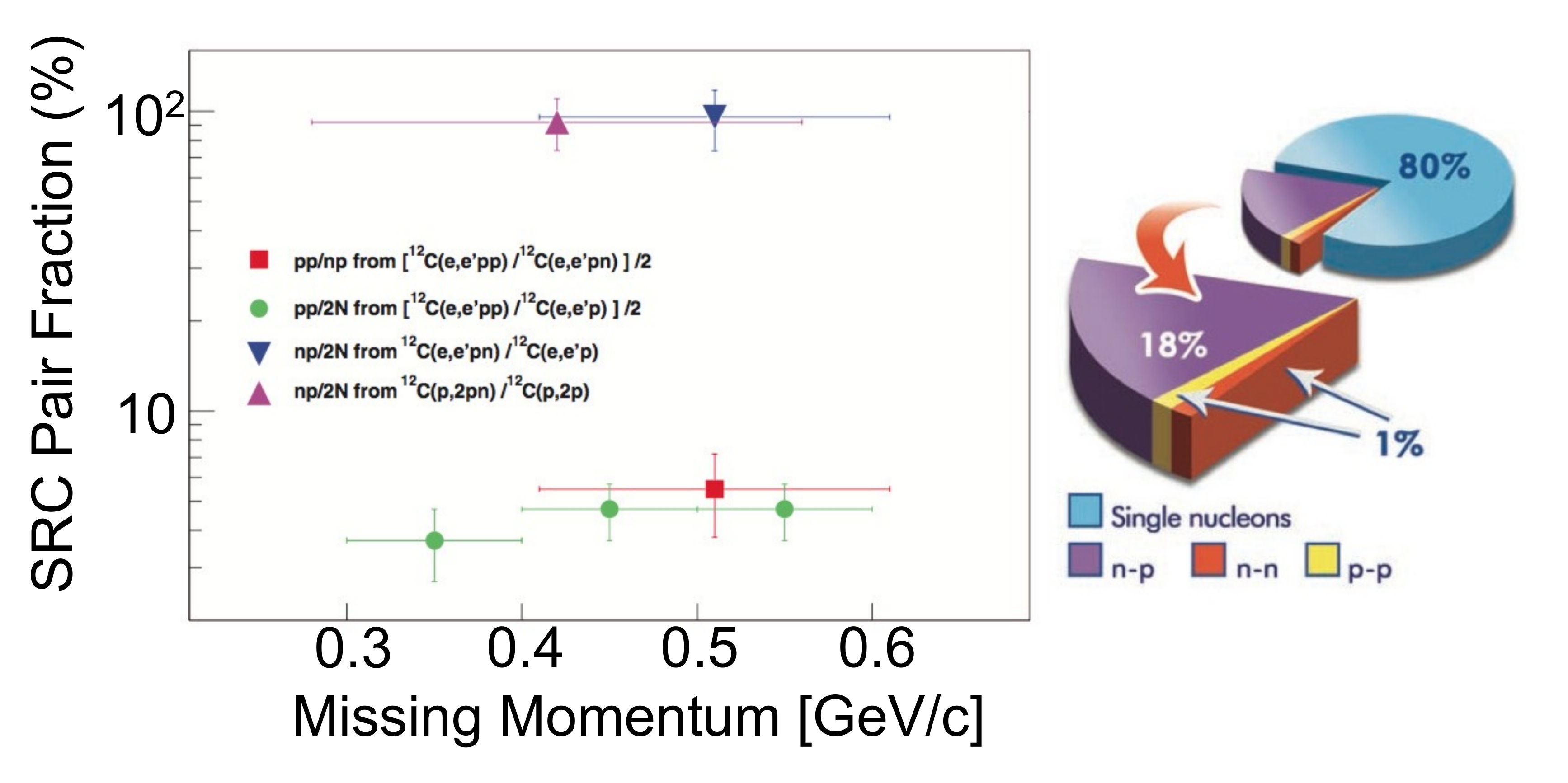} 
\caption{The ratio of $^{12}$C$(e,e'pN)$ double knockout events to $^{12}$C$(e,e'p)$ single knockout events, shown as a function of the reconstructed initial (missing) momentum of the knocked-out proton from the $^{12}$C$(e,e'p)$ reaction. Triangles and circles mark $^{12}$C$(e,e'pn)$and $^{12}$C$(e,e'pp)$ events, respectively. The square shows the $^{12}$C$(e,e'pp)/{}^{12}$C$(e,e'pn)$ ratio.  A clear dominance of ${}^{12}$C$(e,e 'pn)$ events is observed, evidence of the tensor nature of the nucleon-nucleon interaction in the measured momentum range. The pie chart on the right illustrates our understanding of the structure of $^{12}$C, composed of $80\%$ mean-field nucleons and $20\%$ SRC pairs, where the latter is composed of $\sim90\%$ $np$-SRC pairs and $~5\%$ $pp$ and $nn$ SRC pairs each. Figure adapted from \cite{Subedi:2008zz}. }
\label{fig:HallA_Ratios}
\end{figure}

The main results of the $^{12}$C measurements are shown in Figs.~\ref{fig:OpeningAngle1}, \ref{fig:OpeningAngle2}  and~\ref{fig:HallA_Ratios}. Figs.~\ref{fig:OpeningAngle1} and ~\ref{fig:OpeningAngle2} show the angular correlation between the momentum vector of the recoil nucleons and the reconstructed initial momentum of the knocked-out proton. For the BNL data, the angle is shown as a function of the recoil neutron momentum. Two distinct regions are visible: below the Fermi momentum where no angular correlation is observed, and above the Fermi momentum where a clear back-to-back correlation is seen. The width of the recoil nucleon opening angle distribution allowed extracting the pair c.m. motion; this motion can be described by a Gaussian distribution in each direction, with $\sigma=143\pm17$ (BNL) and $\sigma=136\pm20$ (JLab). These values are also in overall agreement with theoretical calculations~\cite{CiofidegliAtti:1995qe,Colle:2013nna}. The electron and proton reactions are characterized by completely different operators and FSI mechanisms; therefore the agreement of their c.m. momentum distributions validates the consistent treatment of FSI in these measurements. 

For example, for proton induced reactions the effective nuclear density  is smaller than for electron induced reactions due to absorption effects that prefers scattering from the edge of the nucleus. The overall agreement between the results obtained using different high energy hadronic and leptonic probes at very  different momentum transfer (2 GeV$^2$ and 5 GeV$^2$) strongly supports the interpretation that in these reactions the projectiles interact with one nucleon of the SRC. Note also that the saturation of the recoil channels by neutron and protons puts a   strong limit on the admixture of non-nucleonic degrees of freedom in SRCs.

Fig.~\ref{fig:HallA_Ratios} shows the extracted ratio of two nucleon knockout (proton-neutron and proton-proton)  to single proton knockout events and the ratio of proton-neutron to proton-proton two-nucleon knockout events.  The ratios are all corrected for finite acceptance effects and shown as a function of $p_{miss}$, the reconstructed initial momentum of the knocked out protons for $300\le p_{miss} \le 600$ MeV/c.  The ratio of two nucleon knockout  to single proton knockout is directly related to the fraction of high-momentum protons that are in SRC pairs. As can be seen, within statistical uncertainties of about $10\%$, all single nucleon knockout events at $300\le p_i \le 600$ MeV/c were accompanied by the emission of a recoil nucleon. The proton-to-neutron recoil ratio was found to be approximately 1:10, which corresponds to 20 times more $np$-SRC pairs than $pp$-SRC pairs in $^{12}$C~\cite{Subedi:2008zz}. This observed proton-neutron pair dominance was associated with the dominance of the tensor part of the nucleon-nucleon interaction at these initial moments~\cite{Sargsian:2005ru, schiavilla07}.

\begin{figure}[tbp]
\centering
\includegraphics[width=7.5cm]{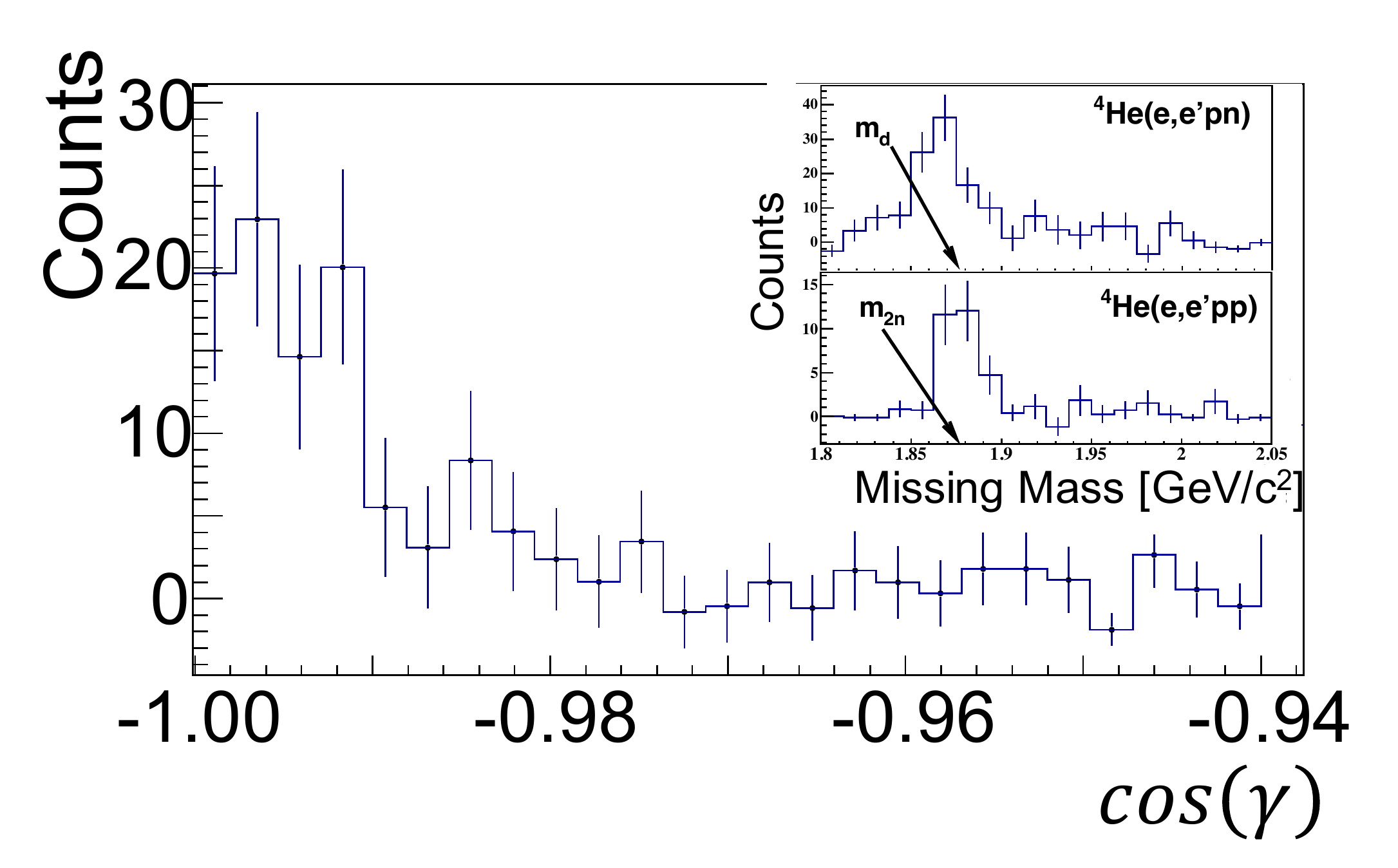}
\caption{(color online) The distribution of the cosine of the opening angle $\gamma$ between $\bfp_{miss}$ and $\bfp_{recoil}$ for the $^4$He$(e,e'pn)$ reaction. The solid curve is a simulation of scattering off a moving pair with a c.m. momentum distribution having a width of 100 MeV/$c$. The insets show the missing-mass distributions. 
Figure reproduced based on \cite{korover14}.}
\label{fig:HallA_4He_angle_mass}
\end{figure}

\begin{figure}[tbp]
\centering
\includegraphics[width=7.5cm]{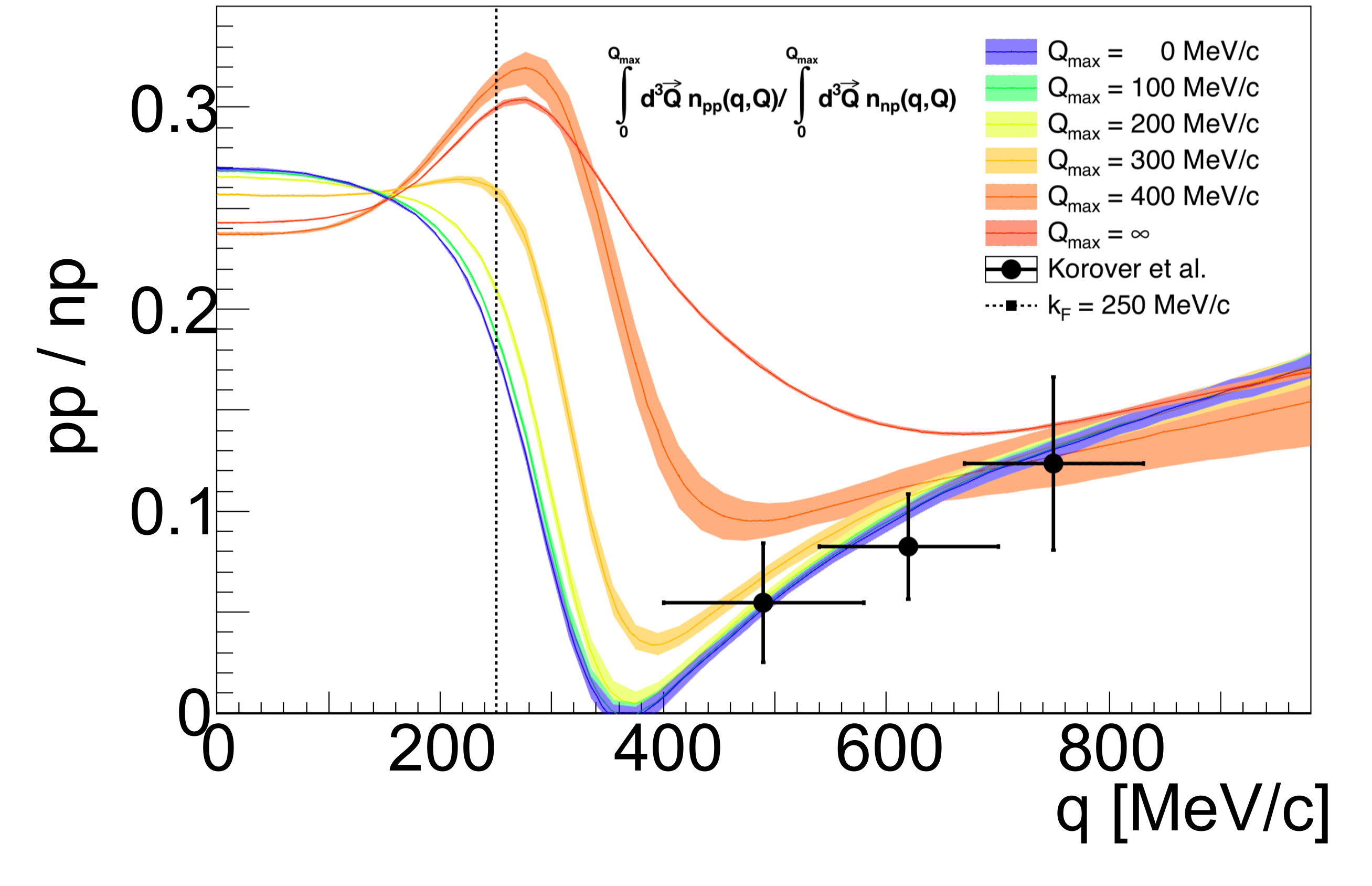}
\caption{(color online)  The measured $pp$ to $pn$ ratio as function of the proton missing momentum (labelled $q$) \cite{korover14} compared to calculations of the two-nucleon momentum distribution \cite{wiringa14} integrated over various ranges of the c.m. momentum \cite{Weiss:2016obx}.  The data is shown as a function of the nucleon momentum and the calculations are shown as a function of the pair relative momentum. The two are equivalent for low c.m. momentumof the pair but differ at large c.m. momentum. Figure adapted from~\cite{Weiss:2016obx}}
\label{fig:HallA_4He_np_pp_ratio}
\end{figure}

A follow-up measurement of $^4$He$(e,e'pN)$ in similar kinematics set out to better constrain the importance of the tensor part of the $NN$ interaction at short distance, and extend the experimental data to larger initial momenta, $400\le p_i \le 800$ MeV/c \cite{korover14}. At these higher momenta, the scalar repulsive core of the nucleon-nucleon interaction is expected to dominate over the tensor part, increasing the fraction of $pp$-SRC pairs.  The $^4$He nucleus was chosen to further reduce FSI and allow for comparisons with detailed ab-initio few-body calculations. The results of this measurement are shown in Figs.~\ref{fig:HallA_4He_angle_mass} and~\ref{fig:HallA_4He_np_pp_ratio}.

The two-nucleon opening angle distribution for $^4$He (see Fig.~\ref{fig:HallA_4He_angle_mass}) is very similar to that for C (see Fig.~\ref{fig:OpeningAngle2}). The reconstructed missing mass distribution peaks at small missing mass for both $pp$- and $np$-SRC pair knockout.  As can be seen, there is a peak at back angle, associated with a breakup of $^4$He into a SRC pair and a residual 2N system with low excitation energy.  As with the $^{12}$C measurements, the width of the opening angle distribution is due to the c.m. motion of the SRC pairs which was found to be consistent with a Gaussian in each direction with a width of $100 \pm 20$ MeV/c.

The extracted $^4$He $pp$/$np$ SRC pairs ratio increases with $p_{miss}$ for $p_{miss}>400$ MeV/c (see Fig.~\ref{fig:HallA_4He_np_pp_ratio}).  The measured ratios are consistent with ab-initio Variational Monte Carlo (VMC) calculations of Ref.~\cite{wiringa14} integrated over  c.m. momentum up to about $300$ MeV/c, which is consistent with the measured width of the c.m. momentum distribution.  At higher c.m. momentum, the two body momentum distribution is dominated by large contributions from un-correlated pairs \cite{Weiss:2016obx}. Similar results were also obtained by different calculations~\cite{ryckebusch15,Alvioli:2016wwp}.

\begin{figure}[tbp]
\includegraphics[width=7.5cm]{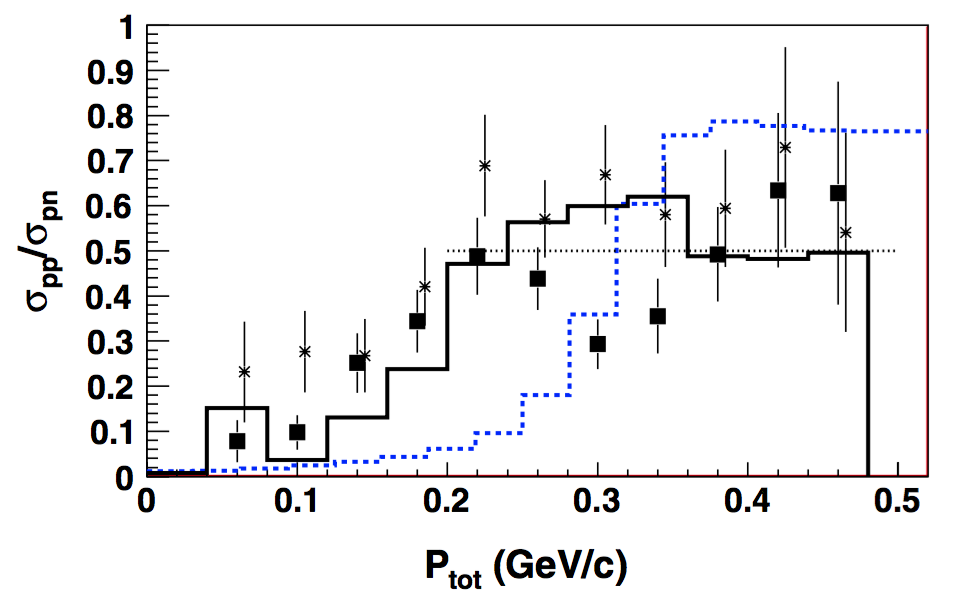}
\caption{\label{fig:3HePpPnRatio} The ratio of $pp$ to $pn$ pairs in $^3$He$(e,e'pp)n$.  The solid and star points show the  ratio for $300 \le p_{rel} \le 500$ MeV/c and for $400 \le p_{rel} \le 600$ MeV/c respectively, as a function of the total (e.g., center-of-mass) momentum of the pair.  The $pp$/$pn$ ratio is much less than one for small $p_{tot}$, increasing to the pair counting ratio of 0.5 at large $p_{tot}$.  The ratio at small $p_{tot}$ is about 0.1 for $300 \le p_{rel} \le 500$ MeV/c, and about 0.25 for $400 \le p_{rel} \le 600$ MeV/c.  The solid line shows a calculation by Golak for $300 \le p_{rel}\le 500$ MeV/c which neglects rescattering of the struck nucleon but includes the reinteraction of the two nucleons in the SRC pair.  The dashed line (blue online) shows the $^3$He momentum distribution integrated over the experimental acceptances.  From \cite{bagh10}. }
\end{figure}

The importance of tensor correlations was further shown by measurements of the $pp$ to $pn$ ratio in $^3$He$(e,e'pp)n$ measured using the CLAS detector at JLab \cite{bagh10}.  They measured the relative and total momentum distribution of $pp$ and $pn$ pairs in $^3$He by detecting events where the virtual photon was absorbed on one nucleon and the other two (spectator) nucleons were also detected.  Fig. \ref{fig:3HePpPnRatio} shows the ratio of $pp$ to $pn$ pairs in $^3$He as a function of the pair total (e.g., center-of-mass) momentum for two pair relative momentum ranges, $300 \le p_{rel} \le 500$ MeV/c and $400 \le p_{rel} \le 600$ MeV/c.  The first range is centered at the $s$-wave minimum at 400 MeV/c where the effects of tensor correlations are expected to dominate; the second is not.  For $p_{rel}$ centered at 400 MeV/c, the $pp$ to $pn$ ratio is very small at $p_{tot}\le 100$ MeV/c and consistent with the $^{12}$C$(e,e'pN)$ measurements discussed above. For $p_{rel}$ centered at 500 MeV/c, the $pp$ to $pn$ ratio at $p_{tot}\le 100$ MeV/c is significantly larger, consistent with the expected decreased dominance of tensor correlated pairs at this higher relative momentum.  At large $p_{tot}$, the $pp$ to $pn$ ratio is 0.5, consistent with simple pair counting.  The points at $300 \le p_{rel}\le 500$ MeV/c are consistent with a calculation by Golak \cite{golak95} which neglects rescattering of the struck nucleon but includes the reinteraction of the two nucleons in the SRC pair.  

The combined results of the $^3$He, $^4$He and $^{12}$C measurements indicate that for $300\le p_i \le 500$ MeV/c, nucleons are predominantly part of $pn$-SRC pairs as predicted by dominance of
the tensor part of the NN interaction at short distances. At higher initial momentum, the contribution of $pp$-SRC pairs seems to increase by a factor of 2 -- 3, possibly due to larger contributions from the scalar repulsive core of the NN interaction.

\begin{figure}[tbp]
\centering
\includegraphics[width=7.5cm]{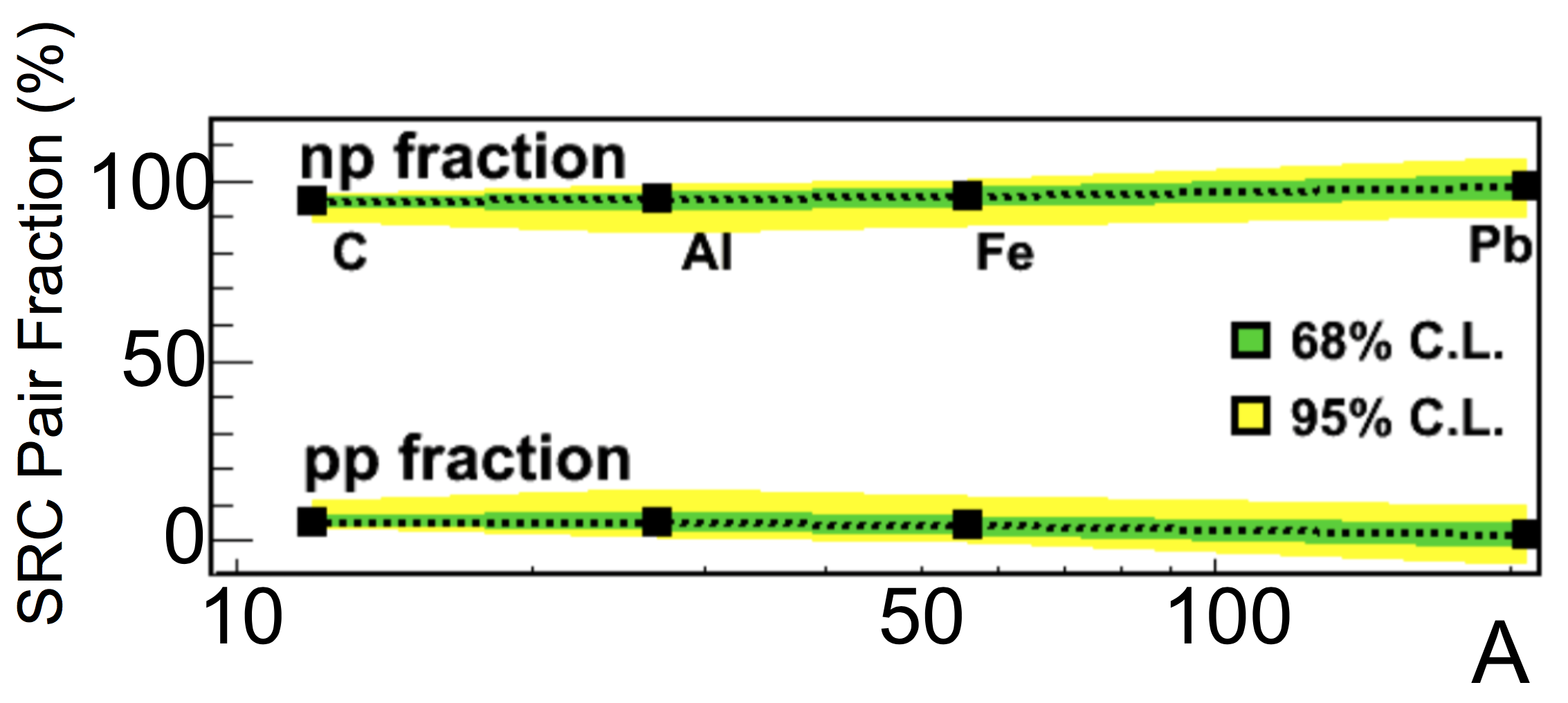}
\caption{The relative fraction of $np$ and $pp$ SRC pairs (excluding $nn$ pairs) derived from $A(e,e'p)$ and $A(e,e'pp)$ measurements on a range of nuclei. From \cite{Hen:2014nza}.}
\label{fig:HallB_DataMining_np_pp_ratio}
\end{figure}

Encouraged by these results, the latest exclusive measurements extended to medium and heavy nuclei ($^{12}$C, $^{27}$Al, $^{56}$Fe, and $^{208}$Pb), where the persistence of $np$-SRC dominance was still unproven \cite{Hen:2014nza}.  In this experiment, the $A(e,e'pp)$ and $A(e,e'p)$ reactions were measured at similar kinematics to the previous $^4$He and $^{12}$C measurements, covering a reconstructed initial proton momentum range of $300\le p_i \le 600$ MeV/c.  The analysis assumed that, in these nuclei, the reaction is still dominated by scattering off SRC pairs and extracted the relative fraction of $np$- and $pp$-SRC pairs. Fig.~\ref{fig:HallB_DataMining_np_pp_ratio} shows that SRC pairs are predominantly $np$-SRC pairs even in heavy neutron rich nuclei.


\subsection{Inclusive Scattering \label{Inclusive}}
Inclusive quasi-elastic electron scattering allows probing the momentum distribution of nucleons in the nucleus.
Elastic scattering from a nucleon at rest occurs at fixed kinematics, $\nu =
\frac{Q^2}{2M}$.  This corresponds to $\xB=1$. If all of the struck nucleons in
a nucleus were at rest, the cross section would show a pronounced narrow  peak-- the   quasi-elastic peak.

This peak is broadened by nucleon motion for electron scattering from
bound nucleons.  In order to study nuclear momentum distributions,
experiments typically focus on the low energy transfer side of the QE
peak, or $\xB\ge 1$.  In this case the initial
momentum of the struck nucleon must be in the opposite direction from
the momentum transfer so that the final momentum of the struck nucleon
${\bf p_f} ={\bf q} +{\bf p}_{\rm miss}$ (in the absence of final state
interactions or FSI, ${\bf p}_{rm miss} ={\bf p}_{rm init}$) is less than the momentum transfer. As the energy
transfer decreases, the final momentum of the struck nucleon must
decrease and therefore the minimum initial momentum of the struck
nucleon must increase.

\begin{figure}
\includegraphics[width=3.5in]{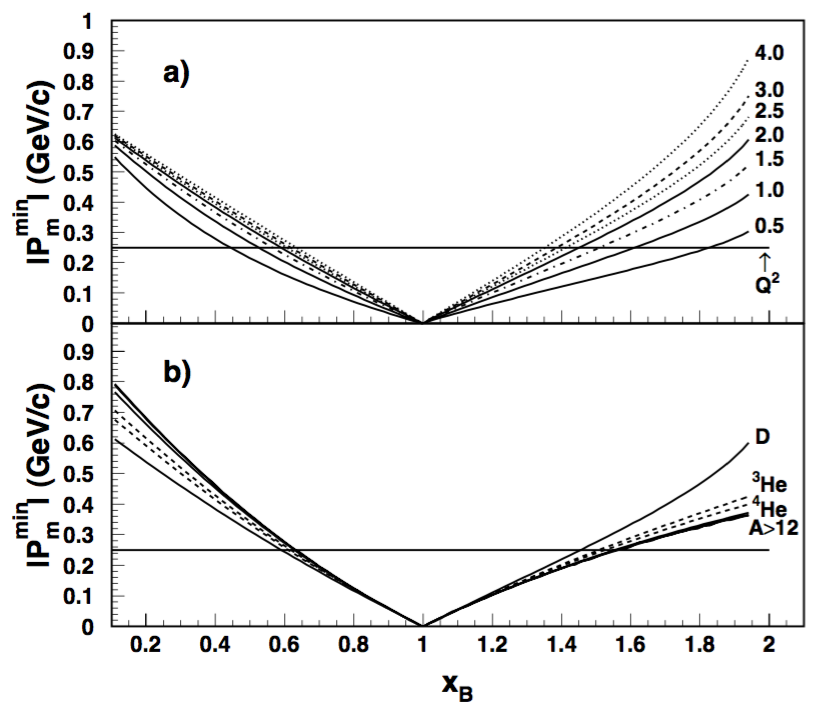}
\caption{\label{fig:pmin} The minimum momentum of the struck nucleon
  in inclusive $(e,e')$ scattering
  as a function of $\xB$.   The top panel shows the minimum momentum
  for deuterium for a variety of momentum transfers  and the bottom
  panel shows the minimum momentum for a variety of  nuclei at $Q^2=2$
  GeV$^2$. The residual $A-1$ system is assumed to be in its ground state.   Figure from \cite{egiyan03}.}
\end{figure}

\begin{figure}
\includegraphics[width=3in]{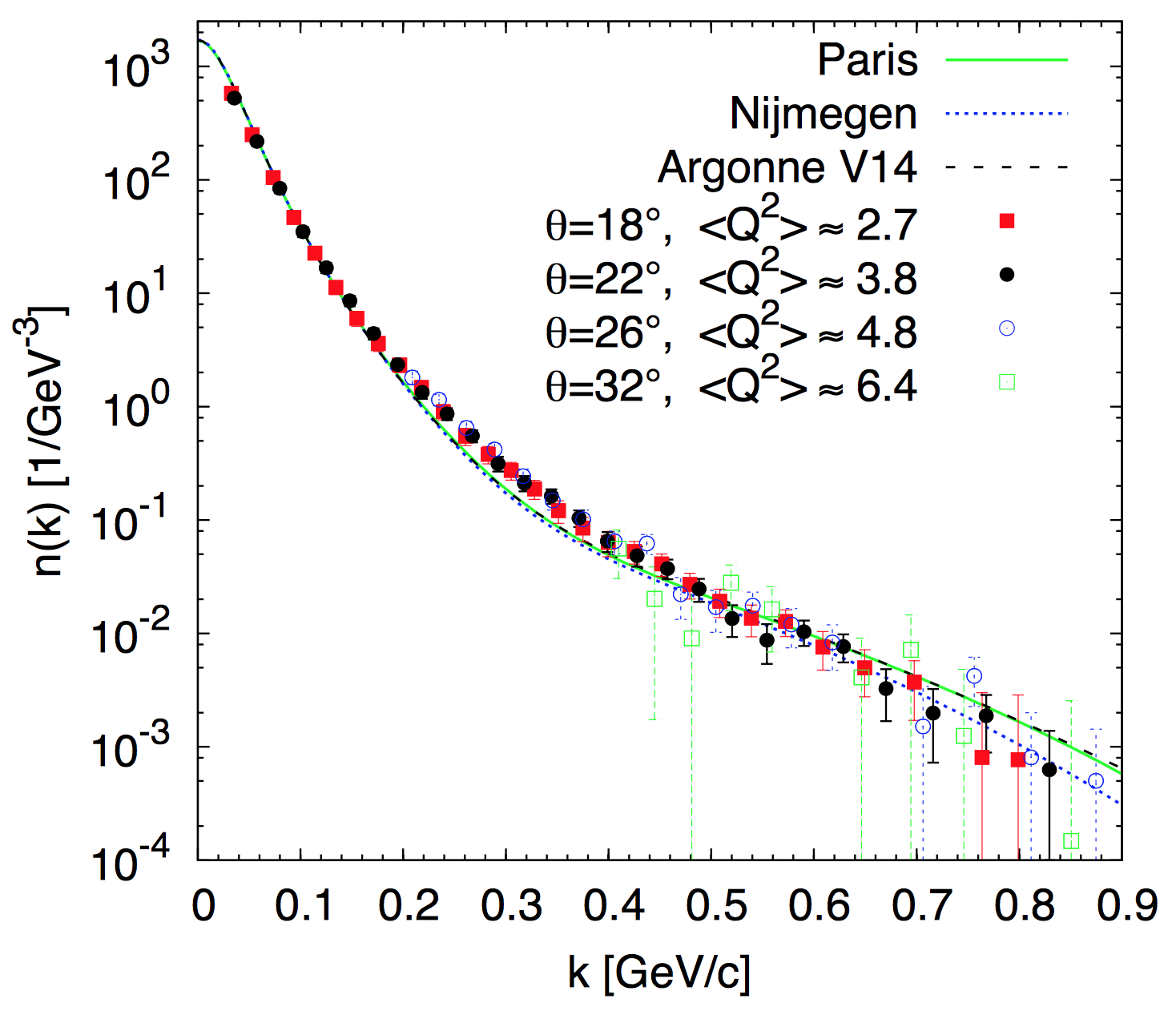}
\caption[Deuteron Momentum Distribution from
$(e,e')$]{\label{fig:dMomDist} Momentum distribution of the deuteron.
Points show the results extracted from the experimental scaling
function $F(y)$ at four different momentum transfers \cite{fomin12}.
Curves show the calculated momentum distributions using three
different $NN$ potentials Paris~\cite{Paris81}, Nijmegen~\cite{Nijmegen94} and Argonne V14~\cite{wiringa95}.  Figure
 from \cite{fomin12}, which uses $k$ for momentum instead of
$p$.}
\end{figure}

The quasielastic inclusive electron scattering $(e,e')$ cross section
can be written in terms of a function $F$ that depends on $(Q^2,y)$
rather than $(Q^2,\nu)$ \cite{Day87}:
\begin{equation}
\frac{d^2\sigma(q,\nu)}{d\nu d\Omega}=F(y,Q^2)
{(Z\sigma_p+N\sigma_n)}\frac{q}{\sqrt{M^2+(y+q)^2}}
\end{equation}
where $\sigma_{p,n}$ are the elastic electron scattering cross sections from a bound nucleon,  the last term is the Jacobian $dy/d\nu$, and $y=y(Q^2,\nu)$ is the minimum
momentum of the struck nucleon (assuming that the residual $A-1$
system is unexcited) \cite{Arrington12,Day90}.

Non-relativistically, $y$ is the component of the struck nucleon's
initial momentum (${\bf p}_{\rm miss}$) in the direction of ${\bf q}$.  The
cross section at fixed $y$ then includes an integral over the
perpendicular components of ${\bf p}_{\rm miss}$.  Relativistically, it is a
little more complicated.  $y$ is determined from energy conservation,
assuming no FSI and that the $A-1$ nucleus recoils with momentum $y$:
\bea
 \nu + M_A = (M^2 + (q+y)^2)^{1/2} + (M^2_{A-1}+y^2)^{1/2} \quad . 
\eea 
At the QE peak, $\nu=Q^2/(2M)$, $\xB=1$, and $y=0$.  As $\nu$
decreases, $\xB$ increases and $y$ decreases.  By selecting $\xB$ or
$y$ (at fixed $Q^2$), we can select the minimum initial momentum of
the struck nucleon (see Fig.~\ref{fig:pmin}).  At large enough $Q^2$ the function $F(y,Q^2)$
scales and depends only on $y$ \cite{cda91b}.  The nucleon momentum
distribution, $n(p=y)$, can be calculated from the derivative of the
scaling function, $dF(y)/dy$, at large $Q^2$:
\begin{equation}
n(p)=\frac{-1}{2\pi p}\frac{dF(p)}{dp} .
\end{equation}
Fig. \ref{fig:dMomDist} shows the deuteron momentum distribution
derived in this manner.

The original $y$-scaling model  discussed here assumes a that the residual $A-1$ nucleus is in a low-lying state.  This procedure neglects the possibly large excitation energy of the residual nucleus, which is an important feature of the spectral function. 
As a result, for scattering by a nucleon in a SRC,  the same internal momenta corresponds to a very different values of $y$  for different nuclei.

Another approach is to  compare the
momentum distributions in different nuclei with reduced uncertainties by taking
ratios of cross sections.
We write the momentum density in terms of the light
cone variable $\alpha_{tn}$ for the interacting nucleon belonging to
the correlated pair,
\begin{equation}
\alpha_{tn} = 2 - \frac{\nu - q + 2M}{2M} (1 + \sqrt{1-4M^2/W^2}) .
\end{equation}
Using this variable, the cross section ratios do not depend on $Q^2$ in the kinematic range of the SLAC  experiments~\cite{PhysRevLett.59.427}.
The onset of the plateaus discussed below occur for the same values of $\alpha_{tn}$ but for slightly different values of $\xB$.

Then  the
ratios of cross sections can be expressed in terms of the light-cone spectral function  at
large $Q^2$ and $1.5<\xB < 2$ as \cite{frankfurt93}:
\begin{equation}
\frac{\sigma_{A_1}(\xB,Q^2)}{\sigma_{A_2}(\xB,Q^2)} = 
   \frac{\int\rho_{A_1}(\alpha_{tn},p_t)d^2p_t}{\int\rho_{A_2}(\alpha_{tn},p_t)d^2p_t}\,\approx  {n_A(p)\over n_D(p)}.
   \label{rats}
\end{equation}
Thus this ratio of cross sections should be a function of
$\alpha_{tn}$ only, which, since it is a function of $(Q^2,\xB)$, is
directly related to $y$, the minimum momentum of the struck
nucleon. The approximate equality shown in \eq{rats} holds for $1.3\le
\a_{tn}\le 1.7$ and $p>p_F$. The second approximate equality appearing in \eq{rats} is obtained using the relation $|{\bf p}|\approx M{|1-\a_{tn}|\over\sqrt{\a_{tn}(2-\a_{tn}))}}$.   Measured ratios should be less sensitive to the influence of  final
state interactions, as discussed below.  Nevertheless, the accuracy of
replacing cross section ratios by ratios of densities, as shown in
\eq{rats}, needs to be studied further.   Furthermore, as yet there is
no separate calculation of the   numerator   term of \eq{rats}, {\it i.e.,}  the basic  nuclear cross section for the $(e,e')$
reaction at large values of $\xB$.

\begin{figure}
\includegraphics[width=1.3in]{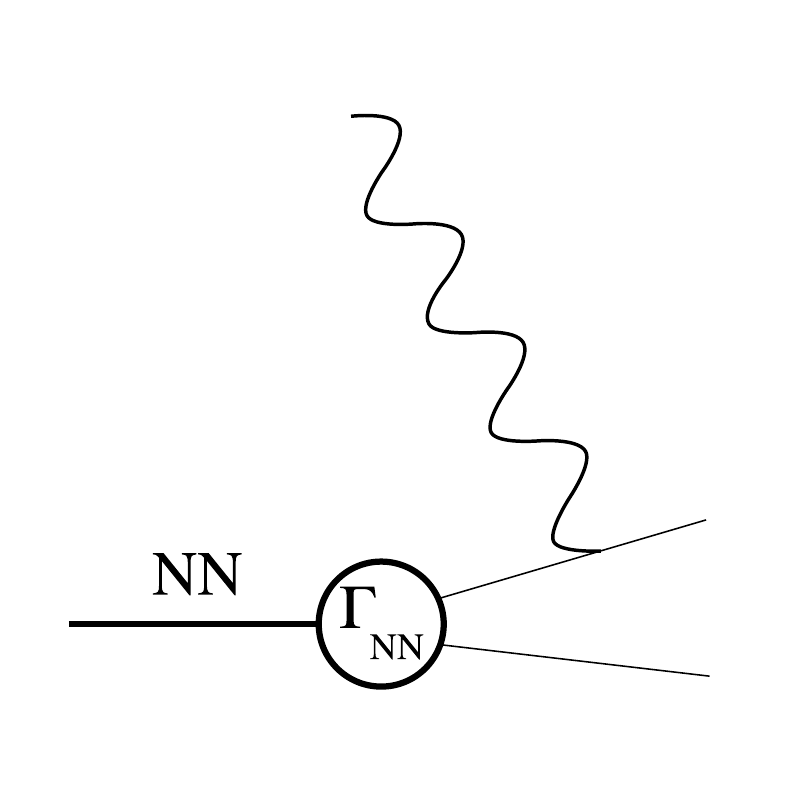}
\includegraphics[width=1.7in]{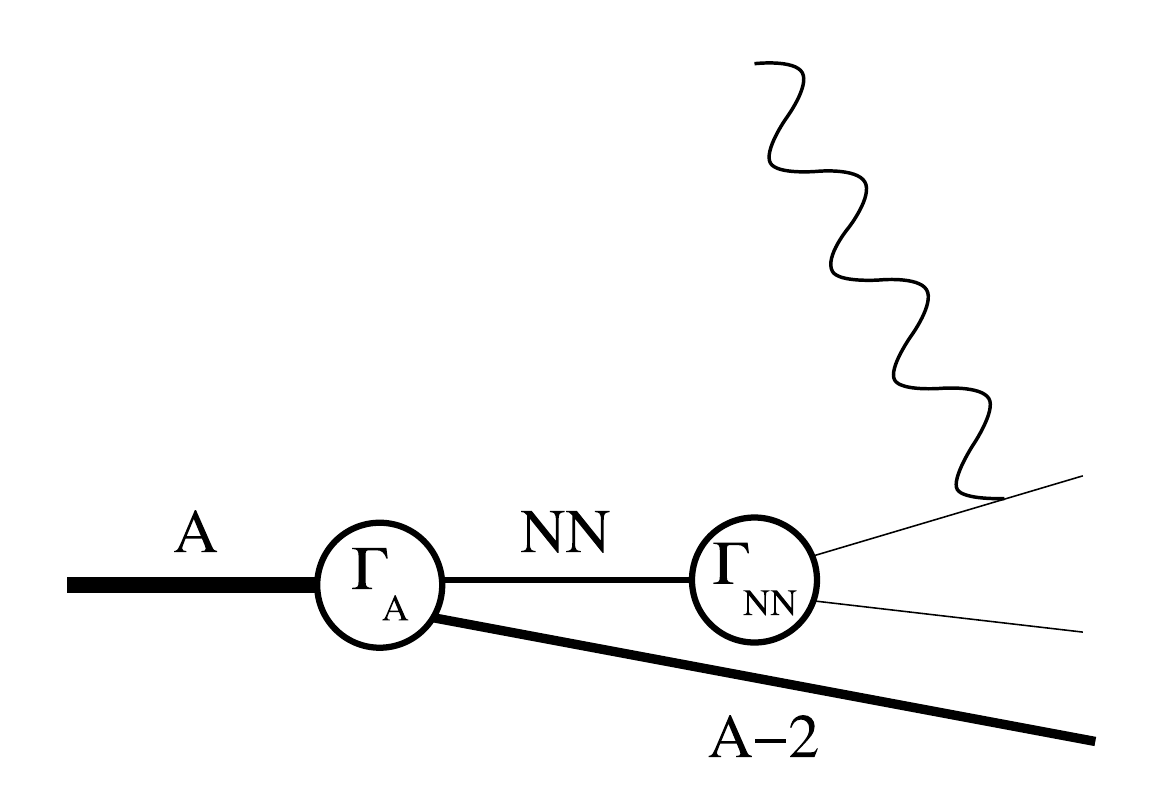}
\caption[Inclusive Cross Section Ratios]{\label{fig:eecartoon}  A
  cartoon of electron quasielastic scattering from a nucleon in
  deuterium (left) and from a nucleon in a SRC pair in a heavier nucleus (right).  The labels $\Gamma_{NN}$ and $\Gamma_A$ refer to the deuteron and nuclear vertex functions respectively.}
\end{figure}

{\bf Physics at large values of $\xB$.} The next step is to use the
inclusive $(e,e')$ cross section to look for the effects of SRC pairs
in nuclei by choosing kinematics where mean field nucleons cannot
contribute to the reaction. This is done by using $\xB>1$. Just as
conservation of four-momentum ensures that $\xB=1$ is the kinematic limit for
scattering from a single nucleon, $\xB=2$ is the kinematic limit for
scattering from a cluster of two nucleons  and $\xB =3$ is the
kinematic limit for scattering from a three-nucleon cluster.


As a result,  we can expand  the
$(e,e')$ cross section 
into
pieces due to electrons scattering from nucleons in 2-, 3- and
more-nucleon SRC \cite{Frankfurt81,Frankfurt88,frankfurt93}
\begin{equation}
\sigma(\xB,Q^2) = \sum_{j=2}^A a_j(A) \sigma_j(\xB,Q^2),
\label{eq:a2def}
\end{equation}
where $\sigma_j(\xB,Q^2)=0$ for $\xB>j$ and the $\{a_j(A)\}$ are
proportional to the probability of finding a nucleon in a $j$-nucleon
cluster. This is analogous to treating nuclear structure  in terms of  independent nucleons,
independent nucleon pairs, etc. This expression  is based on the lack of
interference between amplitudes arising from scattering by clusters of
different nucleon number that occurs because the important final states are different. Its importance lies in the
fact that in a given kinematic region the ratio of cross sections can be used to determine information
about short-ranged correlations.

 If we consider only  the $a_2$ term, then we can write
\begin{equation}
a_2(A) = \frac2A \frac{\sigma_A(\xB,Q^2)}{\sigma_d(\xB,Q^2)} .
\label{eq:EeRatio}
\end{equation}
This approximation should be valid for $1.5<\xB\le 2$. The effect of
neglecting clusters of three or more nucleons has never been
studied. 

If the momentum distribution for $\vert y\vert >p_{fermi}$ is dominated
by nucleons in SRC pairs, then we expect that the momentum
distributions for nucleus $A$ and for deuterium should be almost
identical.  This similarity should show up as a plateau in the
per-nucleon cross section ratio of the two nuclei.
Fig. \ref{fig:eecartoon} shows a cartoon of this process. 

\begin{figure}
\includegraphics[width=2.8in]{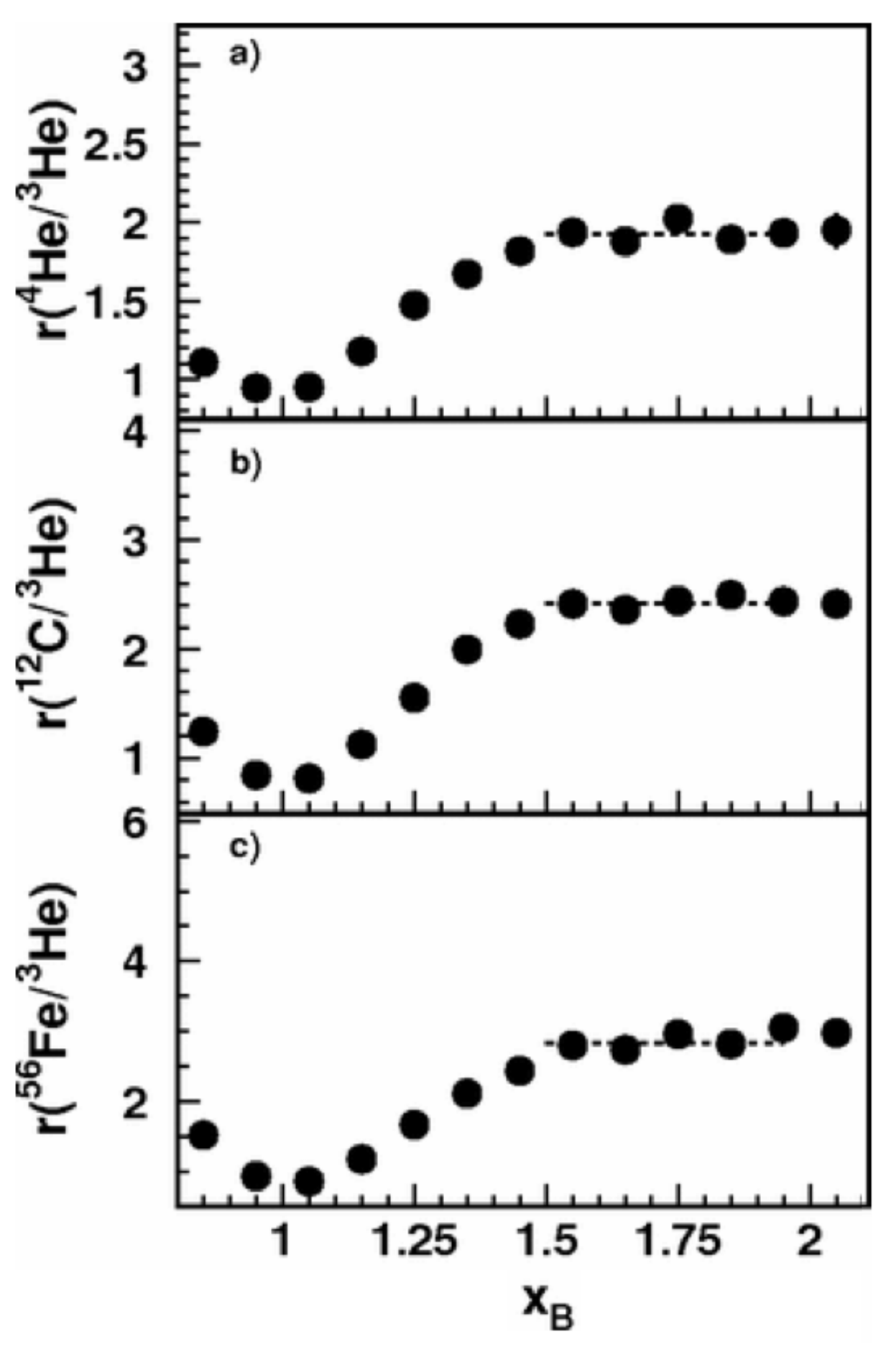}
\includegraphics[width=2.8in]{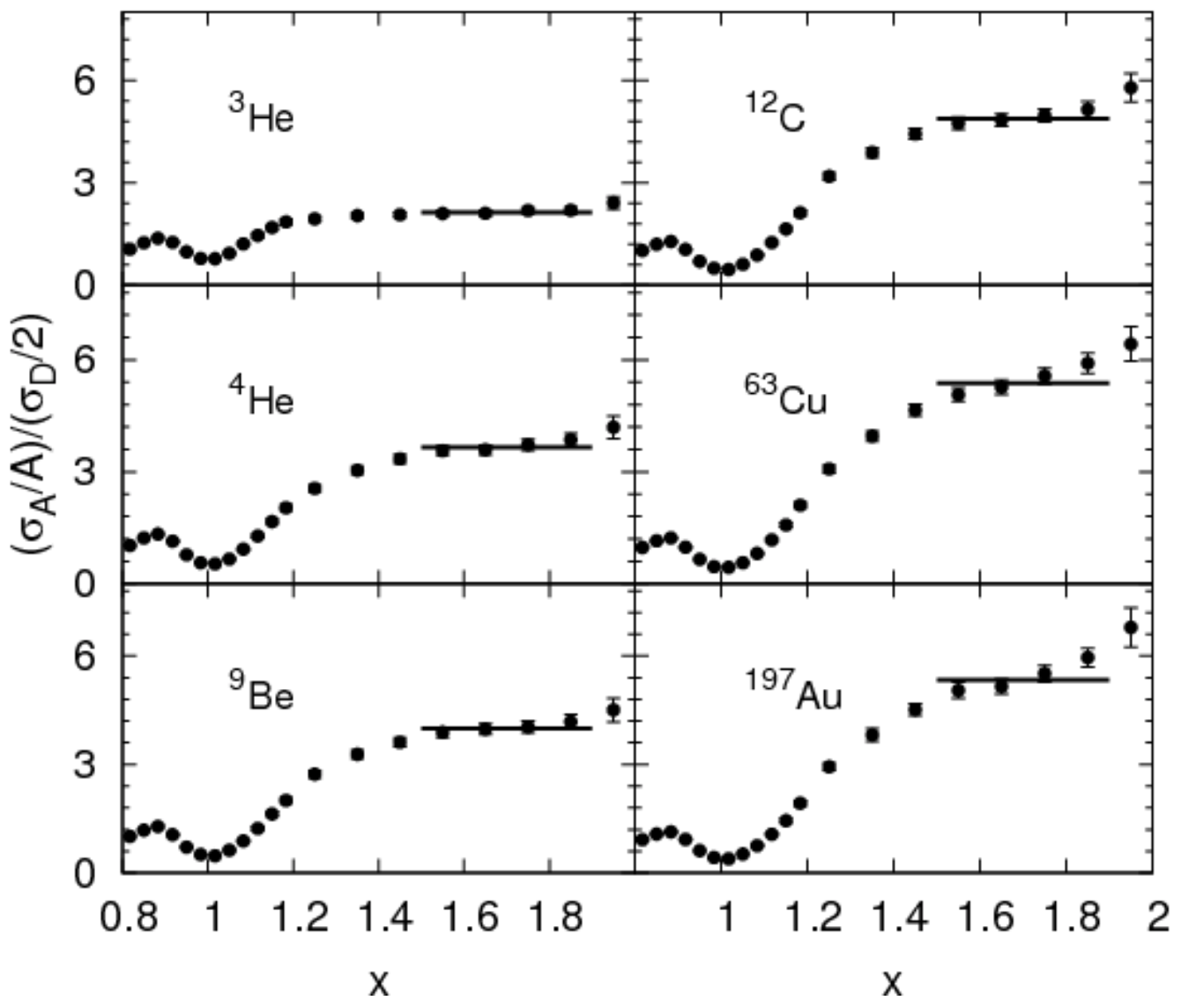}
\caption[Inclusive Cross Section Ratios]{\label{fig:eeratios} 
  Inclusive per-nucleon cross section ratios of (top) nuclei to
  $^3$He from \cite{egiyan06} at $1.4 < Q^2 < 2.6$ GeV$^2$ and (bottom)
  nuclei to deuterium at $Q^2 = 2.7$ GeV$^2$ \cite{fomin12}. Figures adapted from~\cite{egiyan06} (top) and from \cite{fomin12} (bottom).}
\end{figure}

The cross section ratio of nucleus $A$ to deuterium or to $^3$He has
been measured at SLAC \cite{frankfurt93} and at Jefferson Lab
\cite{egiyan03,egiyan06,fomin12}.  They have all observed a plateau in
the cross section ratio at $Q^2>1.4$ GeV$^2$ and from $1.5\le \xB\le
1.9$.  See Fig. \ref{fig:eeratios}.  This corresponds to $y\ge
p_{thresh} = 275\pm25$ MeV/c, which is slightly larger than the Fermi momentum
in medium and heavy nuclei. The value of $Q^2$ is large enough to ensure that contributions from uncorrelated nucleons
(with momentum governed by the size of the nucleus) are negligible.

However, in order to relate these observed plateaus to the ratio of
momentum distributions in the different nuclei, we need to take into
account the final state interactions (FSI) of the nucleon with its correlated partner 
and with the residual system.  For $Q^2>1$ GeV$^2$ and $0.35 < \nu < 1$ GeV, 
the space-time physics~\cite{frankfurt93,Frankfurt:2008zv} of the
inclusive process tells us that final state interaction effects occur
predominantly within the two-nucleon correlation. Such effects are
independent of the nuclear target, and should be small for large values
of $\n$. The relevant values of $\n$ are large enough so that final
state interactions within the pair are not very
important~\cite{Frankfurt81,Frankfurt88}.  Therefore, the effects of
FSI will be approximately the same for high momentum nucleons in
deuterium and in heavier nuclei and will predominantly cancel in the
cross section ratios.

Some measurements, e.g., \cite{egiyan03,egiyan06}, applied isoscalar
corrections to the ratios of Eq.~\ref{eq:EeRatio} (i.e., they
corrected for the unequal electron-proton and electron-neutron cross
sections).  Since the discovery of $pn$-dominance in SRC pairs (see
Section \ref{Exclusive}), these corrections are no longer applied
\cite{fomin12}.
 
The flatness of the cross section ratio plateau  at $Q^2>1.4$
GeV$^2$ and from $1.5\le \xB\le 1.9$, and its approximate independence of $Q^2$ in this region where SRC effects dominate  indicates the similarity of the
momentum distributions in the two nuclei for $p>p_{thresh}$ and the
validity of the expansion in Eq. \ref{eq:a2def}.  The onset of the plateau at $\xB=1.5$ for $Q^2
>1.4$ GeV$^2$ indicates that the momentum distributions become similar
at a threshold momentum of $p_i=p_{thresh}=275\pm25$ MeV/c
\cite{egiyan03}.  The height of the plateau, $a_2(A)$, indicates the relative
probability that a nucleon in nucleus $A$ has high momentum
($p>p_{thresh}$) relative to a nucleon in deuterium.

In a naive model, this relative probability for a nucleon to have high
momentum equals the relative probability that it belongs to an $NN$
SRC pair.  However, even if all nucleons with $p>p_{thresh}$ belong to
an $NN$ SRC pair as evident from the exclusive measurement (see Section \ref{Exclusive}), we still need to
consider the effects of pair motion.  The high momentum $NN$ pair in
the deuteron has center of mass momentum $p_{cm}=0$.  The non-zero
center of mass momentum distribution of the pair in heavier nuclei
will smear the high-momentum tail of the nucleon momentum
distribution, increasing the cross section ratio in the plateau
region~\cite{cda91,fomin12,vanhalst12}.  This was found to be about a 20\% effect in Fe.  
Thus, while the ratio of the proportion of high-momentum nucleons in Cu to deuterium 
is $a_2(A)= 5.4\pm0.1$, the ratio of the number of SRC $NN$ pairs in Cu to deuterium 
(using the Fe correction factor) is about 20\% less, $R_{2N}=4.3\pm0.3$  \cite{fomin12}.

Multiplying the $4\%$ probability for a nucleon in deuterium to have
momentum $p>p_{thresh}$ by the measured ratios in the plateau region
($a_2(A)$),  as indicated by \eq{rats}, gives us the probabilities for a nucleon to have high
momentum in $^4$He, C, Fe/Cu and Au to be 14\%, 19\%, 21\% and 21\%
respectively \cite{fomin12,hen12}.

Thus, the existence of a plateau in the measured per-nucleon cross
section ratios of various nuclei to deuterium or $^3$He at $Q^2>1.4$
GeV$^2$ and $1.5\le \xB\le 1.9$ shows that the momentum distributions
of all nuclei at high momentum are similar and are thus dominated by
2N-SRC, that the threshold for ``high momentum'' is
$p_{thresh}=275\pm25$ MeV/c, and that the probabilities for nucleons
in nuclei to have high momentum range from 4\% in deuterium to 21\% in
heavy nuclei.

While the inclusive scattering cross section ratios of carbon and iron
to $^3$He measured by Egiyan are flat for $1.5<\xB<2$, the ratios of
carbon, copper and gold to deuterium measured by Fomin appear to slope
upwards slightly.  This is not due to the choice of nucleus in the
denominator, since the ratio of $^3$He to deuterium measured by Fomin
is flat. This is also probably not due to c.m. motion effects as these are 
simililar for $^4$He and $^{12}$C \cite{korover14,tang03,shneor07}, which do
show flat ratios, and are expected to be the same for heavy nuclei \cite{Colle:2013nna,CiofidegliAtti:1995qe}.
This might be due to differences in kinematics. The Egiyan data
covers $1.4 < Q^2 < 2.6$ GeV$^2$  (concentrated at the lower values), 
 while the Fomin data was taken at $Q^2=2.7$ GeV$^2$.  At
$Q^2=1.5$ GeV$^2$ and $1.5\le \xB\le 2$, the minimum momentum of the
struck nucleon ranges from 250 to 500 MeV/c, covering the expected
region of tensor force dominance.  However, at $Q^2=2.7$ GeV$^2$, the
minimum momentum of the struck nucleon ranges from 320 to 700 MeV/c,
where central correlations could become important.  It would be useful
to measure the $Q^2$ dependence of the cross section ratios in future
SRC measurements.

\subsection{Universal Properties of Short Range Correlations in
  Nuclei}  
\begin{figure}[tbp]
\includegraphics[width=7.5cm]{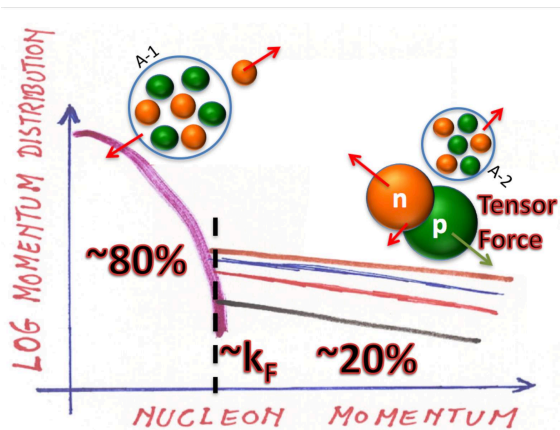}
\caption{A qualitative sketch of the dominant features of the nucleon momentum distribution in nuclei.  At $k<k_F$, the nucleon momentum is balanced by that of the other $A-1$ nucleons and can be described by mean field models.  At $k>k_F$, the nucleon belongs to a $pn$-SRC pair and its momentum is balanced by that of one other nucleon.}
\label{fig:MomentumDistSketch}
\end{figure}

The combined results from the exclusive and inclusive measurements described in Sections \ref{Exclusive} and \ref{Inclusive} lead to a universal picture of SRC pairs in nuclei. In the conventional momentum space picture, the momentum distribution for all nuclei and nuclear matter can be divided into two regimes, above and below the Fermi-momentum (see Fig.~\ref{fig:MomentumDistSketch}). The region below the Fermi momentum accounts for about $80\%$ of the nucleons in medium and heavy nuclei (i.e., $A \ge 12$) and can be described using mean-field approximations. The region with momenta greater than the Fermi momentum accounts for about 20--25\% of the nucleons (see the pie chart in Fig.~\ref{fig:HallA_Ratios}) and is dominated by nucleons belonging to $NN$-SRC, predominantly $pn$-SRC.  

The SRC dominance of the high-momentum tail implies that the shape of the momentum distributions of all nuclei at high momenta is determined by the short range part of the fundamental $NN$ interaction.  The  average number of SRC pairs is determined  by  global prperties of the nucleus. 

The specific predominance of $pn$-SRC over $pp$- and $nn$-SRC is largely associated with the large contribution of the tensor part of the $NN$ interaction at short-distances~\cite{schiavilla07,sargsian05,Alvioli:2007zz}, implying that the high-momentum distribution in heavier nuclei is approximately  proportional to the deuteron momentum distribution. Experimental and theoretical studies of the latter show that, for $300 \le k \le 600$ MeV/c, $n(k)\propto 1/k^4$ \cite{hen15b}. This specific functional form follows directly from the dominance of the tensor force acting in second order, see Section~\ref{appendix:Understanding_np} for details.

The predominance of $np$-SRC pairs implies that, even in asymmetric nuclei, the ratio of protons to neutrons in SRC pairs will equal 1.  This, in turn, implies that in  neutron rich nuclei, a larger fraction of the protons will be in an SRC pair~\cite{sargsian14,Hen:2014nza}, i.e.,  
that a minority nucleon (e.g., a proton) has a higher probability of belonging to a high-momentum SRC-pair than a majority nucleon (e.g., a neutron). This effect should grow with the nuclear asymmetry and could possibly invert the kinetic energy sharing such that the minority nucleons move faster on average then the majority.
 This asymmetry could have wide ranging implications for the NuTeV anomaly~\cite{zeller02,zeller03} (see Sects~\ref{MMF},\ref{pvdis}), the nuclear symmetry energy and neutron star structure and cooling rates~\cite{hen15,Hen:2016ysx}, neutrino-nucleus interactions~\cite{Weinstein:2016inx,Acciarri:2014gev} and more. The study of the nuclear asymmetry dependence of the number of SRC pairs and their isospin structure is an important topic that could be studied in future high-energy radioactive beam facilities.

\section{ Deep Inelastic Scattering (DIS) and the EMC effect \label{EMC}}

Basic models of nuclear physics describe the nucleus as a collection
of unmodified nucleons moving non-relativistically under the influence of
two-nucleon and three-nucleon forces, which can be treated approximately
as a mean field.  In such a picture, the partonic structure functions
of bound and free nucleons should be identical. Therefore, it was
generally expected that, except for nucleon motion effects, Deep
Inelastic Scattering (DIS) experiments which are sensitive to the
partonic structure of the nucleon would give the same result
for all nuclei.

Instead, the measurements~\cite{aubert83,Arneodo:1992wf,Geesaman95,Piller:1999wx,Norton03,Hen13,malace14,Frankfurt:2011cs}  show a reduction of the structure function
of nucleons bound in nuclei relative to nucleons bound in deuterium in the valence quark region. We term this reduction 
the EMC effect. Since its discovery, over 30 years ago, a large
experimental and theoretical effort has been put into understanding
the origin of the effect. While theorists have had no difficulty in
creating models that qualitatively reproduce nuclear DIS data by
itself, there is no  generally accepted model. This is because the
models are either not consistent with or do not attempt to explain
other nuclear phenomena.   

The nuclear deep inelastic scattering data  also  show a reduction in the small $\xB$ region of the structure function, known as the shadowing region.
The physics of  shadowing  has been well-reviewed~\cite{Frankfurt:2011cs} recently, and is not a subject of the present review.


Section \ref{NeedForSrc} showed that  the 
nucleon-nucleon interaction leads to 
the existence of Short-Range Correlated (SRC) pairs in nuclei and Section \ref
{HardScattering} showed the evidence for and our knowledge of the
properties of these pairs. 

This section will describe Deep Inelastic Scattering and its
relationship to nucleon parton distributions.  The EMC effect and the limitations of
conventional nuclear physics to explain it will then be discussed.  Section IV  will
present the phenomenological relationship between the number of SRC
pairs in a nucleus and the strength of the EMC effect and 
use that relationship  to gain new insight into the origin of the EMC effect.

\subsection{DIS and nucleon structure functions \label{DisStrucFun}}
\begin{figure}[h]
\includegraphics[scale=0.2]{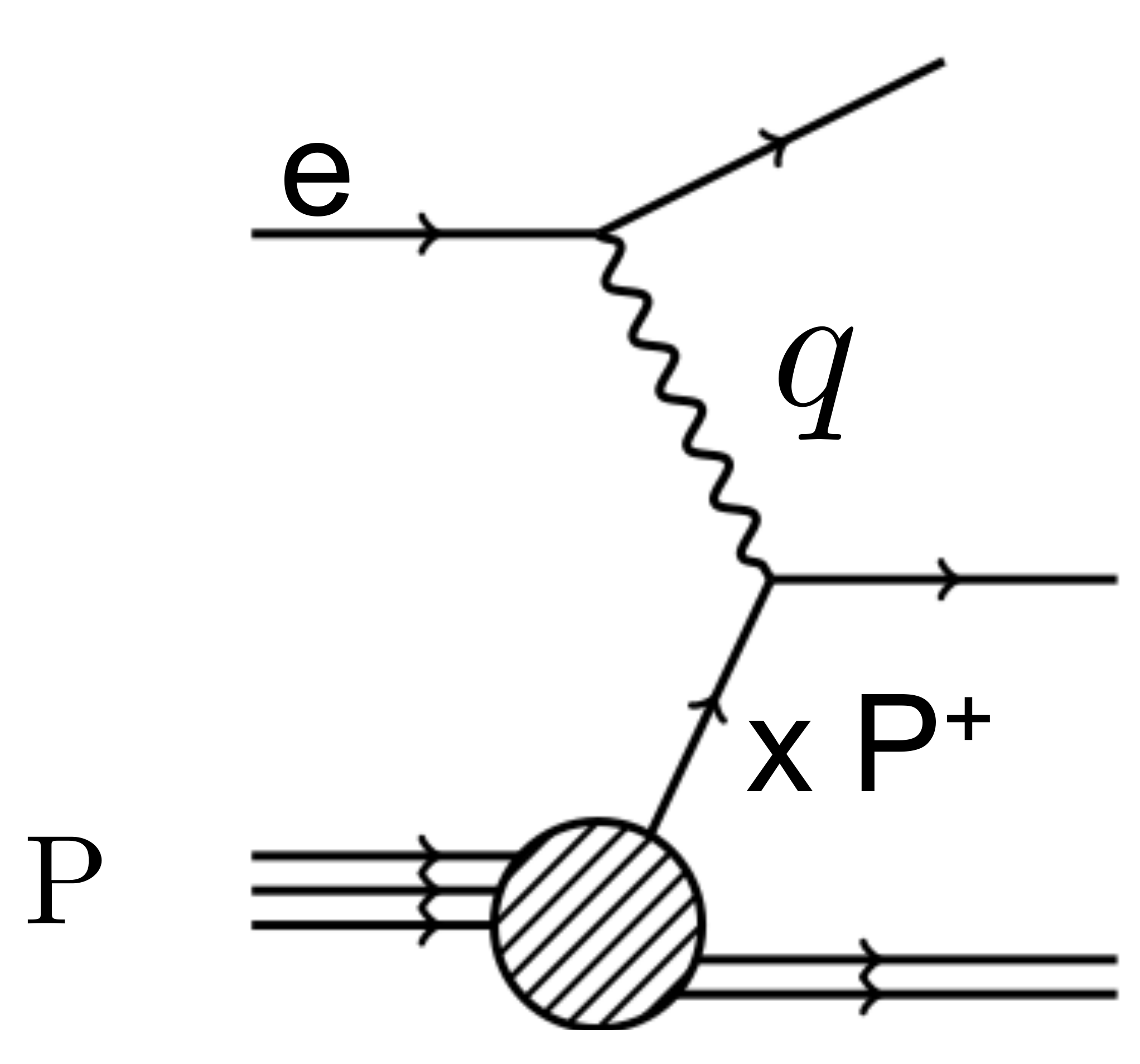}
\caption{\label{fig:dist}  Deep inelastic scattering at large values of $Q^2$.  A lepton (labelled `e') scatters from a nucleon by emitting a  space-like virtual photon with four-momentum $q$, which is absorbed on a single quark with momentum fraction $\xB P^+$.  Only the outgoing lepton is subsequently detected.}
\end{figure}

We begin with a brief description of deep inelastic scattering on a nucleon. See one of the many texts for details, e.g., ~\cite{Close:1979bt,Halzen:1984mc,Roberts:1990ww,Thomas:2001kw,Collins:2011zzd}. The latest information is contained in the Particle Data Group tables~\cite{Agashe:2014kda}.  The inclusive deep inelastic scattering process, $(e,e')$, involves a lepton scattering from a target, with only the final state lepton  being detected.  If spin variables are not observed, the process depends on only two variables, which are traditionally chosen to be  the electron energy loss $\n$ and negative of the four momentum transfer from the lepton to the target $Q^2=\bfq^2 - \nu^2$, see Fig.~\ref{fig:dist}. At large enough values of $\n$ and $Q^2$, conservation of momentum and energy leads to the result that the dynamical information, can be encoded (at a given scale)  in the structure functions $q(\xB)$, which is interpreted as the fraction of the target momentum carried by the struck quark. 

Let's see how this arises. Four-momentum conservation,  the idea that the quark is briefly free after absorbing the high-momentum photon, and ignoring the emissions of gluons  gives 
\bea (k+q)^2=m_q^2\label{cons}\eea
where $k$ is the four-momentum of a  quark in the target, and $m_q$ is the quark mass.
 Let the spatial momentum of the photon  lie in the negative $z$ direction and using the light-front  momentum variables, {\it e.g}  $P^\pm\equiv P^0\pm P^3$, where  $P^\m$ is the target four-momentum,  we have
 $
\, q^-=\n+\sqrt{\n^2+Q^2} = \nu+\vert\bfq\vert, \,
q^+=
 \nu-\vert\bfq\vert, \,
q^-\gg  q^+ 
$, 
so that \eq{cons} can be re-written as 
 \bea k^+={Q^2-k^2- q^+k^-+m_q^2\over \n+\sqrt{\n^2+Q^2}}.\label{kplus}\eea
 If the quark is on its mass shell (as is the case with light-front wave functions) then $k^2=m_q^2$. Furthermore, if the quantity $q^+k^-\ll Q^2$, the numerator becomes simply $Q^2$. Then one defines a dimensionless, Lorentz invariant  variable by dividing the resulting equation by $P^+$, so that 
 \bea 
{k^+\over P^+}\approx {Q^2 \over  P^+(\n+\sqrt{\n^2+Q^2})}\equiv\xi,\label{kpluss}
\eea
 where $\xi$ is the Nachtmann variable. We see that the fraction of target momentum (plus-component)   is simply $\xi$.
  This explains why  deep inelastic scattering shows the scaling phenomenon. The relevant dynamical variable, $k^+\over P^+$,  depends only on one specific combination of $\n$ and $Q^2$. Note that this description is frame-independent. One need not go to the infinite momentum frame to understand scaling or the parton model.

If one further takes the Bjorken limit ($\nu^2\gg Q^2$), then 
 ${k^+\over P^+}= {Q^2\over 2 P^\mu q_\mu}\equiv \xB$. 
 The dominant dependence on $\xB$ is called Bjorken scaling, and its discovery, using hydrogen and deuteron targets (to obtain the neutron information),  was the primary evidence for the existence of quarks within the nucleon. 

Quarks are confined, so they are never on their mass shell. The off-mass shell effects, however, decrease with increasing values of $Q^2$ and are regarded as ``higher twist''. Such effects could be important at Jefferson Lab energies. Effects of final state interactions (which depend on the kinematics of the probing beam)  are not contained in the light-front wave function~\cite{Cosyn:2013uoa}.

 Suppose the struck quark is confined in a nucleon of four-momentum $p^\mu$ that is bound within a nucleus of momentum $P^\mu$. Then we have 
 \bea {k^+\over p^+} {p^+\over P^+}=\xi.\label{convor}\eea 
where a nucleus of momentum $P^+$ contains a nucleon of momentum of $p^+$ which contains a quark of momentum $k^+$.
This is the origin of the convolution model to be discussed in Sect.~\ref{nuconly}.  Therefore, in order to calculate deep inelastic scattering from nuclei we need to know the nuclear wave function, expressed in light-front variables. 

 More formally, one derives the expression for  the momentum distribution  (the probability that a quark has a given value of $k^+/P^+$), known as a quark distribution function,  
 by starting with the  the square of the invariant scattering amplitude. The important part of  this amplitude depends on the hadronic tensor $W^{\m\n}$, which is a matrix element
 of a commutator of electromagnetic current operators. After expanding in terms of the separation $r$ of the spatial variable of the two current operators, 
 the momentum distribution 
 (for  a  specific flavor of quark)  is given in the Bjorken scaling limit (in which the variable $Q^2$ is not explicit)
 by the Fourier transform~\cite{Thomas:2001kw}
 \bea q(\xi)={1\over 2\pi}\int dr^-\,e^{iq^+ r^-}\la P|\psi_+^\dagger(r^-)\psi_+(0)|P\ra_c,\label{qdist}\eea
 where  $|P\ra_c$ is the proton wave function, the subscript $c$ denotes a connected matrix element,  $\psi$ is the quark field-operator, the subscript $+$ denotes multiplication by the projection operator
 $(1+\g^0\g^3)/2$, and $r^-$ is the minus-component of the separation distance.

\begin{figure}[h] 
\includegraphics[scale=0.45]{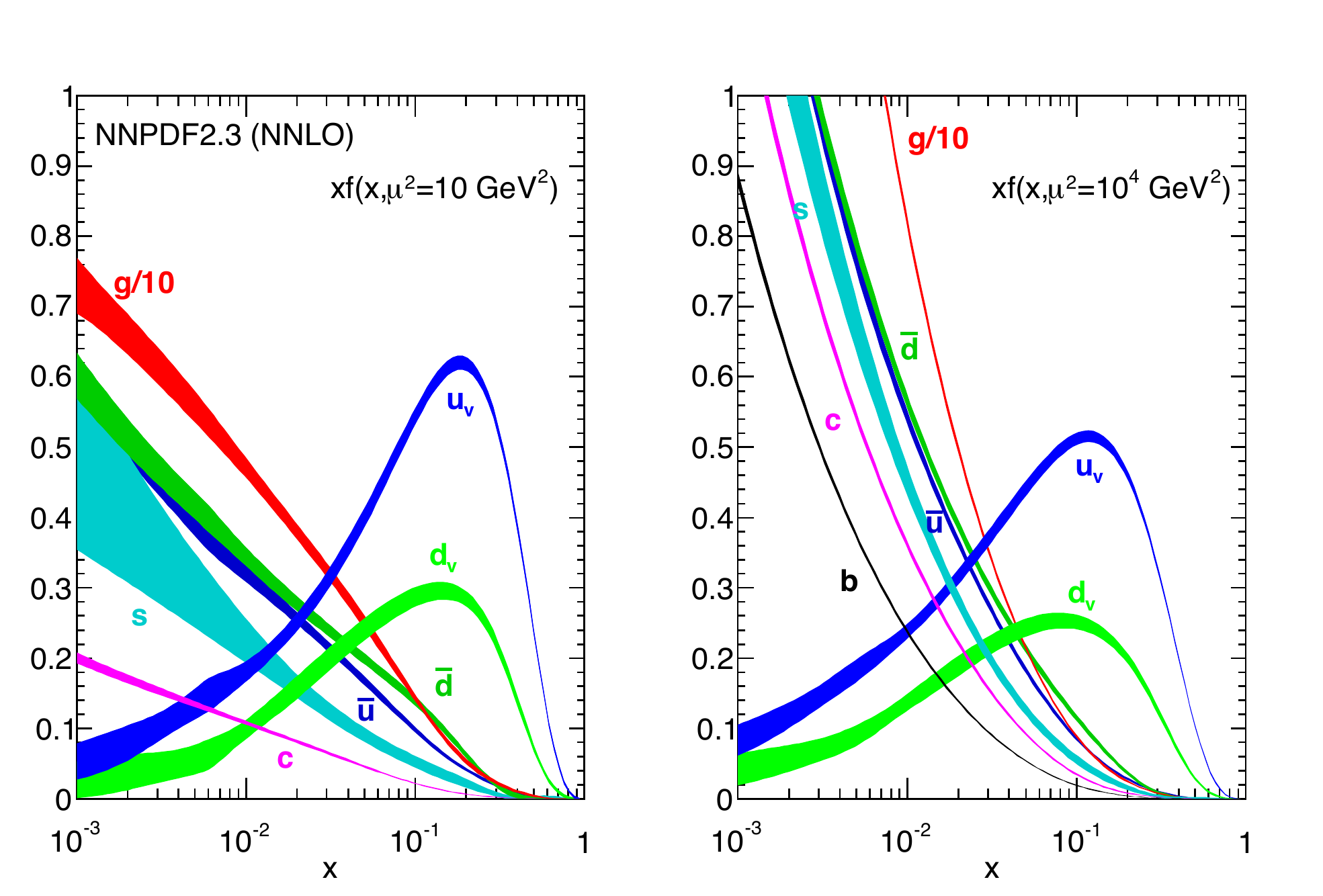}
\caption{\label{fig:pdf} The bands show $\xB$ times the unpolarized parton distributions for the different parton flavors $\{u_v,d_v,u,d,s = \bar{s},c = \bar{c}, b = \bar{b},$ and $g\}$ obtained in NNLO NNPDF2.3 global analysis~\cite{Ball:2012cx}, at  $Q^2 = 10$ (GeV/c)$^2$ and  $10^4$ (GeV/c)$^2$, with $\a_s(M^2_Z) = 0.118$. 
From the PDG~\cite{Agashe:2014kda}. Here $x=\xB$.}
\end{figure}

Parton distributions are needed for a wide variety of applications in  high-energy physics. $q(\xB,Q^2)$ has been determined for various flavors and for a wide range of values of $x$ and $Q^2$.  
Vast amounts of data are now codified as parton distributions, giving the probability  as a function of $Q^2$ that a given flavor of quark  carries a momentum fraction $\xB$, see Fig.~\ref{fig:pdf}. 

This sub-section is concerned with nucleon targets, but  (as mentioned above)  we  need to know how to evaluate a nuclear  version of \eq{qdist},which  would involve nuclear wave functions     expressed in terms of light front variables.
This  difficulty has been handled~\cite{Frankfurt:1988nt, Frankfurt:1981mk,Smith:2002ci},\cite{Miller:2001tg,Blunden:1999gq,Miller:2000kv,Miller:1998tp}. 
One can implement light-front coordinates using a simple transformation. This works because 
the nucleus does not contain a significant $N\bar{N}$ content.
                 
\subsection{The EMC effect}
\begin{figure}[h] \includegraphics[scale=0.4]{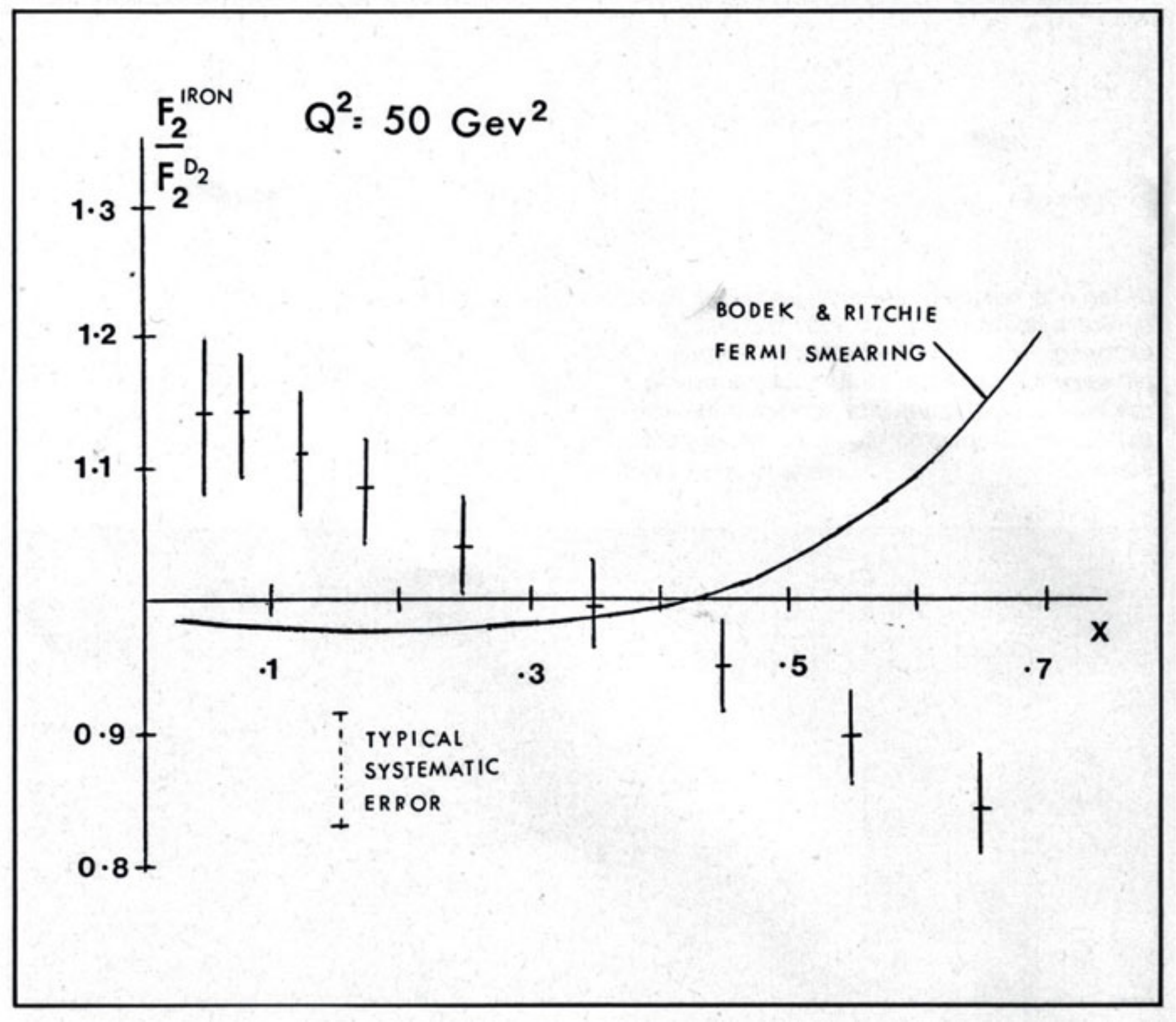}
  \caption{\label{fig:cern} Image of the EMC data as it appeared in the
    November 1982 issue of the CERN Courier. This image nearly derailed
    the refereed publication \cite{aubert83}, as the editor argued that
    the data had already been published.}
\end{figure}

\begin{figure}[htp]
\includegraphics[width=3.33in]{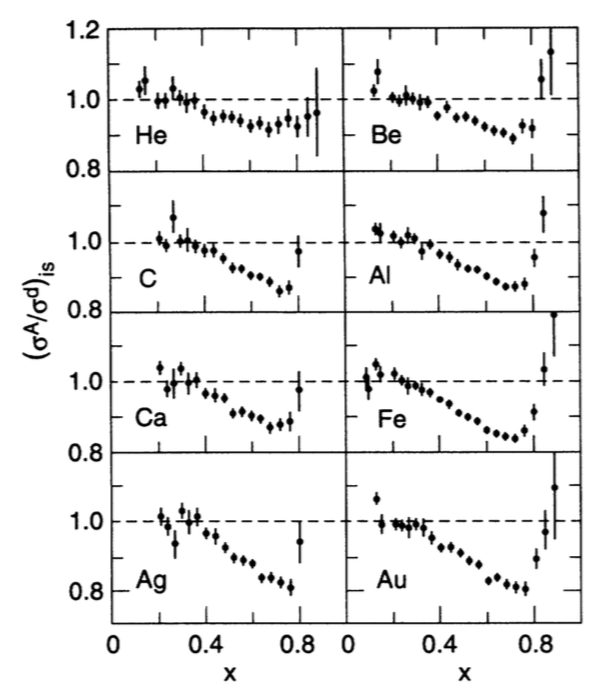}
\includegraphics[width=2.67in]{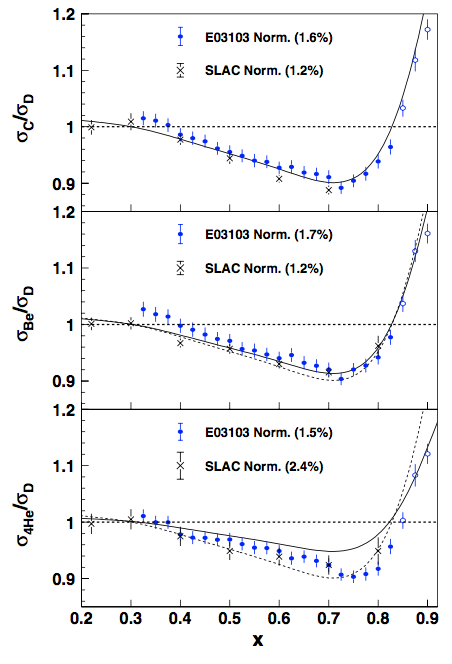}
\caption{\label{fig:Emc} The per nucleon cross section ratio of
  various nuclei to deuterium as measured at (top) SLAC \cite{Gomez94}
  and (bottom)  Jefferson Lab \cite{Seely09}. The solid curves show
  the A-dependent fit to the SLAC data~\cite{Gomez94}, while the
  dashed curve is the SLAC fit to $^{12}$C. Figures from  \cite{Gomez94} (top) and \cite{Seely09} (bottom). }
\end{figure}

As stated, the discovery of Bjorken scaling  was made using hydrogen and
deuterium targets. It occurred to many experimentalists that MeV-scale
nuclear effects should be negligible at GeV-scale momentum and energy
transfers and that therefore they could increase their experimental
statistics by using nuclear targets. Surprisingly, the CERN
European Muon Collaboration (EMC) found that the per-nucleon $(e,e')$ cross
section ratio of iron to deuterium was not unity \cite{aubert83}, see
Fig.~\ref{fig:cern}. This surprising result, now called the EMC Effect,  was confirmed by many
groups, culminating with the high-precision electron and muon scattering
data from SLAC, Fermilab, NMC at CERN, and Jefferson Lab (see Fig.~\ref{fig:Emc}). See one of the many EMC reviews for details
\cite{Arneodo:1992wf,Geesaman95,Piller:1999wx,Norton03,Hen13,malace14}.

\begin{figure}[htp]
\includegraphics[width=3in]{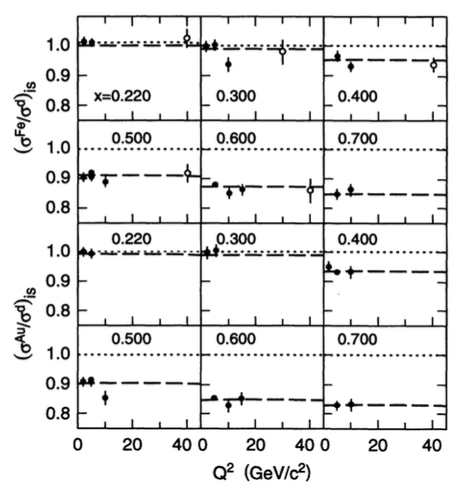}
\caption{\label{fig:EmcQ2} The $Q^2$ dependence of the EMC ratio for iron at various values of \xB{}. From \cite{Gomez94}.}
\end{figure}

\begin{figure}[htp]
  \includegraphics[width=3in]{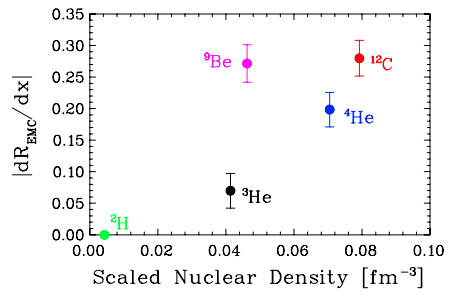}
  \caption{\label{fig:EmcDensity} The slope of the EMC effect for
    $0.35\le \xB \le 0.7$ plotted versus the average nuclear density for
    various light nuclei as measured at Jefferson Lab. From \cite{Seely09}.}
\end{figure}

The conclusion from the combined experimental evidence was that the
effect had a universal shape, was independent of the squared four
momentum transfer $Q^2$ starting from remarkably small values of $Q^2$ (see Fig. \ref{fig:EmcQ2}), increased with
nuclear mass number $A$, and increased with the average nuclear density. An early study~\cite{Bickerstaff:1986fv} of the $Q^2$ dependence of nuclear effects showed that   the nuclear-binding and dynamical rescaling models  predict very little variation with $Q^2$ over the range
from $4 $ to $10^4$ GeV$^2$.

One way to characterize the strength of the EMC effect is to measure the
average slope of the cross section ratio for $0.35\le \xB \le 0.7$.
Plotting this slope versus the average nuclear density for light nuclei
(see Fig. \ref{fig:EmcDensity}) shows that the EMC effect does not
simply depend on average density. Since $^9$Be can be described as a
pair of tightly bound alpha particles plus one additional neutron, it
has been suggested that the local density is more important than the
average density \cite{Seely09}. See an early discussion in~\cite{Frankfurt:1981mk}.

The immediate parton model interpretation of the data at high $x$ is
that the valence quarks of a  nucleon bound in a nucleus carry  less momentum than those of free nucleons.
 This notion seems uncontested, but determining the underlying
origin remained an elusive goal for a long time. The great number of models 
created to explain the EMC effect caused one of us to write in
1988~\cite{Miller:1988hj} that ``EMC means Everyone's Model is Cool''.

\subsection{Why Conventional nuclear physics cannot explain the EMC
  effect}\label{convb}
\subsubsection{Nucleons only}\label{nuconly}
\begin{figure}
\resizebox{0.30\textwidth}{!}{ \includegraphics{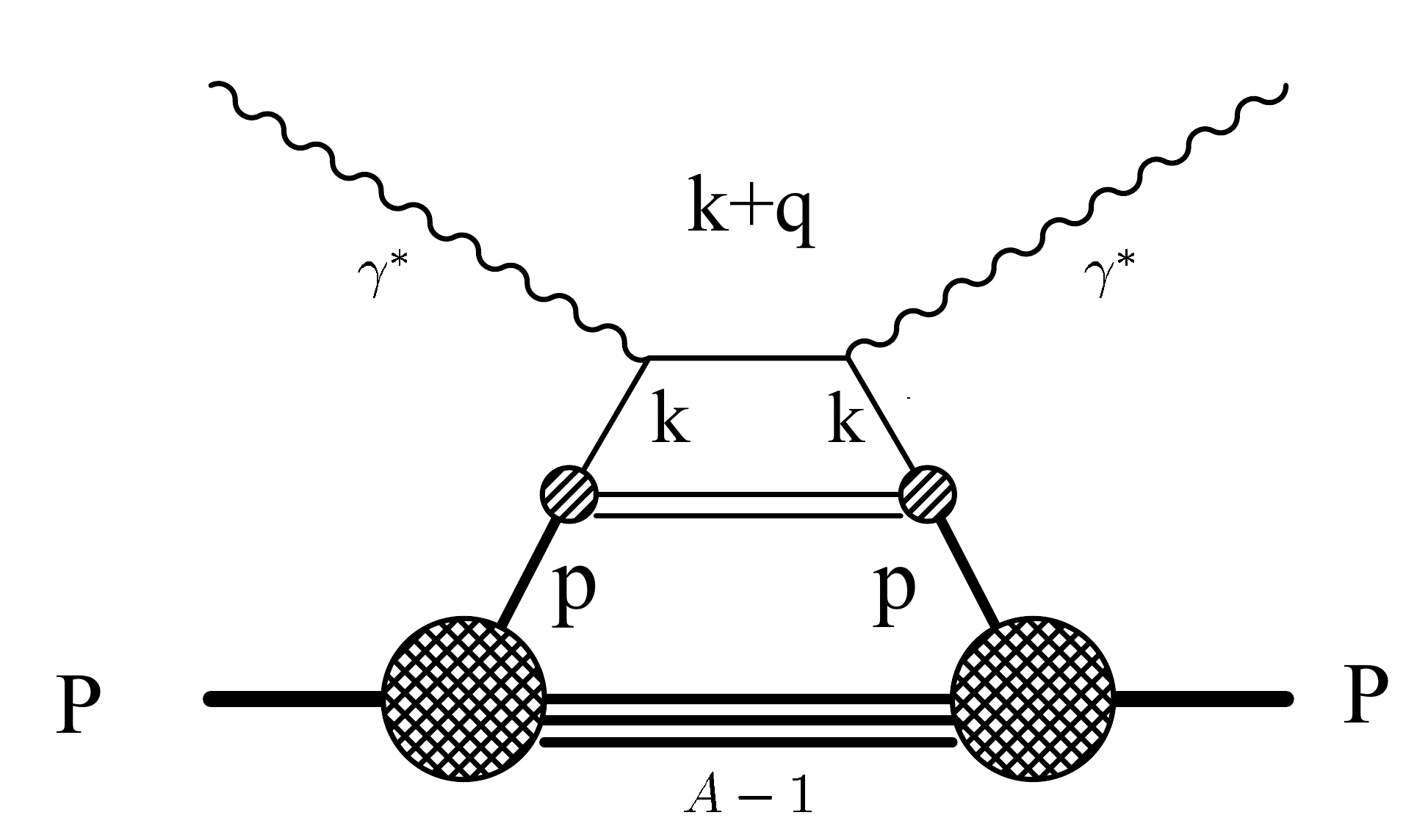}}
\vspace*{.75cm}       
\caption{\label{fig:2} Deep inelastic scattering diagram.  A virtual photon,
  $\gamma^*$, of momentum $q$ is absorbed on a quark of momentum $k$ contained in a
nucleon of momentum $p$ in a nucleus of momentum $P$.  The imaginary
part of this diagram corresponds to the hadronic tensor $W^{\mu\nu}$. Figure adapted from \cite{Miller:2001tg}.}
\end{figure}

One must first try to explain the EMC effect using only the simple kinematic effects
of binding energy and Fermi motion  without modifying the bound
nucleon structure.  If the nucleon structure function is not modified and
is the same on and off the energy shell (nucleon-only hypothesis) 
then evaluation of the diagram of Fig.~\ref{fig:2} 
leads to the simple  convolution  formula:
\bea
{F_{2A}(x_A)\over A}=\int^A_{x_A} dy f_N(y) F_{2N}(x_A/y),\label{obe}
\eea
where $P$ is the total four momentum of the nucleus, and
\be x_A\equiv \frac{Q^2A}{2P\cdot q}=\frac{\xB AM}{ {M}_A}.
\ee 
with $M$ and $M_A$ as the free nucleon and nuclear masses,
respectively.  $x_A$ can be thought of as a version of \xB{} corrected
for the average nucleon binding energy.  The variable $y=Ap^+/P^+$ is the
fraction of the nuclear momentum (per nucleon) carried by a single
nucleon, and $f_N(y)$ is the corresponding probability distribution.
The origin of the convolution formula can be understood using the
simple terms of Sect.~\ref{DisStrucFun}. Suppose the struck quark is
confined in a nucleon (of four-momentum $p$) that is bound within a
nucleus of momentum $P$. Then  from \eq{convor} we have
\bea
\xi \approx {k^+/P^+}=({k^+/p^+})( {p^+/
   P^+})={x_A/ y}.
\eea  
 This accounts for a nucleon in the nucleus of momentum $p^+$  that contains  a quark of momentum $k^+$.
 A proper evaluation of deep inelastic scattering from nuclei therefore involves knowledge of the nuclear wave function, expressed in light-front variables.

 There were many attempts to explain the EMC effect without invoking  medium modifications.  We cite  a few of the references~\cite{Jung:1988jw,CiofiDegliAtti:1989eg,Jung:1990pu,Akulinichev:1990su,Dieperink:1991mw,Benhar:2012nj,Marco:1995vb,Benhar:1999up,Benhar:1997vy}, with others to  be found in the reviews.
 
 The appeal of the nucleon-only idea can be understood using a simple
 caricature of the probability that the nucleon carries a momentum
 fraction $y$.  The width of the function $f_N(y)$ is determined by
 the Fermi momentum divided by the nucleon mass, which is small. In
 the absence of interactions, $f_N(y)$ is peaked at $y=1$. If the
 average separation energy $S\equiv \e M$ (which for
 nuclear matter can be as large as  70 MeV),
 \cite{Dieperink:1991mw,Benhar:2012nj,Benhar:1999up,Benhar:1997vy}
 then $f_N(y)$ is peaked at about $y=1-{\e}$. Taking for
 simplicity a zero width approximation 
\bea
 f_N(y)=\d(y-(1-\e)), \label{del}
\eea 
 then the convolution formula (\eq{obe}) tells us that 
\bea
 {F_{2A}(x_A)\over A}\approx F_{2N}\left({x_A\over   1-\e}\right).
\eea 
As shown in Fig.~\ref{fig:pdf} the structure function falls
 rapidly with increasing $\xB$, so that a slight increase in
 the argument leads to a significant decrease in the structure function. In particular, 
\bea
 {F_{2A}(x_A)\over A F_2N(x_A)}\approx 1+\e {F'_{2N} (x_A)\over
   F_{2N}(x_A)}\approx 1-\g\e,
\eea 
where we have assumed $F_{2N}(\xB)\sim(1-\xB)^\g$ at large \xB{} with
$3\le \g\le 4$.  

Frankfurt and Strikman~\cite{Frankfurt:1985ui}, using a more detailed
calculation found that a value of $\e=0.04$ was sufficient to
reproduce the early EMC data.  However, we will show that the ideas of
shifting the value of $x_A$ based on binding energy or separation
energy considerations violates rigorous~\cite{Collins:2011zzd} baryon
and momentum sum rules, and therefore cannot be a viable explanation
of the EMC effect.  Consider a nuclear model in which nucleons are the
only degrees of freedom. There will be a conserved baryon current and
an energy-momentum tensor expressed in terms of these
constituents. This means that when expressed in terms of the
convolution approach of the previous sub-section we must have the
momentum sum rule: 
\bea 
\int dy y f_N(y)=1,\label{nsr} 
\eea 
where the factor of $y$ represents
the momentum. The use of \eq{del} in \eq{nsr} leads immediately to a
substantial violation of the momentum sum rule: \bea \int dy y
f_N(y)=1-\epsilon.\label{miss} \eea Frankfurt and
Strikman~\cite{Frankfurt:1985ui} also included an important
relativistic correction known as the 'flux factor', which
significantly reduces the effects of nuclear binding.

Going beyond the zero width
 approximation only makes this problem worse~\cite{Miller:1999ap}.  The inclusion of the effects of short-ranged correlations broadens the function $f_N(y)$ leading to a value of the ratio that exceeds unity for small values of $x$, an effect found earlier in~\cite{Dieperink:1991mw}.
 A violation of the sum rule by a few
 percent is actually a huge violation, because the EMC effect itself
 is only a 10-15\% effect. Thus nucleon-only models are logically inconsistent and therefore wrong, even if they can be
 arranged to describe the data.

 One might argue that sum rules can not be applied directly to the
 data because of the need to incorporate initial and final state
 interactions. Nevertheless, in using the convolution formalism in the
 nucleon-only approximation one must use a light-front wave function
 of the nucleus consistent with the conservation of baryon number and
 momentum, as discussed above. There is no way to avoid the
 constraints imposed by the sum rules.

\begin{figure}\resizebox{0.40\textwidth}{!}
{ \includegraphics{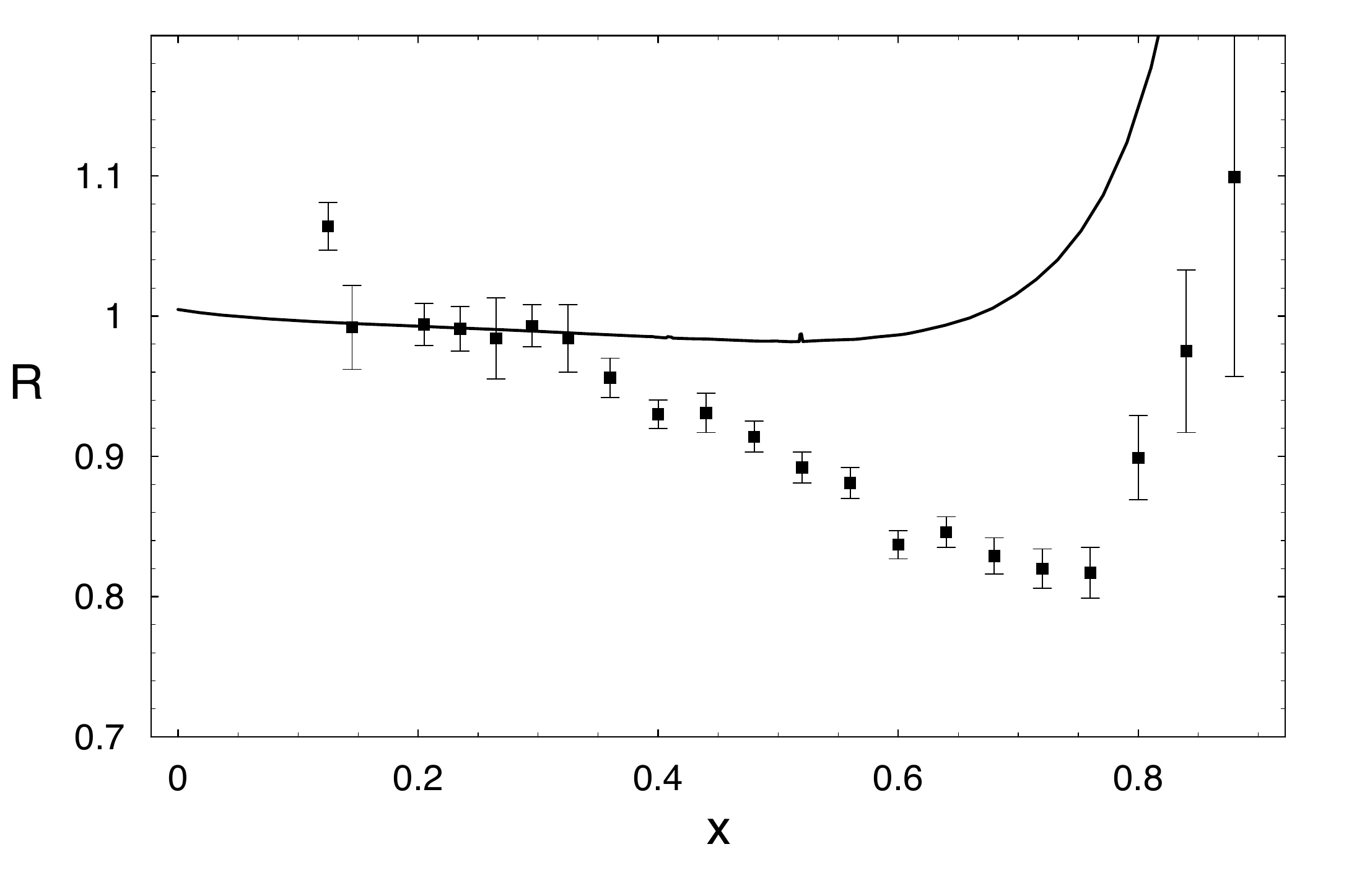}}
\caption{The measured EMC effect in gold \cite{Gomez94} compared to a nucleons-only
  calculation of the EMC effect in lead. From \cite{Smith:2002ci}.}
\label{fig:1}       
\end{figure}

 Indeed, the application of sum rules and simple reasoning shows that
 \eq{obe} leads to the result that the nucleon-only hypothesis can not
 explain the EMC effect. Under the Hugenholz van Hove 
 theorem~\cite{Hugenholtz:1958zz,Miller:2001tg,Smith:2002ci}
 nuclear
 stability (pressure balance) implies (in the rest frame) that
 $P^+=P^-=M_A$. But to an excellent approximation $P^+=A(M_N- 8 \,{\rm
   MeV})$. Thus an average nucleon has $p^+=M_N- 8$ MeV.  As caricatured in \eq{del}, the function $f_N(y)$ is narrowly peaked because the Fermi
 momentum is much smaller than the nucleon mass. This means that the
 value of $y$ in the integral of \eq{obe} is constrained to be very
 near unity. Thus $F_{2A}/A$ is well approximated by $F_{2N}$ and one
 gets no substantial EMC effect this way
 \cite{Miller:2001tg,Smith:2002ci}.  This is shown as the solid curve
 in Fig.~\ref{fig:1}. 

\begin{figure}[htbp]
\includegraphics[width=6 cm]{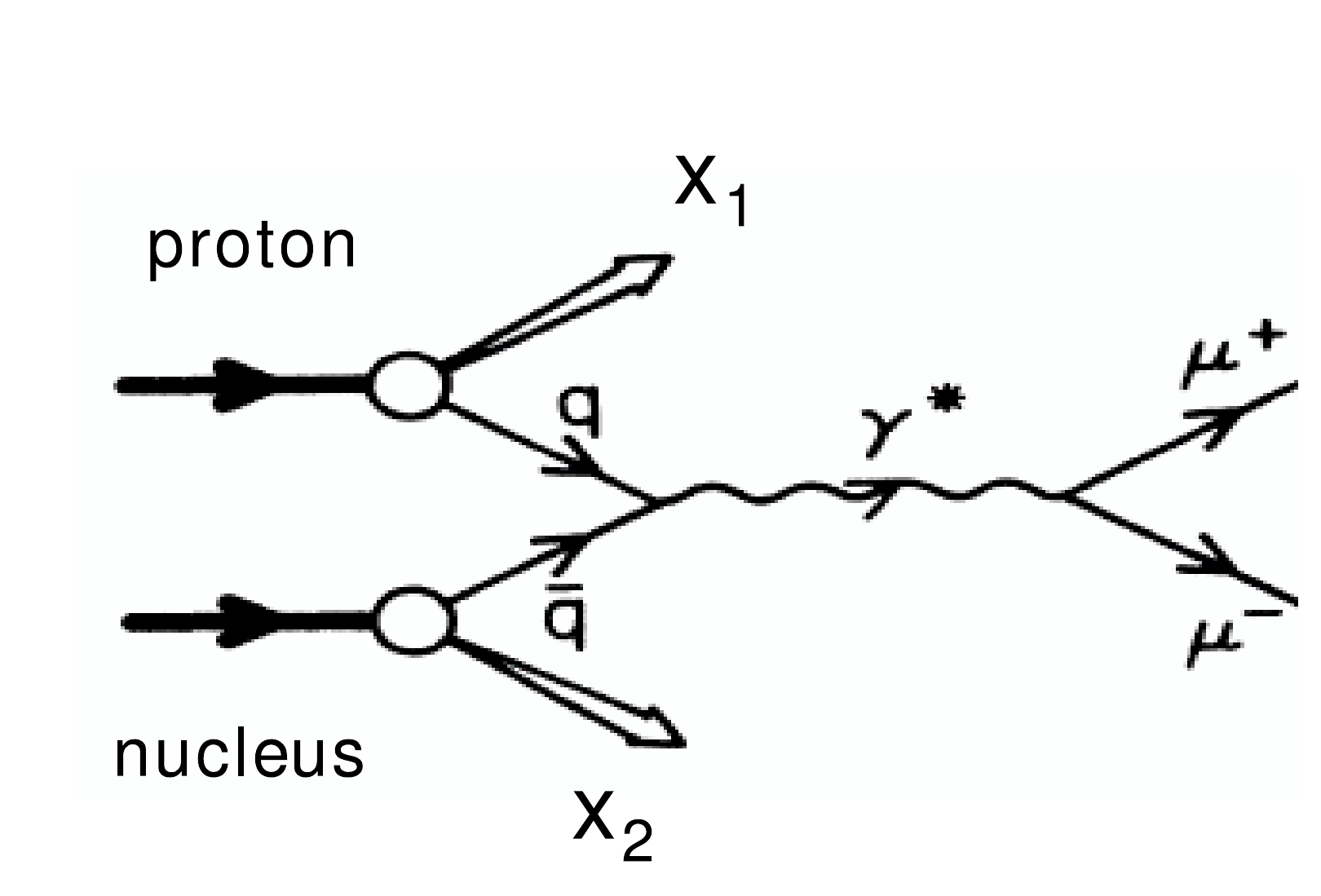}
\caption{The Drell-Yan process.  A quark with momentum fraction $x_1$ from
  the incident proton annihilates with an anti-quark from the nuclear
  target with momentum fraction $x_2$ to form a time-like virtual
  photon which decays to a $\mu^+\mu^-$ pair.  Figure adapted from \cite{Bickerstaff:1985da}.}
\label{dy}       
\end{figure}
\begin{figure}\resizebox{0.45\textwidth}{!}
{ \includegraphics{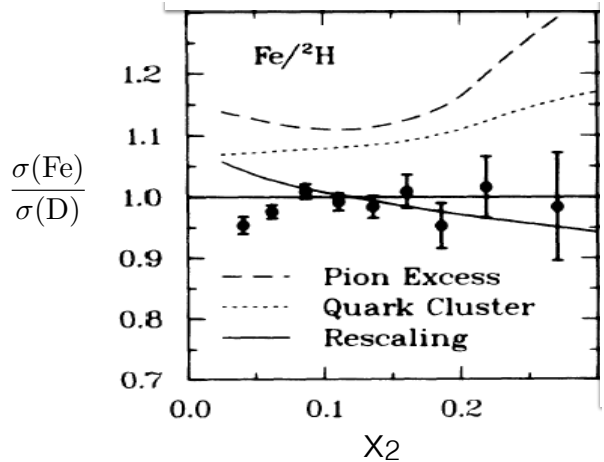}}
\caption{Drell-Yan experimental results~\cite{Alde:1990im}. Ratio of Drell-Yan cross sections as a function of the momentum fraction $x_2$ of a quark in the nucleus. The version of the rescaling model shown in this figure does not reproduce the nuclear deep inelastic scattering data~\cite{Bickerstaff:1985da,Bickerstaff:1985ys}. Figure adapted from~\cite{Alde:1990im}.}
\label{dyexp}       
\end{figure}

\subsubsection{Nucleons plus pions}
Nucleons-only models fail,  but it was natural to consider the
idea that the missing momentum $\epsilon
$ of \eq{miss} is carried by non-nucleonic degrees of freedom, e.g., pions \cite{LlewellynSmith:1983vzz,Ericson:1983um}. In this case,
\bea P^+=P^+_N+P^+_\pi=M_A.
\eea 
Many authors, see the
reviews~\cite{Arneodo:1992wf,Geesaman95,Piller:1999wx,Frankfurt88}
found that using $P^+_\pi/M_A=0.04$ is sufficient to account for the
EMC effect.  However, if nuclear pions carry 4\% of the
nuclear momentum (in the rest frame the plus component of momentum is
the nuclear mass) then there should be more nuclear sea quarks
(i.e., both quarks and anti-quarks).  This
enhancement should be observable in a nuclear Drell-Yan
experiment~\cite{Bickerstaff:1985ax,Bickerstaff:1985da,Ericson:1984vt}.
The idea, see Fig.~\ref{dy}, is that a quark from an incident proton
(defined by a large value of $x_1$) annihilates an anti-quark from the
target nucleus (defined by a smaller value of $x_2$). A significant
enhancement of pions would enhance the anti-quarks and enhance the
nuclear Drell-Yan reaction. But no such enhancement was
observed~\cite{Alde:1990im} as shown in Fig. \ref{dyexp}. This caused
Bertsch {\it et al.} \cite{Bertsch:1993vx} to announce ``a crisis in
nuclear theory'' because conventional theory does not work. This
statement is the verification of the title of this subsection.

The reader might ask at this stage, if the two-pion exchange effects discussed in the Appendix and Sects. I \& II lead to a significant pion content and an enhanced sea in the nucleus. Explicit calculations show that the pionic content associated with
the tensor potential is very small~\cite{Miller:2013hla}.

Subsequent  work has confirmed that an intrinsic modification of the nucleon structure function is needed to explain the EMC effect~\cite{kulagin10,Hen13,Kulagin:2004ie,Kulagin:2014vsa,Frankfurt:2012qs}.  This result had been expected for some time, as stated explicitly ``The change of the structure functions in nuclei (EMC effect) gives direct evidence for the modification of quark properties in the nuclear medium"~\cite{Walecka:2001gs}.
The following sections discuss specific proposals for such modifications.

\subsection {Beyond Conventional Nuclear Physics: Nucleon Modification  \label{SRCvsMF}}

The failure of the nucleon-only or nucleon+pion models to explain the EMC and Drell-Yan data indicates that the structure of a nucleon bound in a nucleus significantly differs from that of a free nucleon.  The medium modifies the nucleon.
  
\begin{figure}
\includegraphics[width=3.in]{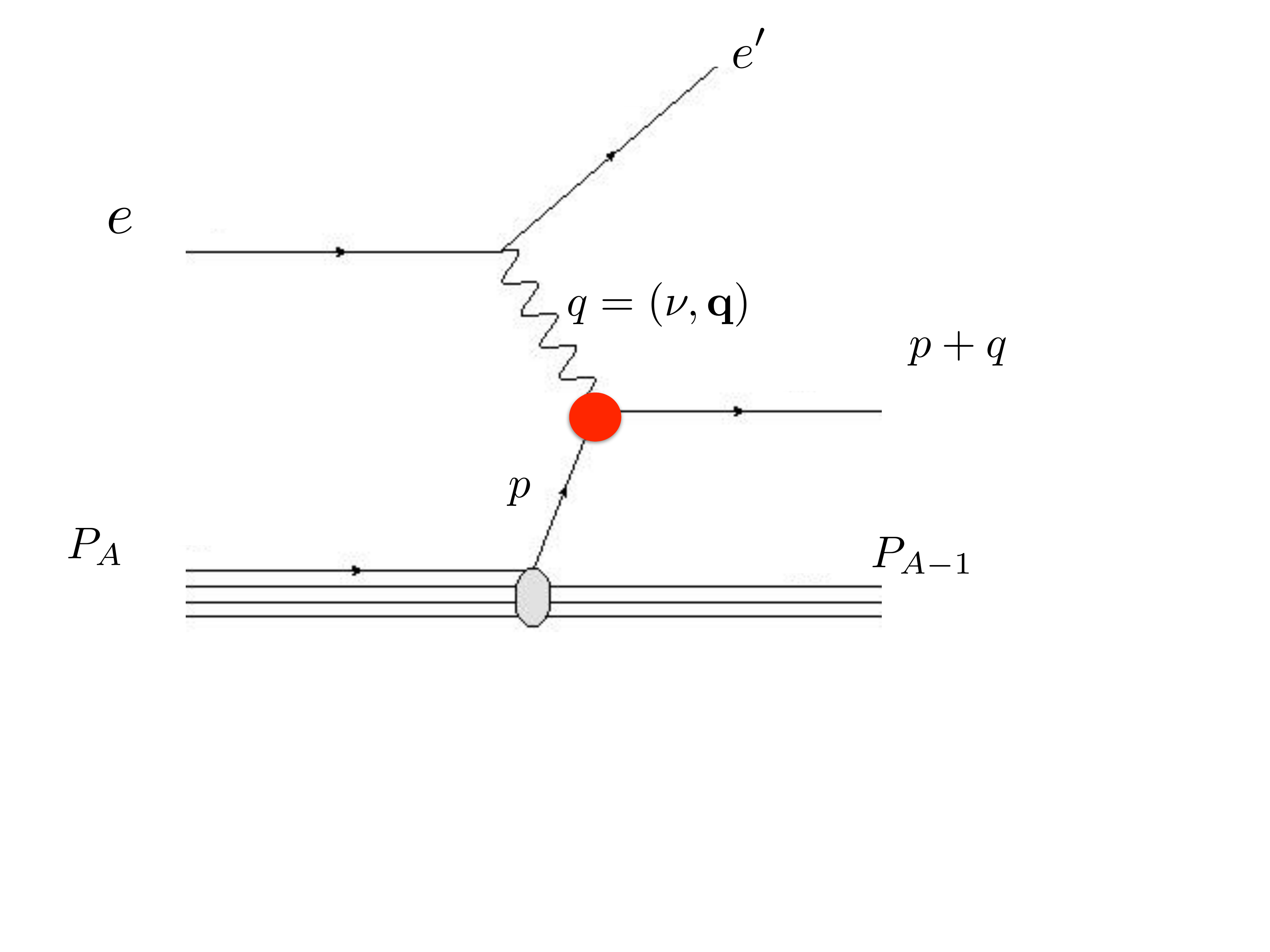}
\caption{\label{fig:pwia} The $A(e,e'p)$ reaction in the Plane Wave Impulse Approximation.  A nucleus of four-momentum $P$ emits a nucleon of four-momentum $p$ that absorbs a virtual photon of four-momentum $q$ to make a nucleon of four momentum $p+q$, with $(p+q)^2=M^2$, where $M$ is the nucleon mass. The blob represents the in-medium electromagnetic form factors. }
\end{figure}

This is not surprising, as there are evident simple examples.  A free neutron undergoes $\b$ decay, so it can be thought of as having a $|pe^-\nu\ra$ component. When bound in a stable nucleus, the neutron is stable. This ``medium modification'' suppresses the $|pe^-\nu\ra$ component.  Additionally, in the $(e,e'p)$ reaction shown in Fig. \ref{fig:pwia}, four-momentum conservation shows that the square of the initial four-momentum of the struck nucleon, $p$, cannot satisfy $p^2=M^2$. Thus the form factor of a nucleon bound in the nucleus cannot be the same as that for a free nucleon; it  is instead the amplitude for a transition between a virtual nucleon of mass $\sqrt{p^2}$ and a physical nucleon of mass $M$.

Now we must ask: what is the origin of the medium modification?  This question is coupled to the broader questions listed in Sect.~\ref{Intro}, and more deeply
to the very nature of confinement.   

\begin{figure}\resizebox{0.40\textwidth}{!}
{ \includegraphics{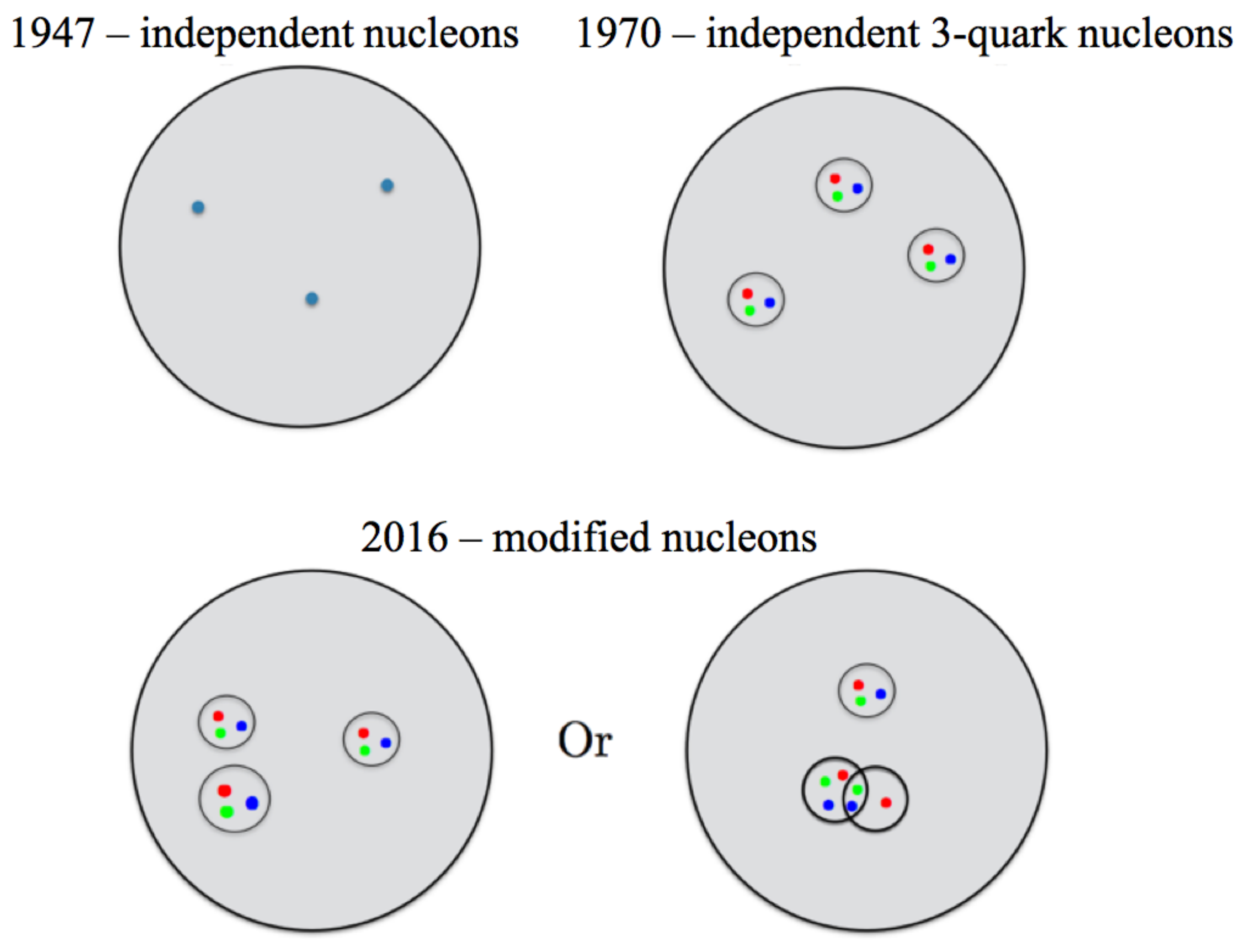}}
\caption{Evolution of nuclear physics from structureless nucleons in the 1940s to independent 3-quark nucleons in the 1970s to the modified nucleons of today, either modified single-nucleons (left) or modified two-nucleon configurations (right).  }
 \label{fig:NpEvolution}      
\end{figure}

The parton model interpretation of the large-\xB{} part of the EMC effect is that the medium reduces the nuclear structure functions for large $\xB$, so  that there are fewer high-momentum quarks in a nucleus than in free space.  This momentum reduction leads, via the uncertainty principle, to the notion that quarks in nuclei are confined in a larger volume than that of a free nucleon.

There are two general ways to realize this simple idea: mean-field effects cause  bound nucleons to  be larger than free ones, or nucleon-nucleon interactions at close range cause nucleon structure to be modified, by including either $NN^*$ configurations or 6-quark configurations that are orthogonal to the two-nucleon wave functions.  All of the   papers seeking to explain the EMC effect using medium modification use one of the two ideas (that are cartooned 
in Fig.~\ref{fig:NpEvolution}).  


Since only about 20\% of nucleons belong to SRC pairs, Fig.~\ref{fig:MomentumDistSketch}, five times more nucleons would be modified by  mean-field effects than by nucleon-nucleon interactions at close range.  Therefore, if nucleons are only modified at short range, then the modifications needed to explain the EMC effect would have to be five times larger than if all nucleons were modified by mean-field effects.

A phenomenological assessment of this idea in which the mean-field and SRC  related  origins of the EMC effect were treated phenomenologically was made in Ref.~\cite{Hen13}.  The separation of the spectral function \eq{eq:01} into terms arising from low-lying excited states $P_0$ and higher-energy continuum states related to short-ranged correlations, $P_1$ was used. In the mean field model, a nucleus-independent modification of $F_2$ was included in the contribution to the nucleon distribution function, $f_N(y)$ \eq{obe} arising from $P_0$. In the alternate model a  much larger nucleus-independent modification of $F_2$ was included in the contribution to $f_N(y)$ arising from $P_1$. Both approaches gave reasonably good descriptions of the nuclear DIS data.

We next describe  specific   models associated with the two different mechanisms.
  
 
    %


\subsubsection{Mean field}
In mean-field models of nucleon modification, the interaction between
nucleons occurs by the exchange of mesons between quarks confined in
different nucleons. Four general models of the quarks
confined in the nucleon have been used for this. 
The earliest model (quark meson coupling, QMC)
used the MIT bag model to represent the three confined quarks in the
proton ~\cite{Guichon:1987jp,Guichon:1995ue,Stone:2016qmi}. Later work
used the QMC model with more general confinement
mechanisms~\cite{Blunden:1996kc}, the covariant NJL
model~\cite{Cloet:2006bq,Cloet:2009qs,Cloet:2015tha} and the chiral
quark soliton model
\cite{Diakonov:1996sr,Smith:2003hu,Smith:2004dn,Smith:2005ra}. 
 In these models
the attraction needed to produce a bound state is generated by the
exchange of scalar quantum numbers (either by a scalar
meson~\cite{Guichon:1987jp,Guichon:1995ue,Stone:2016qmi} or by pairs of
pions \cite{Smith:2003hu,Smith:2004dn,Smith:2005ra}) and the repulsion
needed to obtain nuclear saturation is caused by exchange of vector
mesons. 
Within these mean field models the exchanged mesons are treated as classical static fields, as such  these mesons do not interact with the  photon probe.

\begin{figure}[tbp]
\centering
\includegraphics[width=7.5cm]{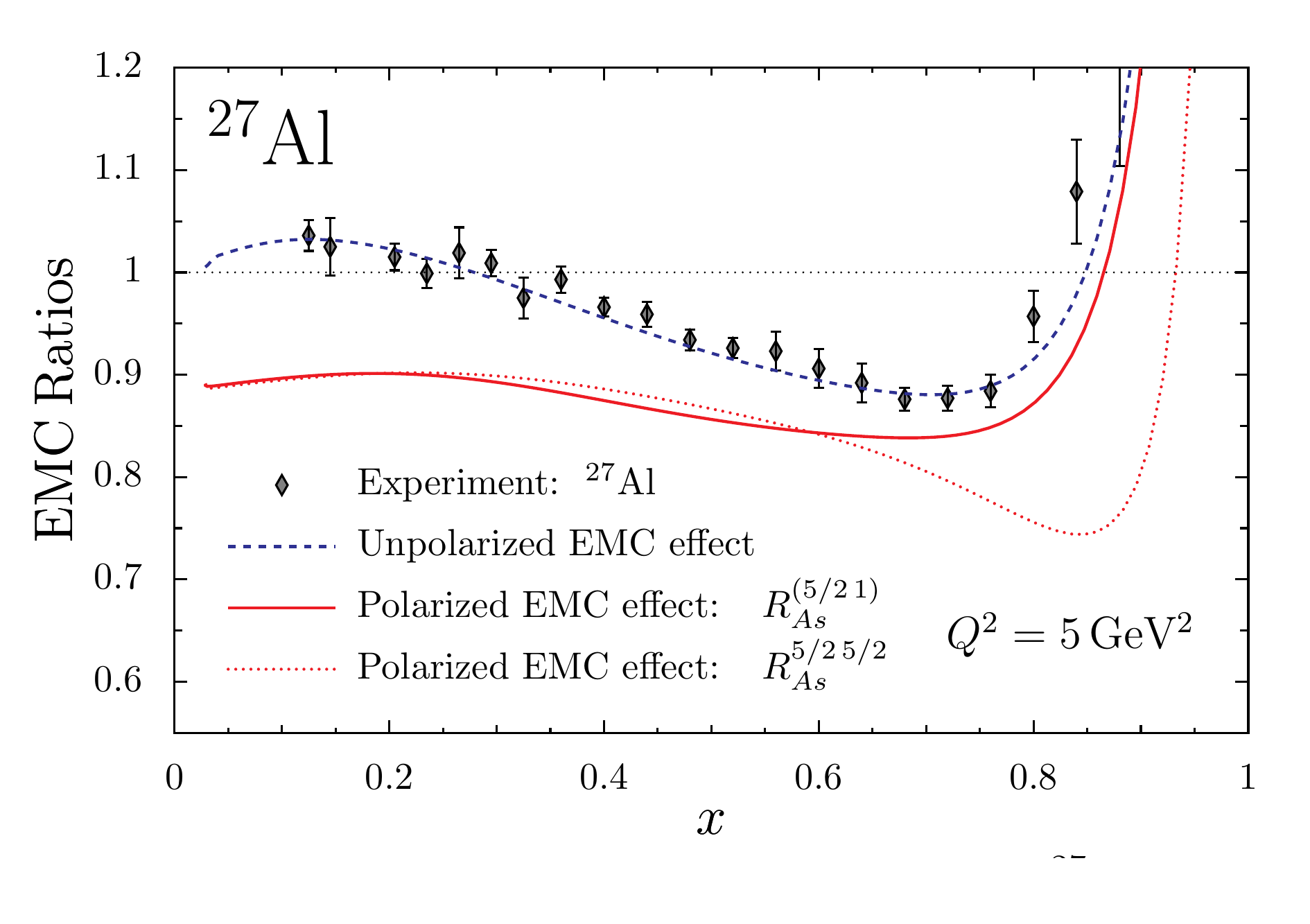}
\caption{(color online) The measured EMC effect for $^{27}$Al
  \cite{Gomez94} compared to QMC calculations of both the regular and
  the polarized EMC effect of Ref. \cite{Cloet:2006bq}. Figure from~\cite{Cloet:2006bq}.  }
\label{fig:EmcQmc}
\end{figure}

We next explain two classes of models.
The chiral quark soliton model (CQSM) 
 is based on the instanton-dominated nature of the vacuum 
\cite{Negele:1998ev}. The coupling of quarks to  vacuum instantons 
spontaneously generates a  constituent quark mass of about 400 MeV. These quarks interact with
pions through an effective CQSM Lagrangian. 
This model reproduces
nucleon properties well, including structure functions which vanish at
$\xB=0$ and 1  \cite{Diakonov:1996sr}.  

Nuclei are formed by collections of such nucleons exchanging scalar
and vector mesons \cite{Smith:2003hu,Smith:2004dn,Smith:2005ra}.
Excellent saturation properties were obtained. The dominant effect of
the medium is a slight broadening of the effective potential that
binds the quarks in the nucleon. The use of the medium modified wave
function to compute structure functions allows one to account for the
EMC effect, while still agreeing with the Drell-Yan data. This
indicates that the sea is not very modified.

The next model places an NJL-model nucleon in the medium (NJLMM) which
is a relativistic extension of the earlier QMC including the effects
of spontaneous symmetry breaking.  Here the external scalar field
enhances the lower component of the quark's Dirac wave function by
about 15\%.  This model describes the EMC effect well (see
Fig.~\ref{fig:EmcQmc}).  It also predicts an enhancement of the EMC
Effect for spin
structure functions \cite{Cloet:2005rt} in nuclei which could be
measured at Jefferson Lab (see Section \ref{PolEmc}).


The NJLMM predicts the effects of having different numbers of
neutrons $N$ and protons $Z$.  Cloet and Thomas~\cite{Cloet:2009qs}
  explained
that a neutron or proton excess in nuclei leads to an isovector-vector
mean-field which, through its coupling to the quarks in a bound
nucleon, causes the quark distributions to be evaluated at a shifted
value of the Bjorken scaling
variable~\cite{Mineo:2003vc,Detmold:2005cb}.  In relativistic
mean-field models, the effect of a vector field is to shift the energy
and therefore the value of the plus component of momentum of the
single particle state.  The isovector-vector mean field is represented
by the $\rho^0$, and in this work its strength is chosen to reproduce
the nuclear symmetry energy.  In a nucleus like $^{56}$Fe or
$^{208}$Pb where $N > Z$, the $\rho^0$ field causes the $u$-quark
to feel a small additional vector attraction and the $d$-quark to feel
additional repulsion.  This effect  leads to 
a significant correction to the NuTeV measurement of $\sin^2\Theta_W$~\cite{zeller02,zeller03}.
The sign of this correction is largely model independent, and it 
 accounts for approximately two-thirds of the NuTeV anomaly. Thus 
  the  NuTeV measurement   provides further evidence
for the medium modification of the bound nucleon wave function.

Both sets of mean field models predict modification of nucleon electromagnetic form factors.
The 
QMC model predicts modifications to both $G_E$ and $G_M$
~\cite{Lu:1998tn}, while the chiral quark soluton model only modifies
$G_E$~\cite{Smith:2004dn}.  Both models predict the same ratio
$G_E/G_M$. Note that electron-nucleus quasi-elastic data was used~\cite{Sick:1986pt} to put a limit of between 3 and 6\% on the possible increase of the nucleon radius in nuclear matter.
None  of the mean field models discussed here  violate this limit. 

The QCD eigenstates of a free nucleon form a complete set. Thus  
the medium modified nucleon can be regarded as a superposition the nucleon and all of its excited states. 

Despite the general success of mean-field models it must be noted that none 
 predicts significant extra
high-momentum strength in the nuclear momentum distribution. Therefore, it is
very difficult to see how they could reproduce the plateaus observed
in the cross section ratios
at $\xB \ge 1.5$  seen in Section \ref{Inclusive}. 


\begin{figure}\resizebox{0.40\textwidth}{!}
{ \includegraphics{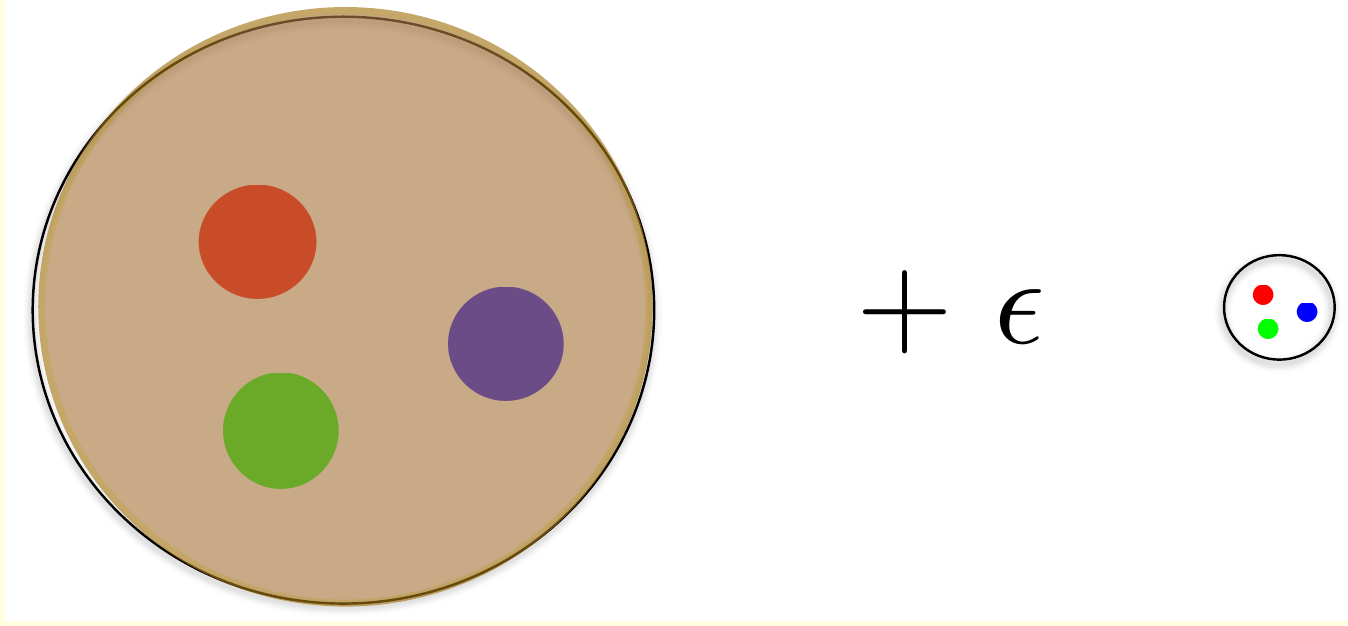}}
\caption{ Two component nucleon model: normal-sized component plus
  point-like configuration component.}
\label{plcsup}       
\end{figure}

\label{MMF}

\subsubsection{Suppression of point-like configurations}
\label{SSplc}

We can also make a more general model of the nucleon as a
superposition of various configurations or Fock states, each with a different
quark-gluon structure.  Fig.~\ref{plcsup} shows a two-component
nucleon where one component is ``blob-like'' (BLC) with the normal nucleon
size and the other is ``point-like'' (PLC).  The BLC can be thought of as an object that is similar to a nucleon.
The PLC is meant to represent a  three-quark system of small size that is responsible for the high-$x$ behavior of
the distribution function. The smaller the number of  quarks, the more likely one can carry a large momentum fraction. 
Furthermore, because the PLC is smaller than the BLC, the uncertainty principle tells one that 
quarks confined in the PLC have higher momentum. The small-sized configuration (with its small number
of $q\bar{q}$ pairs) is very different than a low lying nucleon excitation.

When placed in a nucleus, the blob-like configuration feels the
regular nuclear attraction and its energy decreases. The
point-like-configuration feels far less nuclear-attraction because the
effects of gluons emitted by small-sized configurations are cancelled
in low-momentum transfer processes.  This effect is termed color
screening and has been verified in several different
reactions~\cite{Frankfurt:1994hf,Dutta:2012ii}.  The nuclear
attraction increases the energy difference between the BLCs and the
PLCs, therefore reducing the PLC
probability~\cite{Frankfurt:1985cv}. The PLC is suppressed.
Reducing the probability of PLCs in the nucleus reduces the
quark momenta, in agreement with the EMC effect.

\begin{figure}\resizebox{0.40\textwidth}{!}
{ \includegraphics{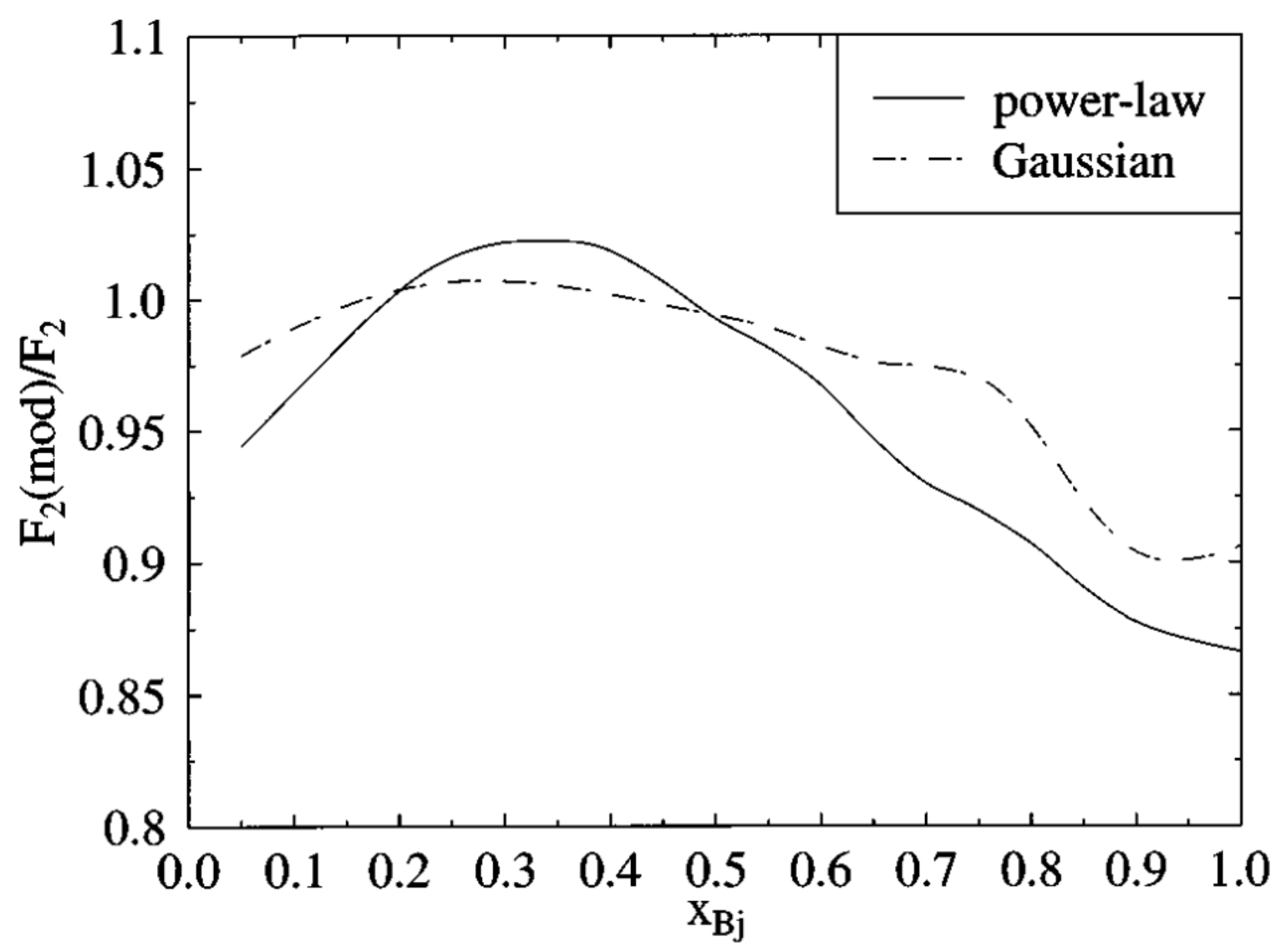}}
\caption{The ratio of $F_2$ in the nucleus to the free $F_2$ (the EMC
  ratio) in the
  point-like configuration suppression model.
 From \cite{Frank:1995pv}.  }
\label{plcemc}       
\end{figure}

This idea was studied \cite{Frank:1995pv} using a relativistic constituent quark model for
the nucleon~\cite{Schlumpf:1992ce,Schlumpf:1992vq}. 
A nucleon is placed  in the nucleus and therefore subject to a mean field that
vanishes for configurations in which the three quarks are close
together. The quark momentum distribution decreases for
$\xB>0.3$, see  Fig.~\ref{plcemc}.  The effects of nucleon motion are not included, so there is no rise for large values of $\xB$, and
the dip  at low values of $\xB$ would be removed by such effects.  This model gives only a 2.5 \% enhancement at $\xB=0.5$ because the enhancing effects of
large virtuality discussed below were not included. The PLC model,
being a modification at large values of $\xB$, does not contradict the nuclear Drell-Yan data, Sect.~\ref{EMC}.

The notion that different constituents of the nucleon have different
sizes and therefore different interaction strengths is directly
related to medium modifications of all kinds.  The main features of this 
  idea can be understood using  a simple schematic 
two-component model of the nucleon with a dominant normal-sized blob-like constituent
(denoted by $B$) and a very small point-like constituent (denoted by $P$).
 The Hamiltonian is given by the matrix 
\bea H_0=
\left[ \begin{array}{cc}
    E_B & V \\
    V & E_P \end{array} \right], \eea 
where $E_P\gg E_B$. Because of the hard-interaction potential, $V$, that connects the two
components, the eigenstates of $H_0$ are $|N\ra$ and $|N^*\ra$  rather
than $|B\ra$ and $|P\ra$.  In
lowest-order perturbation theory, the 
eigenstates are given by 
\bea 
|N\ra&=&|B\ra  +\e|P\ra, \\
|N^*\ra&=&-\e|B\ra +|P\ra,  
\eea
with
$
\e=V/(E_B-E_P). \label{eps} 
$ 
We assume $|V|\ll E_P- E_B$, so that  the nucleon is mainly $|B\ra$ and its excited
state is mainly $|P\ra$, and also take $V>0$. We use the notation $|N^*\ra$ to denote the state that is mainly a PLC, but the PLC, as discussed above,  does not resemble a low-lying baryon resonance.

 Now suppose the nucleon is bound to a nucleus. The nucleon feels an attractive nuclear potential $H_1:$ 
 \bea H_1= \left[ \begin{array}{cc}
U & 0 \\
0 & 0  \end{array} \right]
 \eea
to represent the idea that  only the large-sized component of the nucleon feels the influence of the nuclear attraction. The treatment of the nuclear interaction, $U$,  as a number is clearly a simplification. 
The interaction varies with the relevant kinematics, and our model
will include this dependence explicitly. Our model is similar to 
   the model of~\cite{Frankfurt:1985cv}, with the important difference that the medium effects will enter as an amplitude instead of as a probability.
In~\cite{Frank:1995pv}  the PLC is subject to a non-zero, but
small, attractive potential that fluctuates with the nucleon configurations. The complete Hamiltonian $H=H_0+H_1$ is now  given by
\bea H= \left[ \begin{array}{cc}
E_B-|U| & V \\
V & E_P  \end{array} \right],
 \eea
in which the attractive nature of the nuclear  binding potential is
emphasized. Then interactions with the nucleus increase the energy difference between
the BLC and the PLC, which  decreases the PLC
probability.

The medium-modified nucleon and its excited state,
$|N\ra_M$ and $|N^*\ra_M$, are now (using first-order perturbation theory)
\bea 
|N\ra_M &=& |B\ra +\e_M|P\ra\ \label{mod1}\\
|N^*\ra_M&=& -\e_M|B\ra +|P\ra, 
\eea
where
\bea
\e_M={V\over{E_B-|U|-E_P}}= \e{E_B-E_P\over E_B-|U|-E_P}
\label{epsm}
\eea
so  that the PLC probability in the medium is suppressed. Both $\e_M$ and $\e$ are less than zero, so that $\e_M-\e>0.$

The medium modified nucleon  $|N\ra_M$ may be expressed in terms of
the unmodified eigenstates $|N\ra,|N^*\ra$ as
\bea
|N\ra_M\approx|N\ra+(\e_M-\e)|N^*\ra. 
\label{modnuc}\eea
Within this model the  medium-modified nucleon contains a component that  is an excited state of  a free nucleon. The amount of modification, $\e_M-\e$, which gives a  deviation of the EMC ratio from
unity, is controlled by   the
  potential $U$.   
An initial pioneering qualitative
description of the EMC effect was obtained~\cite{Frankfurt:1985cv}   (at $\xB=0.5$, where effects of   Fermi motion are small) using  $U=-40$ MeV and
$E_P-E_B \sim  500 $ MeV.   The present treatment instead calculates
the effects of the medium  on the amplitude instead of the
probability,  so that the effects are generally larger.   We will explore this further in Section \ref{EMC_SRC_Theory}.

 The PLC suppression model also predicts changes to the elastic
 electric and magnetic form factors $G_{E,M}$.    The electromagnetic form factor in free-space is obtained  as 
\bea F= {1\over 1+\e^2}\left(\la  B| J|B\ra+2\e\la B|J|P\ra+\e^2\la P|J|P\ra\right),\,\label{ff}
\eea
where momentum and spin labels have been  suppressed.

 \begin{figure}\resizebox{0.40\textwidth}{!}
{ \includegraphics{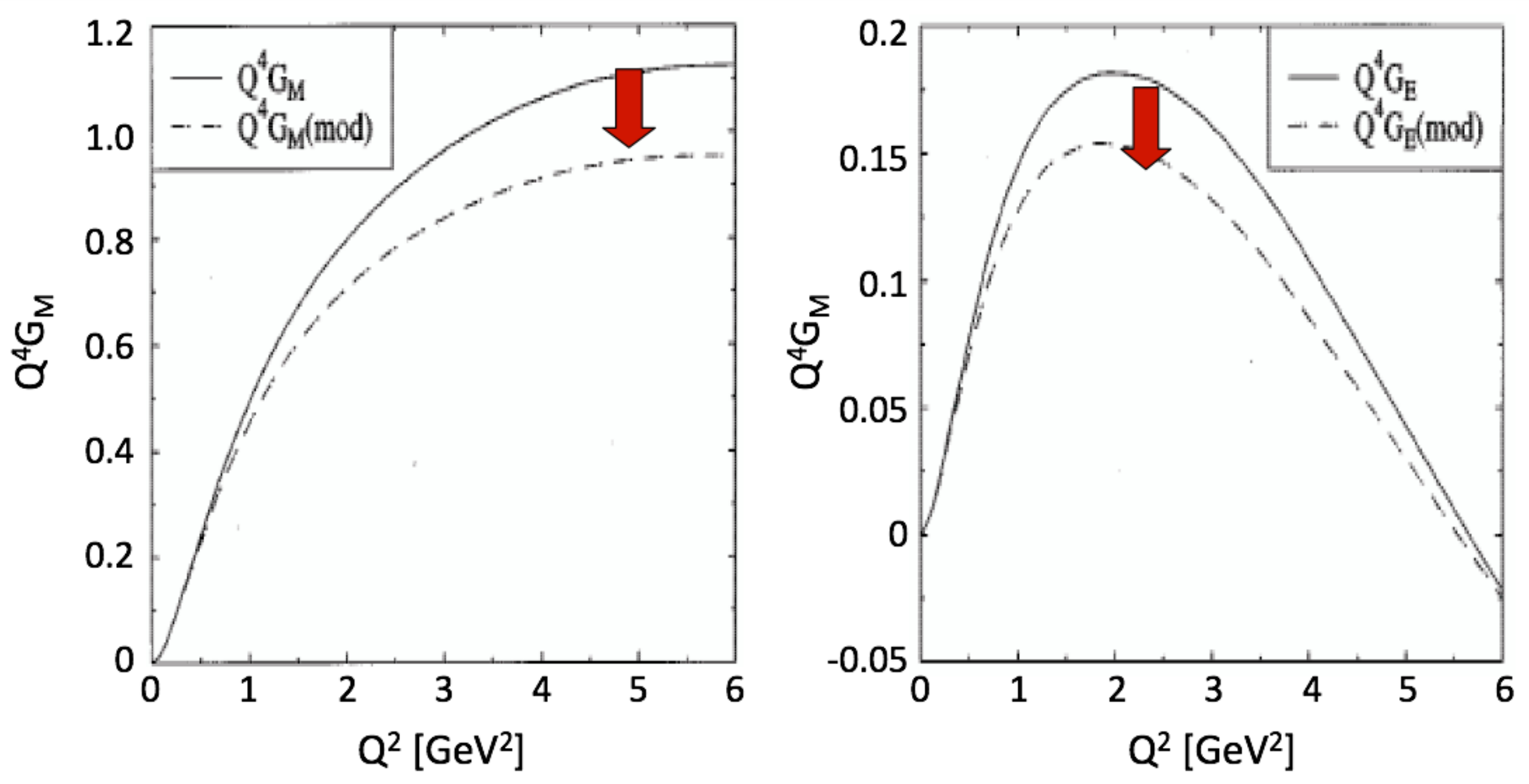}} 
\caption{Medium modification of form factors. Figure adapted with slight modifications. Adapted from~\cite{Frank:1995pv}.  }
\label{modss}       
\end{figure}  

It is instructive to examine what to expect at both high and low momentum transfer. At low momentum transfer the first term dominates so that the spatial extent of the nucleon and its modification in the medium are important.
Frankfurt and Strikman~\cite{Frankfurt:1985cv}  estimated the value of $\la r^2\ra$. Assuming that only the blob-like configuration $|B\ra$ contributes to this long-ranged observable, one finds
\bea \la r^2\ra={\la  B| r^2|B\ra\over 1+\e^2}.\label{rsq}\eea
In the medium the potential $U$
acts, so the value of   $\e$ is changed to $\e_M$.
Since $\vert\e_M\vert<\vert\e\vert$, immersion of the nucleon in the medium suppresses the point-like components and increases $\la r^2\ra$.  The effect is of order $\e (\e_M-\e)$,  which was estimated to be between 2 and 5\%.

At high momentum transfer, the term $2\e \la B|J|P\ra$ becomes
dominant. 
Then the change in the form factor is of order $\e-\e_M$, which is a larger effect. 

The application of the  PLC-suppression idea presented in the present two-state model  is schematic: it does not distinguish between the
electric, $G_E$,  and magnetic, $G_M$, form factors.

A more detailed evaluation was included by~\cite{Frank:1995pv}. 
 Medium modifications  of the proton  form factors  were predicted  as shown in
Fig.~\ref{modss}. The important modifications shown by the red arrow occur at larger
values of momentum transfer than  currently accessible experimentally. Fig~\ref{modss} shows fairly significant effects, greater than about
10\% (consistent with our present  analysis)
 for the individual form factors. Experimentally it is easier to measure 
the medium modifications of the ratio
$G_E/G_M$. The figure shows that since both $G_E$ and $G_M$ are
decreased, the change in the ratio $G_E/G_M$ is expected to be
smaller.

In addition to the medium modifications, 
\cite{Frank:1995pv} also predicted the  more spectacular
decrease~\cite{Jones:1999rz,Gayou:2001qd,Punjabi:2005wq} in the
free-proton ratio $G_E/G_M$ with increasing values of $Q^2$. 

\subsubsection{Six-quark bags and the EMC Effect}

One of the earliest attempts to understand the EMC
effect~\cite{Carlson:1983fs,Bickerstaff:1985ax,Jaffe:1982rr} was to
hypothesize that part of the time one nucleon is part of a six-quark
configuration~\cite{Pirner:1980eu} (who predicted the existence of
plateaus in $(e,e') $ cross section ratios) that is orthogonal to any
two-nucleon wave function. Because a
six-quark configuration is larger than a nucleon,  the quarks
are partially deconfined. Larger confinement volumes are associated
with lower momenta, and therefore with a suppression of the structure
function.  The idea was usually implemented through the MIT bag model,
or by guessing the related structure functions. Several reviews
discuss this idea
\cite{Frankfurt88,Arneodo:1992wf,Geesaman95,Sloan:1988qj,Norton03,Berger:1987er,Mulders:1990xw,Miller:1984em}. It
was relatively easy to use this idea to compute a wide variety of
nuclear phenomena
\cite{Miller:1985un,Miller:1984em,Guichon:1983hi,Miller:1982sh,Koch:1985yi,Miller:2013hla}, but the calculation of each new observable was accompanied by the need
to incorporate an additional free parameter. The
use of 6-quark models that describe nuclear DIS led to predictions of
large effects in the nuclear Drell-Yan process discussed in
Sect.~\ref{convb}, but little modification was seen, Fig.~\ref{dyexp},
severely limiting the applicability of six-quark bag models.  In addition, in some applications the necessary six-quark bag probability needed to reproduce the EMC effect is so large as to
conflict with knowledge of nuclei~\cite{PhysRevLett.61.686}.

 For a recent study of the possible influence of hidden-color  and short-ranged correlation effects at EIC energies, see ~\cite{Miller:2015tjf}.

     %

%




\section{The EMC - SRC Correlation} 
\subsection{Experimental Overview} 
\begin{figure}[t]
  \centering
    \includegraphics[width=9cm, height=8cm]{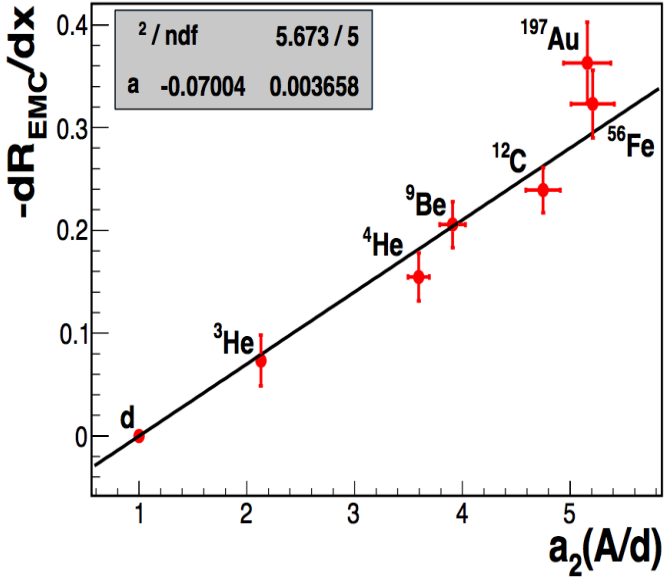} 
    \caption{
      The slope of the EMC effect 
      ($R_{\rm EMC}$, ratio of nuclear to deuteron cross section)  for $0.35 \le x_A \le 0.7$ plotted vs. $a_2(A)$, the SRC scale factor (the relative probability that a nucleon belongs to an SRC $NN$ pair) for a variety of nuclei~\cite{Hen:2013oha}.  The fit parameter, $a=-0.070\pm0.004$ is the intercept of the line constrained to pass through the deuteron (and is therefore also the negative of the slope of that line). From~\cite{Hen:2013oha}. }
     \label{fig:emcsrc} 
\end{figure}

While there is no obvious connection between DIS scattering from quarks in the nucleus at $0.3\le \xB\le0.7$ and QE scattering from nucleons in the nucleus at $1.5\le\xB <2$, analysis of world data showed a remarkable correlation (see Fig.~\ref{fig:emcsrc}) between the magnitude of the EMC effect in nucleus $A$ and the probability that a nucleon in that nucleus is part of a $2N$-SRC pair \cite{weinstein11,hen12}.

\begin{table*}[t]
\centering
\caption{A compilation of world data on SRC scaling factors, $a_2(A)$ and EMC slopes $dR_{EMC}/dx$. Columns 2 through 4 show the SRC scaling factors extracted from various measurements. Column 5 shows the SRC scale factor prediction of~\cite{weinstein11} based on the EMC-SRC correlation. Column 6 shows the world average of the EMC effect slope as compiled by~\cite{weinstein11}, using the data of~\cite{Gomez94,Seely09}. See text for details.}
\begin{tabular}{|l| l | l | l | l | l |}
    \hline
                 										 &\cite{frankfurt93}		&\cite{egiyan06}		& \cite{fomin12}  					& \cite{weinstein11}  			& \cite{weinstein11} \\
                 										 &						&  						& [excluding the CM 				& EMC-SRC Prediction			& EMC Slope    \\
   Nucleus 										 & 	$a_2(A)$				&        $a_2(A)$               		& motion correction]  				&     $a_2(A)$          				& [$dR_{EMC}/dx$]   \\
    \hline
    column $\#$      								 &\multicolumn{1}{c|}{2}	& \multicolumn{1}{c|}{3}	&\multicolumn{1}{c|}{4}     			& \multicolumn{1}{c|}{5} 			& \multicolumn{1}{c|}{6} \\
    \hline
    $^3$He          								 & $1.7 \pm 0.3$		& $1.97 \pm 0.10$      	& $2.13 \pm 0.04$ 					&                            				& $-0.070\pm0.029$    \\
    $^4$He          								 & $3.3 \pm 0.5$		& $3.80 \pm 0.34$      	& $3.60 \pm 0.10$ 					&                            				& $-0.197\pm0.026$    \\					
    $^9$Be          								 &						&                            	    	& $3.91 \pm 0.12$					& $4.08 \pm 0.60$				& $-0.243\pm0.023$    \\
    $^{12}$C        								 & $5.0 \pm 0.5$		& $4.75 \pm 0.41$     	& $4.75 \pm 0.16$ 					&								& $-0.292\pm0.023$    \\
    $^{56}$Fe($^{63}$Cu) 						 & $5.2 \pm 0.9$		& $5.58 \pm 0.45$ 		& $5.21 \pm 0.20$					&								& $-0.388\pm0.032$    \\
    $^{197}$Au     								 & $4.8 \pm 0.7$		&					   	& $5.16 \pm 0.22$					& $6.19 \pm 0.65$				& $-0.409\pm0.039$    \\
    \hline
    \hline
    EMC-SRC slope        	         				&						& $0.079 \pm 0.006$	& $0.084 \pm 0.004$  				&								&      \\
    $\frac{\sigma(n+p)}{\sigma_d}|_{x_B=0.7}$	&						& $1.032 \pm 0.004$     & $1.034 \pm 0.004$				&    							& 		      \\
    $\chi^2/ndf$    								&						& $0.7688 / 3$		  	& $4.895 / 5$						&     							&		 \\								
    \hline
\end{tabular}
\label{tab:EMC_SRC}
\end{table*}

The strength of the EMC effect for nucleus $A$ is characterized as the slope of the ratio of the per-nucleon deep inelastic electron scattering cross sections of nucleus $A$ relative to deuterium, $dR_{EMC}/dx$, in the region $0.35\ge x_B\le0.7$ \cite{Seely09}.  This slope is proportional to the value of the cross section ratio at $\xB\approx 0.5$, but is unaffected by overall normalization uncertainties that merely raise or lower all of the data points together.  
Table \ref{tab:EMC_SRC} shows data from the $x_A$ corrected EMC data base of~\cite{Hen:2013oha} which used the EMC data of \cite{Seely09,Gomez94}.

The SRC scale factors were determined from the isospin-corrected per-nucleon ratio of the inclusive $(e,e')$ cross sections on nucleus $A$ and $^3$He or deuterium.   Columns two through four of Table \ref{tab:EMC_SRC} show the SRC scale factors measured by \cite{frankfurt93, egiyan06,fomin12}.  The large uncertainties in the SRC ratios of \cite{frankfurt93} are due to extrapolating data from different experiments measured at different kinematics.  
The SRC ratios measured by~\cite{egiyan06} were used in the original EMC-SRC analysis of \cite{weinstein11}.  The later results of~\cite{fomin12} include $^{63}$Cu rather than $^{56}$Fe; the SRC scaling factor of $^{63}$Cu is assumed to be the same as that of $^{56}$Fe.  The values of $^9$Be and $^{197}$Au in the fifth column are those predicted by Ref.~\cite{weinstein11} based on the measured EMC effect and the linear EMC-SRC correlation.  These predictions are in remarkable agreement with the later results of~\cite{fomin12}.  Following~\cite{hen12}, the~\cite{fomin12} results are shown without the center of mass motion correction (i.e., including inelastic, radiative, and coulomb corrections only).  Applying the SRC-pair center of mass motion correction decreases the ratios by 10\% to 20\%.

The EMC effect correlates imperfectly with other $A$-dependent quantities (see~\cite{Seely:2009gt, Arrington12} and references therein).  In general, nuclei with $A\ge 4$ fall on one straight line but deuterium and $^3$He do not.  This is true when the EMC effect is plotted versus $A$, $A^{-1/3}$, or the average nuclear separation energy.  When plotting the EMC effect versus average nuclear density, $^9$Be is a clear outlier (see Fig.~\ref{fig:EmcDensity}).  This indicates that the excellent correlation with the SRC scale factor is not just a trivial byproduct of their mutual $A$-dependence.  
 
The correlation between the EMC effect and the SRC scale factor is robust~\cite{hen12}.  It applies to both SRC data sets of~\cite{Egiyan:2006} and~\cite{Fomin:2012}.  The quality of the correlation also does not depend on the corrections applied to the SRC data.  These corrections include 
isoscalar cross section corrections, center-of-mass motion corrections, and isoscalar pair-counting corrections.  The isoscalar correction to the SRC scale factors accounts for the different elementary electron-neutron and electron-proton cross sections.  This has a negligible effect on the fit quality and the extracted fit parameter.  Fomin \etal{} did not apply this correction, arguing that short range correlations are dominated by $np$ pairs.  Fomin \etal{} also argued that the SRC scale factors measured the relative probability of finding a high-momentum nucleon in nucleus $A$ relative to deuterium and that these scale factors needed to be corrected for the center-of-mass (cm) motion of the pair in order to determine the relative probability that a nucleon in nucleus $A$ belongs to an SRC pair.  As shown in both \cite{hen12} and \cite{Arrington12}, including the pair c.m. motion correction improves the EMC-SRC correlation only slightly.

This EMC-SRC correlation gives new insight into the origin of the EMC effect. As discussed in Sect.~\ref{EMC}, many different explanations of the EMC effect have been proposed since 1983.  
After accounting for the standard nuclear effects of binding energy and Fermi motion, explanations for the EMC effect fall into two general categories, those that require modifications 
of mean-field nucleons and those that require modifications of high-momentum (large virtuality) nucleons.  

{\bf{
The linear correlation between the strength of the EMC effect and the SRC scale factors indicates that possible modifications of nucleon structure occurs in nucleons belonging to SRC pairs.  
This implies that the EMC effect, like short range correlations, is a short-distance, high virtuality, and high density phenomenon. 
}}

\subsection{Theory Overview \label{EMC_SRC_Theory}}
%
%
\subsubsection{High momentum nucleons and PLC suppression}

Next we try to use the EMC-SRC correlation to better understand the relationship between
short-ranged correlations measured in the $A(e,e')$ reaction and deep
inelastic scattering reactions.  Both processes involve a probe that
strikes a nucleon of four-momentum $p$ in the nucleus,
Fig.~\ref{fig:pwia}. It is natural to expect that the medium
modification depends on the virtuality
$v(\bfp,E)$ of the struck nucleon ~\cite{Ciofi07}:
\bea
 v\equiv p^2-M^2=(P_A-P_{A-1})^2-M^2.
\label{eq:virtuality}
\eea

  In the $(e,e'p)$ reaction in PWIA (see Fig.~\ref{fig:pwia}), the
  nucleon initial momentum 
opposes the $A-1$ recoil momentum $\bfp=-\bfP_{A-1}$. Using the recoil mass
  $M_{A-1}^*=M_A-M+E,$ where $E>0$ represents the excitation energy of
  the spectator nucleus 
(known as the removal energy~\cite{cda96}), we find
  \begin{eqnarray}
 v(\bfp,E)&=&\left(M_A-\sqrt{(M_{A-1}^*)^2 +\bfp^2}\,\right)^2
 -\bfp^2-M^2
 \end{eqnarray}
which reduces to
  \begin{eqnarray}
 v(\bfp,E)&\approx& -2M\left({A\over A-1}{\bfp^2\over 2M}+E\right),  
\label{virt}
\end{eqnarray}
 in the non-relativistic limit. The   magnitude of the virtuality, $v(\bfp,E)$
 increases with both the $A-1$ excitation energy and the  initial momentum of the struck nucleon.

\cite{Frankfurt:1985cv,Ciofi07} obtained a relation between
the potential $U$ of Section \ref{SSplc} and the virtuality
$v(\bfp,E)$ by using the extension of the Schroedinger equation to an
operator form: 
\bea {\bfp^2\over 2M_r}+U=-E,
\eea 
where $M_r= M
(A-1)/A$, and $U$ is the interaction that both binds the nucleon to
the nucleus and modifies its structure.  The simple idea behind this
equation is that, if the nucleon binding energy is fixed, then the
$NN$ interaction energy, $U$, must become more negative as the kinetic energy
becomes more positive. In this work the modification of nuclear properties was found to be proportional to $v(\bfp,E)$ for moderate values of the virtuality. It should be noted that the short ranged correlations give a dominant contribution to the average nucleon virtuality, which naturally leads to an approximate proportionality of the EMC effect to $a_2$. 

Comparing this equation with \eq{virt} one finds
 \bea U=\frac{v(\bfp,E)}{2M_r},\label{uu}
\eea
so that    the modification of the nucleon due to the PLC suppression is proportional to its virtuality. Potentially large values of the virtuality  greatly enhance the difference between  $\e_m$ and $\e$. 

Now we need to understand how the structure function changes in the
medium. In principle one needs to calculate the hadronic tensor
$W^{\m\n}$ and $q(x)$ for the medium modified nucleon of \eq{modnuc}
by replacing the state $|P\ra$ in \eq{qdist} by the state $|N\ra_M$.  To
leading order, the change in the structure function will be linear in
$\e_M-\e$. The hadronic part is an off-diagonal matrix element between
a free physical nucleon, $|N\ra$ and a free physical state $|N^*\ra$.
Thus the modification is the product of a coefficient that depends on
the medium and a term that is independent of the medium.

These hadronic matrix elements have not yet been calculated.  Instead we adopt a phenomenological
  approach, based on the suppression of point-like configurations~\cite{Frankfurt:1985cv,Frank:1995pv} where
the medium modified quark structure function is given by the expression 
\bea  q_M(x)= q(x) +{(\e_M-\e) } f(x) q(x),
\eea
with  the suppression of point-like components manifest by the condition $df/dx<0$.
so that the ratio of structure functions is given by $R(x)=q_M(x)/q(x)$, so that 
\bea 
{dR\over dx}&=& {(\e_M-\e)  }{df\over dx}.\label{fxx}
\eea
This expression is only meaningful for $\xB<0.7$ where Fermi motion effects can be ignored. Given that $\e_M-\e>0$ (as discussed above), \eq{fxx} shows that
the slope of the EMC ratio is negative, consistent with observations.

  \cite{CiofidegliAtti:2007ork} calculated the expected size of the
  modification of~\eq{uu} using the spectral function $P(\bfp,E)$
  of~\cite{CiofidegliAtti:1995qe} (as discussed in Section \ref{phen}).
  The average values of the virtuality are quite large, as can be seen from Table~\ref{tab31}. 
  The average kinetic and removal energies in channel $1$ (high
  excitation final states) are much larger than the corresponding
quantities in channel $0$ (low excitation final states) and the high momentum components are linked to high removal energies~\cite{CiofidegliAtti:1980dbw}.
  ~\cite{CiofidegliAtti:2007ork} shows that these values of the
  virtuality, for reasonable choices of $E_B$ and $E_P$, can account for the EMC effect at
  $\xB\approx 0.5$.
 
\begin{table}[h]
\caption{\label{t2}The virtualities (in MeV) for
  channels 0 and 1 (see~\eq{eq:01})  
  and their sum~\cite{CiofidegliAtti:2007ork}.}
\begin{tabular}{|c|c|c|c|}\hline\hline
 A  & $\la v_0(\bfp,E)\ra/2M$ &
  $\la v_1(\bfp,E)\ra/2M$ & $\la v(\bfp,E)\ra/2M$\\\hline
$^3{\rm He}  $   &  -7.15 & -27.44 & -34.59\\
$^4{\rm He}  $   & -26.82 & -42.58 & -69.40\\
$^{12}{\rm C}$   & -33.17 & -49.11 & -82.28\\
$^{16}{\rm O}$   & -31.40 & -48.28 & -79.68\\
$^{40}{\rm Ca}$  & -35.00 & -49.54 & -84.54\\
$^{56}{\rm Fe}$  & -31.66 & -50.76 & -82.44\\
$^{208}{\rm Pb}$ & -32.87 & -59.33 & -92.20\\
 \hline\hline
\end{tabular}
\label{tab31}
\end{table}
  
This shows that high-momentum nucleons in nuclei can cause the EMC
effect.  Now we need to find a similar relation between these
high-momentum nucleons and the plateaus observed at high $\xB$ in
inclusive $(e,e')$ QE scattering.  We first review the kinematics. We
assume that the virtual photon is absorbed by one of the
baryons contained in an interacting system of two baryons $M_2\approx
M_d$.  The virtual photon hits a baryon of momentum $p$ in a `deuteron' of
momentum $P$, and the second, spectator baryon has momentum $p_s=P-p$. The
struck nucleon has final momentum $p_f = p+q$. Let the plus-component
of $p$ be given as $a M_d $. The light-front fraction $a$ is related
to the Frankfurt-Strikman variable $\a$ by $a=\a{M\over M_d}$.  Then
\bea p_f^-&=&{p_\perp^2+M^2\over aM_d+q^+}>0\label{e1}\\
p_s^-&=&{p_\perp^2+M^2\over(1-a) M_d}>0.\label{e2}
\eea 
In our
convention $q^+<0$ so that \eq{e1} tells us that $a>0$ and \eq{e2}
tells us that $a<1.$ Conservation of energy tells us that $p_f^+
+p_f^- +p_s^+ +p_s^-=2(M_d+\n)$, which leads to a quadratic equation
for $a$: 
\bea 
(a M_d+q^+)(1-a)M_d={M_d+q^+\over M_d+q^-}(p_\perp^2+M^2). \label{aquad}
\eea 
The
condition that this equation for $a$ has real roots leads to limits on
the value of $p_\perp$.

 \begin{figure}
 \includegraphics[width=7cm]{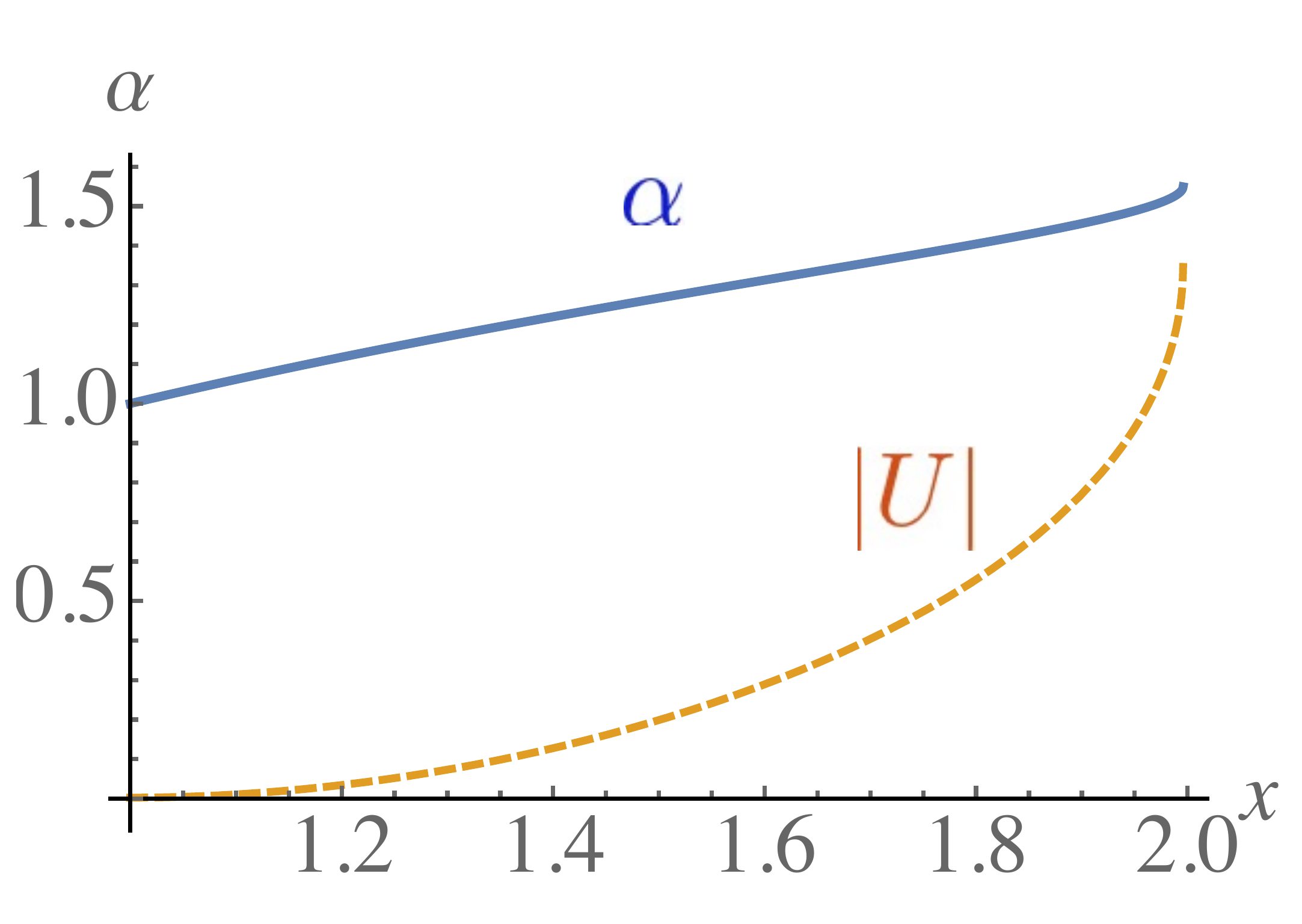}
 \caption{(color online) \label{fig:vkin} $\a$ solid, $|U|$ dashed for $Q^2=2.7$ GeV$^2$ and $p_\perp=0$. The quantity $|U|$ is presented in units of the nucleon mass $M$ and is proportional to the virtuality $v(\bfp,E)$ via \eq{uu}}
 \end{figure}

Fig.~(\ref{fig:vkin}) shows the  results of a specific example using $Q^2=2.7$ GeV$^2$ and $p_\perp=0$. 
Solving \eq{aquad}  gives the resulting values of $\a$ as a function of $\xB$. We see that  
$\a$ is considerably greater than one
 for  $1.5<\xB<1.8$, corresponding to the plateau region of Fig.~\ref{fig:eeratios}. 
 Using the displayed values of $\a$ we can calculate $v(\bfp,E=p^0)$:
 \bea v(\bfp,p^0)=p^+p^--p_\perp^2-M^2
\eea
where
$
 p^+=a M_d, 
 p^-={M^2+p_\perp^2\over a M_d+q^+}-q^- 
 .$
 Then the use of \eq{uu} gives the values shown in  Fig.~(\ref{fig:vkin}):
 Thus, for $1.5<\xB<1.8$, we have 
\bea 270 \,{\rm MeV} <|U| <600 \,{\rm MeV} .
\eea
 Such large values of $|U|$ can only arise from hard interactions of
 two nucleons, i.e., at short range.
 
 Thus $(e,e')$ at high $\xB$ is associated with short-ranged
 correlations. Next we relate the virtuality to the observed
 plateaus in the cross section ratios.  ~\cite{CiofidegliAtti:1995qe}
 showed that, for large values of $|\bfp|$: \bea n_A(\bfp)\approx
 n_A^{(1)}(\bfp)\approx a_2(A) n_D(\bfp).\label{ddd}\eea
This relation is explained in Sect.~\ref{Appendix}.

 To summarize: there is a consistent picture in which short-ranged
 correlations are involved with significant modification of the
 nuclear quark distribution function by suppressing the point-like
 configurations.  The key feature is that larger values
 of the nuclear excitation energy $E$, associated with the short-ranged correlations, correspond to
 larger values of virtuality and therefore to more significant
 deformations of the nucleon.  These very same short-ranged
 correlations are also responsible for the validity of \eq{ddd} for
 large values of momentum (where the virtuality is large), which via
 the logic of
 \cite{Frankfurt81,Frankfurt:1988nt,frankfurt93} is
 responsible for the cross section ratio plateaus. The spectral function $P(\bfp,E)$
 contains the information necessary to compute both the virtuality
 needed to understand the DIS EMC effect and the momentum probability
 $n_A(\bfp)$ needed to understand the plateaus.  

\subsubsection{Effective Field Theory  }
 
It is not necessary  that  the suppression of point-like configurations for off-shell nucleons be 
the sole origin of the EMC effect. Indeed 
another dynamical idea  could also account for  the experimental findings. 
 For example, the presence of non-nucleonic 6-quark
clusters (Sect.~III D) in nuclei could be important.
 A more general approach, using effective field theory (EFT),
which is not specific as to the underlying mechanism of medium
modification has been presented~\cite{Beane:2004xf,Chen:2004zx,Chen:2016bde}.  The  authors~\cite{Chen:2016bde} show
that the empirical linear relation between the magnitude of the EMC
effect in deep inelastic scattering on nuclei and the short range
correlation scaling factor $a_2$ extracted from high-energy
quasi-elastic scattering at $\xB\ge 1$ is a natural consequence of scale
separation and derive the relationship using effective field theory.

   \newcommand{\xn}[0]{\langle \xB^n\rangle\ }
\newcommand{\xm}[0]{\langle \xB^m\rangle\ }
\newcommand{\xnA}[0]{\langle \xB^n\rangle_A}
\newcommand{\xmA}[0]{\langle \xB^m\rangle_A}
\def\dfrac#1#2{{\displaystyle {#1 \over #2}}}

\newcommand{\FIXME}[1]{[\textcolor{red}{FIXME: #1}]}
\newcommand{\JW}[1]{[\textcolor{blue}{JW: #1}]}
\newcommand{\Blue}[1]{\textcolor{blue}{#1}}
\newcommand{\JL}[1]{[\textcolor{joelred}{JL: #1}]}

\newcommand{\ket}[1]{\left|#1\rangle\right.}
\newcommand{\bra}[1]{\langle#1|}
\newcommand{\braket}[2]{\langle#1\vphantom{#2}|#2\vphantom{#1}\rangle}
\newcommand{\ketbra}[2]
{|#1\vphantom{#2}\rangle\!\ \!\langle#2\vphantom{#1}|}
\newcommand{\matrixel}[3]
{\left\langle#1\right.\left|#2\right|
\left.#3\right\rangle}
\newcommand{\ev}[1]{\langle#1\rangle}

\newcommand{\gvec}[1]{\boldsymbol{#1}}
\newcommand{\up}{\uparrow}
\newcommand{\down}{\downarrow}

Their EFT Analysis proceeds by studying the dominant (leading-twist)
parton distributions  determined by target matrix elements of
bilocal light-cone operators. Applying the operator product expansion,
the Mellin moments of the parton distributions,
\begin{equation}
\langle \xB^n\rangle_A(Q) = \int_{-A}^{A} \xB^n q_A(\xB,Q) d\xB, 
\end{equation}
are determined by matrix elements of local operators.
Each  of the QCD operators is matched to hadronic operators
\cite{Chen:2004zx}. The relative importance of the hadronic operators
in a nuclear matrix element can be systematically estimated from EFT 
power counting. The nuclear matrix element is given by
\begin{equation} \label{xA}
\langle \xB^{n}\rangle _{A}(Q)
=\langle \xB^{n}\rangle _{N}(Q)
\Bigl[A+\alpha_{n}(\Lambda,Q)\langle A|(N^{\dagger}N)^{2}|A
\rangle_\Lambda\Bigr],
\end{equation}
where $\alpha _{n}$ depends on $\Lambda$ but not $A$ and is
completely determined by the two-nucleon system. This relation is valid
for all $n$, so after an inverse Mellin transform, the isoscalar PDFs
satisfy
\begin{eqnarray}
\label{qA}
{1\over A}F_2^{A}(\xB,Q)=F^N(\xB,Q)+g_{2}(A,\Lambda )\tilde{f}_{2}(\xB,Q,\Lambda),\,
\label{g00}\end{eqnarray}
where 
\begin{align}\label{g2}
g_{2}(A,\Lambda)={1\over A}\bigl\langle A|\left(N^{\dagger}N
\right)^{2}|A\bigr\rangle_\Lambda, 
\end{align}%
and $ f_2(\xB,Q,\Lambda)$ is an unknown function independent of $A$.
This feature is similar to that of our \eq{fxx}. Indeed, Equation~(\ref{qA}) was also obtained phenomenologically in
Ref.~\cite{Frankfurt81,Frankfurt:1988nt} using  the impulse
approximation. The equation \eq{g00} appears also in~\cite{kulagin10,Hen13,Kulagin:2004ie,Kulagin:2014vsa}.
Note that $g_2$ receives dominant contributions from the single nucleon density.

The factorization scale of the PDF is $\mu_f=Q$, while $\Lambda$
is the nuclear physics ``ultraviolet'' cut-off that separates the high
energy parton physics from  lower energy hadronic and nuclear effects.
The two scales must be significantly separated for the  EFT description
to be valid.

The second term on the right-hand side of Eq.~(\ref{qA}) is the nuclear
modification of the  structure function. The shape of
distortion, i.e., the $\xB$ dependence of $f_2$, which is due to physics
above the scale $\Lambda$, is $A$ independent and hence universal among
nuclei. The magnitude of distortion, $g_2$, which is due to physics
below the scale $\Lambda$, depends only on $A$ and $\Lambda$.

At smaller  values of $Q^2$, the  previous analysis was generalized to apply to the $(e,e')$ cross section at 
large $\xB$, so that 
\begin{eqnarray}\label{sigmaA}
\sigma_{A}/A=\sigma_{N}+g_{2}(A,\Lambda)\sigma_{2}(\Lambda),
\end{eqnarray}
where the $E_0$ (incident electron energy), $\xB$ and $Q^2$ dependence
of $\sigma_i$ is suppressed. 
With $\sigma_{N}$ vanishing for $\xB > 1$, for both DIS and QE,
\begin{equation}\label{sigmax}
a_{2}(A,\xB>1)=\frac{g_{2}(A,\Lambda )}{g_{2}(2,\Lambda)}.
\end{equation}
In principle, $a_2$ could depend on $E$, $\xB$ and $Q^2$.  However the EFT
factorization shows that this dependence cancels at this order yielding a plateau in $a_{2}$ as observed experimentally at
$1.5<\xB<2$.  (The influence of Fermi motion
extends the contribution of the single nucleon PDF to $\xB$ above
1, pushing  the onset of the plateau   to larger values of   $\xB$.)
The function  $a_2$  was also computed using the Green's Function
Monte Carlo  method~\cite{Carlson:2014vla} and it agrees well with the data.

Eq.~ (\ref{qA}) and the definition $R(A,\xB)\equiv F_2^A/(AF_2^N)$, lead to the result 
that 
\begin{eqnarray} 
\frac{dR(A,\xB)}{d\xB}
=C(\xB)\left[a_{2}(A)-1 \right],
\label{EMC-SRC}
\end{eqnarray}
has a linear relation with $a_2$,
with $C(\xB) = g_{2}(2) [f_{2}^{\prime}F_2^N-f_2
F_2^{N^{\prime}}]/[F_{2}^N+g_{2}(2) f_2]^2$ independent of $A$ and
$\Lambda$ 
(here,
$f^\prime=df/d\xB$).  
This means  that  EFT naturally accounts for the linear relation between the EMC slope and the height of the plateau. However, the sign of the EMC effect is not explained.

\subsubsection{The Isovector EMC Effect}

This SRC-related PLC suppression model also leads to an explanation of
the NuTeV anomaly~\cite{Sargsian:2013gea}.  We have discussed the
dominance of the $pn$ SRCs, relative to the $pp$ and $nn$
correlations, for nuclear internal momenta between 300 and 600
MeV/c, that is caused by the effects of the tensor force.  The $pp$
and $nn$ components of the $NN$ SRC are strongly suppressed since they
are dominated by the central $NN$ potential with relative $ L =
0$. The resulting picture for nuclear matter consisting of protons
and neutrons at densities in which inter-nucleon distances are about
1.7 fm is rather unique: it represents a system with suppressed $pp$
and $nn$ but enhanced $pn$ interactions.  Using this idea Sargsian
\cite{ Sargsian:2012sm} predicted two new properties for the nuclear
momentum distributions for momenta between the Fermi momentum and
about 600 MeV/c.  There is an approximate equality of $p$- and $n$-
momentum distributions weighted by their relative fractions in the
nucleus $x_p = Z/A$ and $x_n =( A-Z)/Z$ : \bea x_p n_p^A(p)\approx x_n
n_n^A(p)\eea with $\int d^3p n^A(p)=1. $ The probability of a proton
being in a high momentum $NN$ SRC is inversely proportional to its
relative fraction, $x_p$, and can be related to the momentum
distribution in the deuteron $n_D(p)$: \bea n^A_p(p)={1\over
  2x_p}a_2(A,N)n_D(p)\label{m2}\eea and similarly for neutrons.  The
main prediction of \eq{m2} is that high momentum protons and neutrons
became increasingly unbalanced as the ratio $(N-Z)/(N+Z)$
increases. Using this equation one can calculate the fraction of the
protons having momenta greater than the Fermi momentum as \bea& P_p(A,
N) \approx {1 \over 2x_p}a_2(A,N) \int d^3p\thinspace
n_D(p)\Theta(p-k_F),\nonumber\\& \eea and similarly for neutrons.  For
example in Iron, $P_p=23\%,$ and $P_n=20\%$.

The energetic protons in neutron rich nuclei will result also in the
stronger nuclear modification of $u$-quarks as compared to $d$-quarks
and the effect grows with $A$. 
The
predicted effects also can be checked in parity violating deep
inelastic scattering off heavy
nuclei~\cite{Cloet:2012td,Souder:2016xcn} (see Section \ref{pvdis}).

\subsubsection{Summary} 

In summary, driven by the short-range
correlations between two nucleons, the strong connection between the
EMC effect and the plateaus observed in $(e,e')$ scattering at high
$\xB$ is both a natural consequence of the impulse approximation of
scattering theory, and also of effective field theory. In the impulse
approximation the relevant ratio is that of momentum-space densities;
in the EFT the relevant ratio is that of coordinate space
densities. Sect.~\ref{Appendix} shows that ratios of these are the same as
long as large values of momenta are used in the impulse approximation
and small values of relative distance are used in the EFT. This means that
the relation shown in Fig.~\ref{fig:emcsrc} is derived using two very
different techniques.  The fact that using two different
technical approaches, each driven by short-range physics, leads to the same conclusion, gives significant credence to the interpretation that the
same short-range physics accounts for both the EMC effect and the QE
cross section plateaus.

The underlying mechanism of the distortion of the nucleon structure is
not yet established, and could occur from PLC suppression or from
other mechanisms. Nevertheless, it is very clear that the relation
shown in Fig.~\ref{fig:emcsrc} is no accident.  There is a true
underlying cause of the EMC effect and the observed plateaus in ratios
of $(e,e')$ scattering cross sections.

      %


\subsection{Are the nucleons in the correlated pair really nucleons?}
According to the logic presented here, most of the correlated pair consists of nucleons, but the part that is responsible for the EMC effect consists of non-nucleonic configurations. 
This conclusion is valid for both classes of models of the EMC effect: the mean-field based or SRC based.   The non-nucleonic configurations could be a medium  modified single-nucleon, or  $NN^*$ or $N^*N^*$ configurations, or even  more complex 6-quark configurations. 

\subsection{Determining the structure function of a free neutron}
Determining the structure function of the neutron is challenging
because a free neutron target does not exist. Experimentalists have
therefore used deuteron or $^3$He targets to extract the neutron
structure.  This implies that our knowledge of the neutron structure
function is intimately connected with medium effects in light
targets. As we shall see, medium effects in the deuteron must be
accounted for accurately if one hopes to correctly understand the free
neutron structure function.

\subsubsection{The Deuteron IMC Effect}
The deuteron In Medium Correction (IMC) effect   refers to the difference between the DIS
cross section for the deuteron and the sum of the DIS cross sections for a
free proton and neutron~\cite{Frankfurt:1985cv,Melnitchouk:1994rv}.  The term IMC was introduced in
Ref.~\cite{weinstein11} which showed that one can use the EMC-SRC correlation as a
phenomenological tool to constrain the deuteron IMC effect, and thus
extract the free neutron structure function. Following Weinstein
\etal{} \cite{weinstein11}, we can extrapolate the linear fit to the
EMC-SRC correlation to the limit of $a_2(A)\rightarrow0$. This is the limit of 
no correlations, which is equivalent to a free proton-neutron pair. The intersection of this
limit with the $y$-axis is therefore the IMC ratio of the free proton-neutron pair to the deuteron.

The $a_2(A)\rightarrow0$ extrapolation to the $y$-axis of the EMC-SRC correlation gives
  $dR_{\rm EMC}/dx_{a_2(A)=0}=-0.070\pm0.004$.  Since the EMC effect is linear
  for $0.3\le x_A \le0.7$ for all nuclei with $A>2$, we assume that
  the EMC effect is also linear in this region for the
  deuteron. This implies that the EMC effect for the deutereon relative to a free proton plus neutron can be written as:
\[
\frac{\sigma_d}{\sigma_p+\sigma_n} = 1 - a(\xB-b) \quad \hbox{for } 0.3\le\xB\le 0.7,
\]
where $\sigma_d$ and $\sigma_p$ are the measured DIS cross sections
for the deuteron and free proton, $\sigma_n$ is the free neutron DIS
cross section that we want to extract, $a = \vert dR_{\rm
  EMC/dx}\vert_{a_2(A)=0} =0.070\pm0.004$, and $b = 0.34\pm0.02$ is the
average value of $\xB$ where the EMC ratio is unity\footnote{The
  $x_A$ correction does not significantly change the slope, $a$ , of the EMC-SRC
  correlation, and it increases the $b$ parameter by less than the uncertainty
 reported in Ref.~\cite{weinstein11}}. This implies that
$\sigma_d/(\sigma_p+\sigma_n)$ decreases linearly from 1 to 0.97 as
$\xB$ increases from 0.3 to 0.7.  We can then use this relationship to
extract the free neutron cross section in this $\xB$ range, as shown
in the next section.

\begin{figure}[htb]
\begin{center}
\includegraphics[scale=0.3]{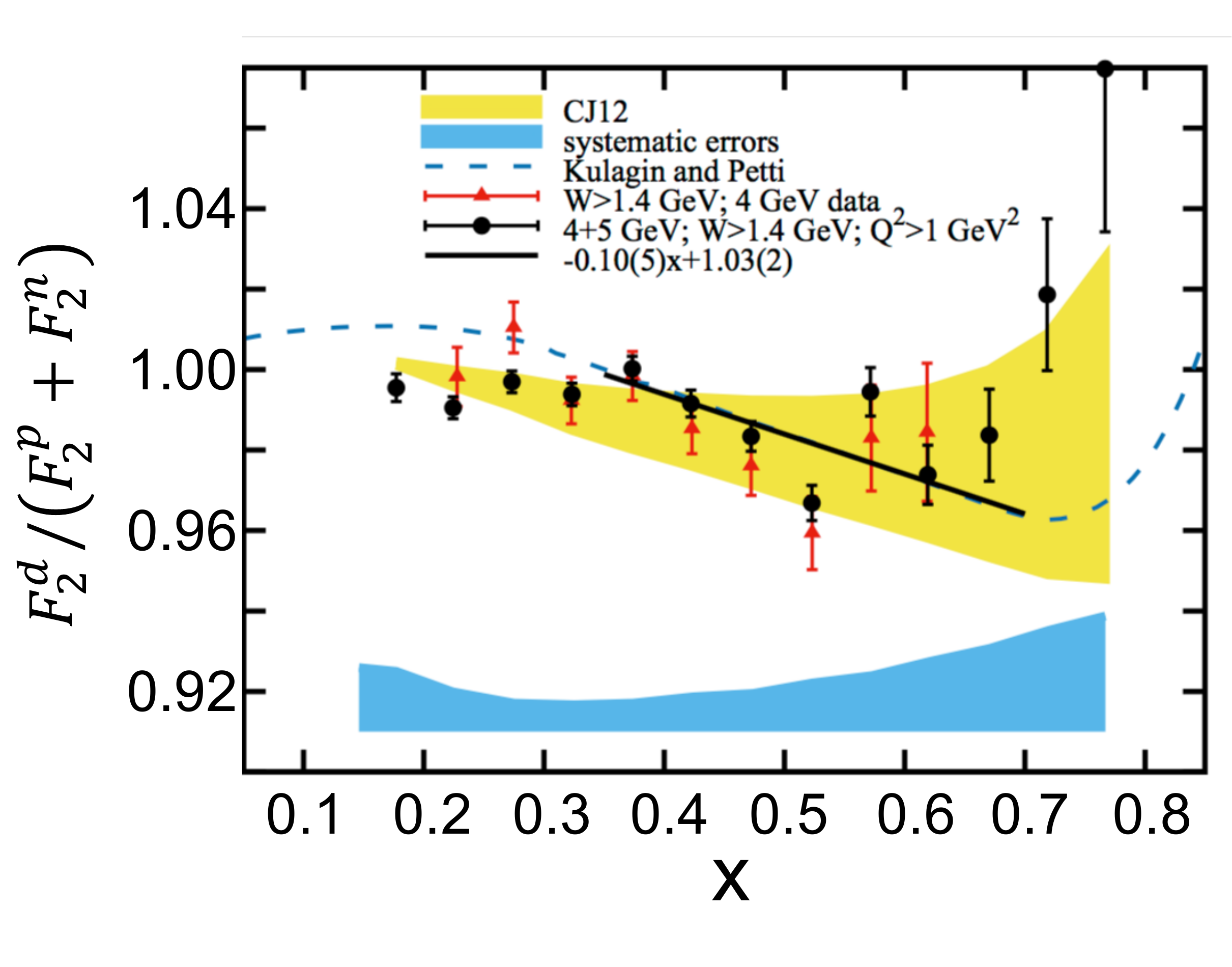}
\caption{ \label{fig:BONUS_IMC} The deuteron IMC ratio
  $R_{EMC}^{d}=F_{2}^{d}/(F_{2}^{n}+F_{2}^{p})$ as extracted from the
  BONuS data. Total systematic uncertainties are shown as a band
  arbitrarily positioned at $0.91$ (blue). The yellow band shows the
  CJ12~\cite{Owens13} limits expected from their nuclear models. The
  black points are the combined $4$- and $5$-GeV data, whereas the red
  points are the $4$-GeV data alone. The dashed blue line shows the
  calculations of Ref.~\cite{KULAGIN2006126}. The solid line (black)
  is the fit to the black points for $0.35<\xB<0.7$.  From \cite{Griffioen:2015hxa}.}
\end{center}
\end{figure}

The uncertainty quoted above for the IMC slope is  due to the EMC and
SRC data and to the fit.  
It does not include any uncertainty due to corrections applied to the EMC and SRC data.  
As stated above, if we include  the proposed correction for $a_2(A)$ due to the c.m. motion of
the correlated pair, then the fit parameter increases by 25\% and so does the free proton plus neutron EMC effect. 
These effects are discussed in detail in \cite{hen12}.

Following the prediction of the IMC effect, the BONuS collaboration \cite{tkachenko14}
published their experimental extraction of the IMC effect, 
measured at $Q^2>1$ GeV$^2$ and $W>1.4$ GeV, see Fig.~\ref{fig:BONUS_IMC}~\cite{Griffioen:2015hxa}. A
linear fit for $0.35<x<0.7$ yields $dR_{EMC}^{d}/dx=-0.1\pm0.05$ where
the uncertainties comes from the  fit. This result is
consistent with the IMC prediction of $-0.07$.  For $x<0.5$ the EMC ratios
$R_{EMC}^{d}$ agree within uncertainties with those obtained using
more stringent cuts in $W$. The ratio for $\xB>0.5$ continues the trend
of the lower-$\xB$ data, with a hint of the expected rise above $\xB=0.7$
as seen in $R_{EMC}^{A}$ for heavier nuclei, but these high-$\xB$ values
are more uncertain because there are fewer data points for resonance
averaging. 

\subsubsection{The Free Neutron Structure Function}

If the structure function $F_2$ is proportional to the DIS cross
section ({\it i.e.,} if the ratio of the longitudinal to
transverse cross sections is the same for $n,p$ and $d$ [see
discussion in \cite{Geesaman95}]),
then the free neutron structure function, $F_2^n(x_B,Q^2)$, can also
be deduced from the measured deuteron and proton structure functions
and from the deuteron IMC effect:
\begin{equation}
F_2^n(x_B,Q^2) = \frac{2F_2^d(x_B,Q^2) -
  [1-a(x_B-b)]F_2^p(x_B,Q^2)}{[1-a(x_B-b)]}
\end{equation}
which leads to
\begin{equation}
\frac{F_2^n(x_B,Q^2)}{F_2^p(x_B,Q^2)} = \frac{2\frac{F_2^d(x_B,Q^2)}{F_2^p(x_B,Q^2)} -
  [1-a(x_B-b)]}{[1-a(x_B-b)]} \quad .
\end{equation}
This is only valid for $0.35\le x_B \le 0.7$.  

\begin{figure}[htb]
\begin{center}
\includegraphics[scale=0.3]{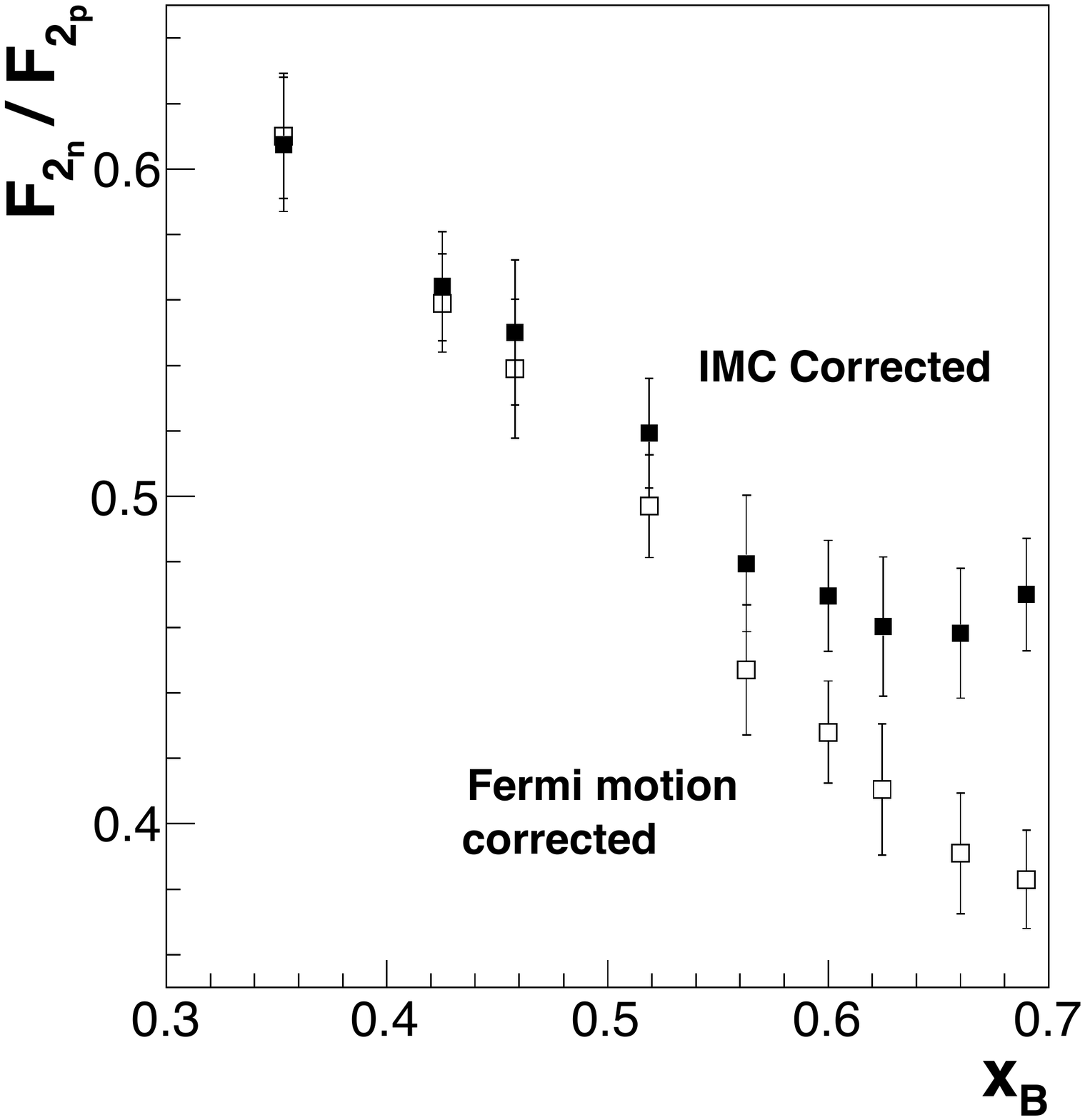}
\caption{ \label{fig:F2} The ratio of neutron to proton structure
  functions, $F_2^n(x_B,Q^2)/F_2^p(x_B,Q^2)$ as extracted from the
  measured deuteron and proton structure functions, $F_2^d$ and
  $F_2^p$. The filled symbols show $F_2^n/F_2^p$ extracted
  by~\cite{weinstein11} from the deuteron In Medium Correction (IMC)
  ratio and the world data for $F_2^d/F_2^p$ at $Q^2 = 12$ GeV$^2$
  \cite{Arrington09}. The open symbols show $F_2^n/F_2^p$ extracted
  from the same data correcting only for nucleon motion in deuterium
  using a relativistic deuteron momentum density \cite{Arrington09}.
  From \cite{weinstein11}.}
\end{center}
\end{figure}

Fig.~\ref{fig:F2} shows the ratio of $F_2^n/F_2^p$ extracted	by~\cite{weinstein11}
using the IMC-based correction and the $Q^2=12$ GeV$^2$ ratio
$F_2^d/F_2^p$ from Ref.~\cite{Arrington09}.  Note that the
ratio $F_2^d/F_2^p$ is $Q^2$-independent from $6 \le Q^2 \le 20$
GeV$^2$ for $0.4\le x_B \le 0.7$ \cite{Arrington09}. 
The dominant uncertainty
in this extraction is the uncertainty in the measured $F_2^p/F_2^d$.
The IMC-based correction increases the extracted free neutron
structure function (relative to that extracted using the deuteron
momentum density \cite{Arrington09}) by an amount that increases with
$x_B$. This is qualitatively similar to the recent extraction of~\cite{Cosyn:2015mha}.
Thus, the IMC-based $F_2^n$ strongly favors model-based extractions of
$F_2^n$ that include nucleon modification in the deuteron \cite{Melnitchouk96}.  

\begin{figure}[htb]
\begin{center}
\includegraphics[scale=0.3]{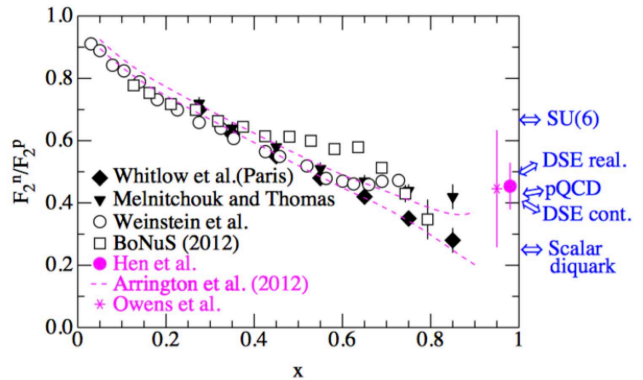}
\caption{ \label{fig:F2n_F2p} $F_{2}^{n}/F_{2}^{p}$ as a function of $\xB$. Results from the IMC and other phenomenological extractions are compared to selected theoretical predictions. From~\cite{Holt:2013pja}. See~\cite{Roberts:2013mja, Holt:2013pja} for details.}
\end{center}
\end{figure}

The IMC based extraction of $F_{2}^{n}/F_{2}^{p}$, extrapolated in the region of $x_B<0.3$, 
is compared in Fig.~\ref{fig:F2n_F2p} to several other experimental and phenomenological extractions 
of this ratio. Also shown are several QCD predictions. see~\cite{Roberts:2013mja, Holt:2013pja} for details. 

\subsubsection{The $d/u$ ratio at large $x_B$}

The ratio of the neutron structure function, $F_{2}^{n}$, to the
proton structure function, $F_{2}^{p}$, is particularly interesting
as it can be related, within the parton model, to the ratio of the
$d$-quark and $u$-quark distributions. The latter provides a unique
opportunity for studying the flavor and spin dynamics of quarks in the
nucleon, with the $d/u$ quark distribution ratio in particular being
very sensitive to different mechanisms of spin-flavor symmetry breaking
\cite{MELNITCHOUK199611,Holt10}.

Historically, proton DIS data placed strong constraints on the $u$-quark
distribution, while neutron structure functions were used to constrain the 
$d$-quark distribution and form the $d/u$ ratio. Specifically, the $d/u$ ratio
in the  valence quark dominance domain (i.e., at large \xB) was extracted 
from the $F_{2}^{n}/F_{2}^{p}$ structure function ratio using: 
\[
F_{2}^{n} / F_{2}^{p} = [1+4(d_{v}/u_{v})]/[4+(d_{v}/u_{v})],
\]
where the absence of free neutron targets meant  
that the neutron structure function was not measured directly, but instead 
extracted from deuterium DIS data. However, uncertainties in the nuclear corrections 
in the deuteron, such as those associated with nucleon off-shell effects and the 
large-momentum components of the deuteron wave function, give rise to significant
uncertainties in the resulting $d/u$ ratio for $\xB \gtrsim 0.5$
\cite{accardi11}. 

To rectify the situation, Hen et al., \cite{hen11} used the  phenomenological IMC 
corrected extraction of $F_2^n/F_2^p$ discussed above as an added constraint on the
extraction of the $d/u$ ratio in the global analysis of the CTEQ-JLab collaboration \cite{accardi11}. 
 
New data on charged lepton and $W$ boson asymmetry measured at the 
 Tevatron \cite{D0:2014kma,Abazov:2013rja,Abazov:2013dsa}
are sensitive to the large-$\xB$ $d/u$ ratio with no nuclear uncertainties \cite{Accardi:2016qay}.
 
\begin{figure}[b]
\includegraphics[width=9cm]{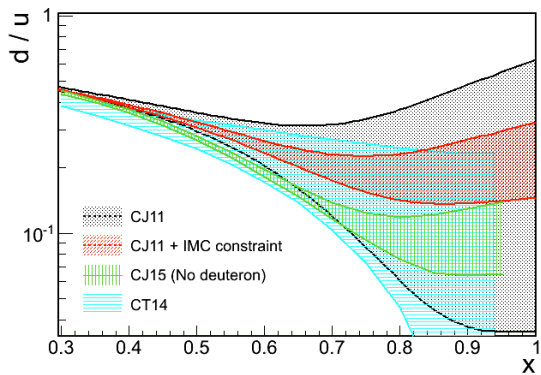}
\caption{$d/u$ ratio at $Q^2 = 12$~GeV$^2$ with the full theoretical
	uncertainty from Ref.~\cite{accardi11} (black) and with the IMC
	constraint at the 90\% C.~L. (red) from \cite{hen11}. Also shown for comparison are recent extractions that do not include nuclear correactions from the CJ15, \cite{Accardi:2016qay} and CT14, \cite{Dulat:2015mca}) PDF extractions.}
\label{fig:duratio}
\end{figure}

Fig.~\ref{fig:duratio} shows the $d/u$ ratio at large-$\xB$ extracted
from a global QCD analysis using DIS data without (CJ11,
\cite{accardi11}) and with (CJ11+IMC, \cite{hen11}) the IMC constraint
and using the new asymmetry data with no nuclear corrections applied
(CJ15, \cite{Accardi:2016qay} and CT14, \cite{Dulat:2015mca}).  As can
be seen, while the various extractions somewhat differ at large-$\xB$,
the IMC constraints and the new asymmetry data both contrain the CJ11
analysis similarly.

To summarize, the use of the IMC-extracted neutron structure
function directly constrains the $d$-quark PDF for $x \lesssim 0.7$,
and indirectly for $x \to 1$.  We find the $d/u$ ratio in the limit
$x \to 1$ to be $0.23 \pm 0.09$ at the 90\% confidence level,
in overall agreement with new extractions using charged lepton and $W$
boson asymmetry data and
in agreement with the models of \cite{farrarjackson75, Cloet:2005pp} which predict intermediate
values of $d/u$ between the SU(6) symmetry and scalar diquark dominance limits.

    %

\section {Existing searches for medium-modified electromagnetic form factors}
%
%

We have shown that the experimental and theoretical evidence indicates that the structure of the nucleon is modified by its immersion in a nucleus. 
The only models that account for the EMC effect, the plateaus of the high $\xB$ (e,e') reaction and the lack of a medium effect in the nuclear Drell-Yan data are those involving short-ranged correlations. Nevertheless, the task of understanding the EMC and SRC effects is not complete. The available models need to be improved (to be discussed in Sect.~\ref{FT}). 
We need models that are sufficiently complete that they can explain both the EMC effect, the nuclear Drell-Yan data and also predict and account for  new independent phenomena. 

 If the nuclear medium modifies the bound nucleon structure functions
 (and thus their wave functions), then it almost certainly will modify
 their electromagnetic form factors.  All of the medium modification models  modification of bound electromagnetic form factors,  see Sect.~\ref{EMC}.  
These effects could  be manifest in 
 quasi-elastic nucleon
 knockout $(e,e'N)$ cross sections and in the inclusive longitudinal $A(e,e')$ response. The influence of
 nucleon modification on the nuclear elastic form factor can not be
 detected because the distribution of nucleons in the nucleus is imprecisely known. 

This section will discuss the experimental evidence for modification
of bound nucleon form factors.

\subsection{Polarization transfer in  the $(\vec e,e'\vec p)$ reaction}
Polarization transfer in the H$(\vec e,e'\vec p)$ reaction was used to measure the ratio of the free proton electromagnetic form factors $G_E/G_M$ with much smaller systematic uncertainties than previous methods ~\cite{Perdrisat:2006hj}.  This technique was then applied to measure the ratio of bound proton electromagnetic form factors using the quasielastic $A(\vec e,e'\vec p)$ reaction~\cite{Dieterich:2000mu,Strauch:2002wu,Paolone:2010qc,Malace:2010ft,Strauch:2012nra}.  The ratio of the 
 longitudinal and transverse polarization transfers is proportional to the ratio of $G_E/G_M$ for the free proton, $P_x'/P_z'\propto G_E/G_M$~\cite{Perdrisat:2006hj}.
For a bound proton, one must also correct for the effects of meson exchange currents, isobar configurations, 
and especially final state interactions.
After using a model to correct for these effects, the polarization double ratio 
\bea 
 R\equiv \left(\frac{P_x'}{P_z'}\right)_A / \left(\frac{P_x'}{P_z'}\right) _{^1H}
\eea
should be sensitive to medium modification of the form factor ratio.  The induced polarization $P_y$ (measured in the $(e,e'\vec p)$ reaction) should be more sensitive to final state interactions and much less sensitive to medium modification effects.


 \begin{figure}\resizebox{0.40\textwidth}{!}
{ \includegraphics{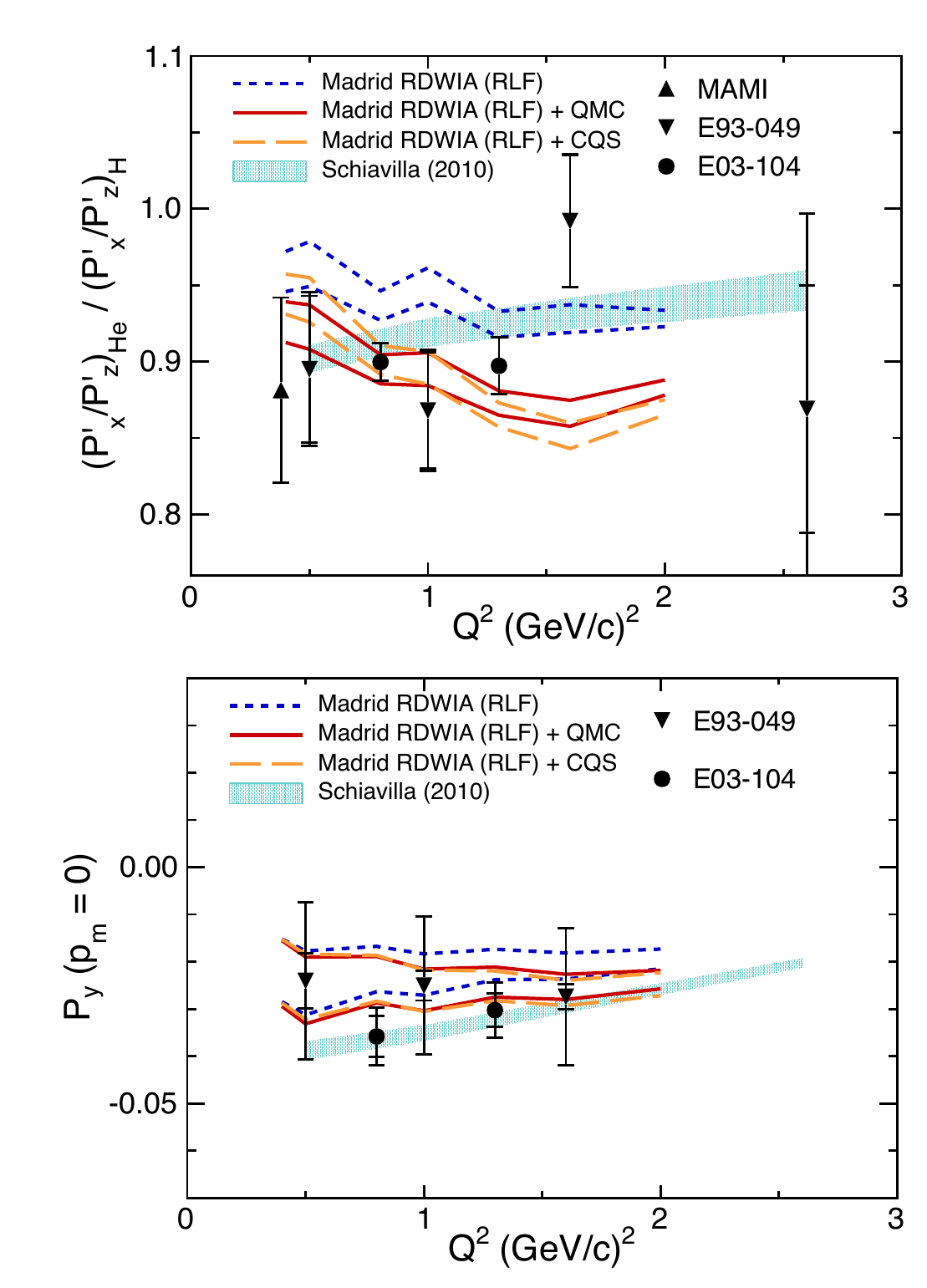}}
\caption{The measured $^4{\rm He}({\vec e}, e'{\vec p})^3$H polarization-transfer double ratio $R$ (upper panel) and induced polarization $P_y$ (lower panel) as a function of $Q^2$: open symbols:~\cite{Dieterich:2000mu,Strauch:2002wu}  and filled circles:~\cite{Paolone:2010qc,Malace:2010ft}. The data are compared to DWIA  calculations from~\cite{Schiavilla:2004xa} (updated in 2010) using unmodified form factors and from the Madrid group~\cite{Udias:1999tm,Caballero:1997gc,Udias:2000ig} using the cc1  (lower set of curves) and cc2 (upper set of curves) off-shell current operators in combination with unmodified (black dashed line), QMC modified (red solid line) and CQS modified (red dashed line) in medium form factors.  See text for details.
 From~\cite{Strauch:2012nra}.}
\label{fig:ffm}       
\end{figure}

Fig.~\ref{fig:ffm} shows the $^4{\rm He} ({\vec e}, e'{\vec p})^3$H double ratio $R$ and the induced polarization, $P_y$, measured at small values of missing momentum ($p_{miss}< 150$ MeV/c) over a range of $Q^2$.
Relativistic distorted-wave-impulse approximation (rDWIA) calculations by the Madrid group~\cite{Udias:1999tm,Caballero:1997gc,Udias:2000ig} can only explain the data if they include medium-modified form factors. They calculated the induced polarization and the polarization transfer ratio using the unmodified but offshell cc1 and cc2~\cite{DeForest:1983ahx} current operators and the optical potentials of 
\cite{Horowitz:1985tw,Murdock:1986fs} to account for final state interactions. No charge exchange effects (photon knocks out neutron, which undergoes a charge exchange reaction)  were included. This unmodified calculation agreed with the induced  polarization data when using the cc1 current operator. However, good agreement with the measured value of $R$ was only achieved by including either the QMC~\cite{Lu:1998tn}   or CQS~\cite{Smith:2004dn} medium modified form factors.

Schiavilla {\it et al.}~\cite{Schiavilla:2004xa} calculated $P_y$ and $R$ using DWIA. They computed the final state interactions  using an optical potential that includes both spin-independent and spin-dependent charge exchange terms. However, they updated their calculation in 2010 with new parameters.       While their calculation describes both $P_y$ and $R$ without medium modified form factors, its significance is decreased because they did not follow the  standard procedure~\cite{Austern}
    of independently constraining the parameters of the optical potential they used to describe the final state interactions. Thus our view is that the results of the nuclear polarization experiments strongly indicate that medium effects
    do influence electromagnetic form  factors. We eagerly await new experiments with improved precision and  at larger values of  $p_{miss}$ which would confirm or rule out this interpretation. 


Experiments performed at the Mainz Microtron (MAMI) using the A1 beam-line \cite{yaron16} measured the polarization transfer ratio $R$ for deuterium and $^{12}$C at lower $Q^2$ ($Q^2 = 0.175$ and 0.4 GeV$^2$) but higher virtuality than at Jefferson Lab. For deuterium, the ratio $R$ decreases significantly with virtuality and is consistent with that previously measured on $^4$He.
This indicates that the effect in nuclei is due to the virtuality of the knocked-out proton and not due to the average nuclear density.
The deuteron calculations \cite{Arenhovel:2004bc} predict this decrease and associate most of it with FSI \cite{yaron16}.  The $\approx10\%$ differences between the data and calculations may indicate the need for
in-medium modifications. The carbon data is still under analysis.  Other double polarization experiments were not sensitive to the effects of nucleon modification \cite{Mayer:2016zmv,Mihovilovic:2014gdi,Passchier:2001uc}.

 Jefferson Lab experiment E12-11-002  will  measure
 polarization-transfer observables as a function of virtuality for both $^4$He and $^2$H and
 will measure the proton recoil polarization at $Q^2=1.8 $ GeV$^2$ to help us better understand the effects of  medium modifications and  FSI. 

 \subsection{Polarization transfer in the $(\vec e, e'\vec n)$ reaction}
 A complementary   experiment would be the measurement
of polarization transfer to the neutron in quasielastic scattering in the $(\vec e,e'\vec n)$ reaction. 
 Clo\"{e}t {\em et al.}  \cite{Cloet:2009tx}  studied
possible in-medium changes of the bound neutron electromagnetic
form-factor ratio with respect to the free ratio, the superratio
$\left(G^*_E/G^*_M\right)/\left(G_E/G_M\right)$.  At small values of
$Q^2$ this superratio depends on the in-medium modifications of the
neutron magnetic moment and the effective electric and magnetic radii.
The superratio of the neutron is dominated by the expected increase of
the electric charge radius in the nuclear medium and is found to be
greater than one.  In contrast, the proton superratio is predicted to
be smaller than one.  A comparison of high-precision measurements of
the reactions $^2$H$({\vec e}, e' {\vec  n})p$ and $^4$He$({\vec e}, e' {\vec n})^3$H would  
test these predictions.

However, a major drawback to nuclear polarization transfer measurements, no matter whether the proton or neutron is detected,  is that medium modifications that affect both $G_E$ and $G_M$ will  cancel in the ratio. See Fig.~\ref{modss}, for example.

\subsection{The $(e,e')$ reaction and the Coulomb Sum Rule (CSR)}
This  sum rule \cite{McVoy:1962zz,DeForest:1966ycn}
states that the integral of the $A(e,e')$ longitudinal response function at fixed momentum transfer over all energy transfers should equal the total charge of the nucleus, $Z$.  The first   CSR experiment \cite{Altemus:1980wt} observed that the sum rule was ``quenched'', i.e., they measured less than $Z$. This indicated that the cross section for scattering from a bound nucleon was significantly less than the free cross section.
Thus,  \cite{Cloet:2015tha}  say that the first hints of QCD effects in nuclei came from quasielastic electron scattering on nuclear targets~\cite{Altemus:1980wt,Noble:1980my,Meziani:1984is}.  However, later work cast doubt on this result.

\def\vp{\varphi}
\def\z{\zeta}
\def\th{\theta}
\def\O{\Omega}
\def\T{\Theta}

\def\pl{\partial}
\def\hs{\hspace}
\def\ms{\mspace}
\def\ol{\overline}
\def\no{\nonumber}
\def\ni{\noindent}

\def\ua{\uparrow}
\def\da{\downarrow}
\def\para{\parallel}

\def\lf{\left}
\def\rg{\right}
\def\la{\langle}
\def\ra{\rangle}
\def\lv{\lvert}
\def\rv{\rvert}
\def\lV{\lVert}
\def\rV{\rVert}
\def\w{\omega}
\def\bfq{{\bf q}}
\def\bfr{{\bf r}}
\def\r{\rho}
\def\lra{\longrightarrow}
\def\Lra{\Longrightarrow}
\def\lla{\longleftarrow}
\def\Lla{\Longleftarrow}
\def\Llra{\Longleftrightarrow}

\def\qqquad{\qquad\quad}
\def\qqqquad{\qquad\qquad}

\newcommand{\vect}[1]{\boldsymbol{#1}}
\newcommand{\ph}[1]{\phantom{#1}}

\newcommand{\sss}[1]{\scriptscriptstyle{#1}}

\newcommand{\sh}[1]{\slashed{#1}}

\font\bb=bbmss10 scaled 1200
\def\ident{\mbox{\bb 1}}

The $(e,e')$  inclusive cross section  can be written as
\begin{equation}
\begin{split}
\frac{d^2\sigma}{d\Omega d\nu} =& \sigma_{Mott} \biggl[ \frac{Q^4}{\lvert{\bf q}\rvert^4} R_L(\nu,\lvert{\bf q}\rvert) \\
&+ \left( \frac{Q^2}{2\lvert {\bf q}\rvert^2 } + \tan^2\frac{\theta}{2} \right) R_T(\nu,\lvert{\bf q}\rvert ) \biggr] 
\label{eq:rlrt}
\end{split}
\end{equation}
where $\sigma_{Mott}$ is the Mott cross section, $R_L$ and $R_T$ are the longitudinal and transverse response functions, and $\theta$ is the electron scattering angle.  In the non-relativistic limit of the impulse approximation~\cite{DeForest:1966ycn,Bertozzi:1972jff} one has
\[
R_L(\w,\bfq) =\la A|\sum_{i=1}^Ze^{i{\bfq\cdot \bfr_i}}\d (\w-H)\sum_{j=1}^Ze^{-i{\bfq\cdot \bfr_i}}|A\ra G_E^2( q^2),
\]
where  $H$ is the nuclear Hamiltonian, the ground state energy is taken as 0, and for simplicity we assume that neutrons do not contribute. The non-relativistic formulation is only valid when $\bfq^2 \approx Q^2$.  Since $R_L$ is proportional to the square of  $G_E$, its sensitivity to medium effects is greater than that of the polarization transfer measurements. 

The Coulomb sum  is the integral over {\it all} values of $\nu$ (including the inaccessible time-like regime where $\nu>|\bfq|$): 
\bea \frac{R_L(q)}{G_E^2(q^2)}=\frac{\int d\nu  R_L(\nu,\bfq)}{G_E^2(q^2)} =\la A|\sum_{i,j=1}^Ze^{i\bfq\cdot(\bfr_i-\bfr_j)}|A\ra.\nonumber\\
\label{csrr}\eea
Splitting \eq{csrr} into terms with $i=j$ and $i\ne j$ we get:
\bea { R_L(q)\over G_E^2(q^2)} = Z+Z(Z-1) \int d^3rd^3r'e^{i\bfq\cdot(\bfr-\bfr')} \r_2(\bfr,\bfr'),\nonumber\\\eea
where  $\r_2$ is the two proton density function, see \eq{rho2def}. At large enough momentum transfer the second term vanishes as $1/\bfq^4$, so that
one finds the CSR:
\bea \lim_{Q^2\to\infty}{ R_L(q)\over G_E^2(q^2)}=Z.\eea

Since electron scattering cannot  measure the cross section in the time-like region, the Coulomb sum is properly defined~\cite{Cloet:2015tha} 
as an integral over $\nu$ from energies just above the   elastic peak to $|\bfq|$:
\begin{align}
S_L(\lf|\vect{q}\rg|) &= \int_{\nu^+}^{\lf|\vect{q}\rg|}d\nu\ 
\frac{R_L(\nu,\lf|\vect{q}\rg|)}{Z\,G_{Ep}^2(Q^2) + N\,G_{En}^2(Q^2)}.
\label{eq:csr}
\end{align}
The quantity $S_L$  can be correctly be compared with  the results obtained from electron scattering. 

The initial motivation to measure the Coulomb Sum Rule~\cite{DeForest:1966ycn}  was to learn about  $\r_2$. However, the recent focus has been to learn about nucleon medium modification at large values of the momentum transfer where  the effect of $\r_2$ is negligible.

~\cite{Cloet:2015tha} discuss the  interesting  history of the theory. Calculations~\cite{Saito:1999bv,Horikawa:2005dh}, in which the internal structural properties of bound nucleons are self-consistently modified by the nuclear medium unsurprisingly predict significant quenching of the CSR. 
However, calculations that assume an unmodified nucleon electromagnetic current~\cite{DoDang:1987zza,Mihaila:1999nn,Carlson:2001mp,Kim:2006iea}, including the state-of-the-art Green function Monte Carlo (GFMC) result for $^{12}$C from Ref.~\cite{Lovato:2013cua,Lovato:2016gkq},  find  modest or no quenching of the CSR.  
Most recently~\cite{Cloet:2015tha}  used an  NJL model in the medium  to find a dramatic  reduction of the Coulomb  sum rule
 for  $|\bfq\gtrsim 0.5$ GeV,    driven by changes to the  bound-proton Dirac form factor. 


The experimental status of the CSR has been unclear.  The initial measurements found quenching of the CSR for $^{12}$C, $^{40}$Ca and $^{56}$Fe \cite{Altemus:1980wt,Meziani:1984is}.  However, a reanalysis of these data \cite{Jourdan:1995np,Jourdan:1996np}, utilizing an alternative prescription for the Coulomb corrections, concluded that there is no quenching. The analysis of the Coulomb corrections in those works was later challenged~\cite{Aste:2005wc,Aste:2007sa,Wallace:2008ev}. These papers  support the conclusion that quenching of the CSR occurs  as reported in Ref.~\cite{Morgenstern:2001jt}.  
New results at high momentum transfer and on a variety of nuclear targets from Jefferson Lab Experiment  E05-110 \cite{Csrexpt05} are eagerly anticipated.  Verification or  disproof of the CSR quenching should  reveal critical aspects of nucleon modification in nuclei.


    %

\section{Future directions in nuclear deep inelastic scattering and detecting short-ranged correlations}

\subsection{Experiment}
There are several different experimental approaches to understanding the EMC-SRC correlation and the origin of the EMC effect.  The most promising approach is to directly test the EMC-SRC correlation by measuring the change in bound nucleon structure function with nucleon momentum  using tagged structure function measurements.

The second approach is to test other predictions of models of the EMC effect by measuring other quantities related to nucleon modification, including the bound ratio of electric to magnetic elastic form factors using polarization transfer $A(\vec e,e'\vec p)$ and the Coulomb Sum Rule.

Lastly, we can learn more about SRC and about the EMC effect individually in several ways.  The first way is to extend EMC and SRC inclusive measurements to more nuclei over a wider range of momentum transfer.  We can also  extend semi-exclusive and exclusive SRC measurements in a similar manner to abteined more detailed information, especially about the potential isospin dependence of the EMC effect, SRCs, and their correlations.  We can select the nucleons we study by measuring the polarized EMC effect and we can measure the isospin dependence of the EMC effect in asymmetric nuclei by measuring parity violating deep inelastic scattering.

\subsubsection{Tagged Structure function Measurements}
The EMC Effect is measured in inclusive $(e,e')$ DIS from a nucleon in a
nucleus.  In order to learn more about the DIS reaction, we can
``tag'' the reaction by detecting
a recoiling nuclear
fragment in coincidence with the scattered electron. By choosing the
nuclear fragment and kinematics wisely, we can restrict the initial
state of the struck nucleon (the nucleon that absorbed the virtual
photon), and thereby learn more about
the microscopic origin of the EMC effect.

The simplest example for such a process is DIS on the deuteron.  If we
can detect a {\bf recoil} nucleon with momentum \bfp{} that did not interact in the
DIS reaction and did not have a final state interaction (i.e., a
spectator), then we know that the struck nucleon had initial momentum
$-\bfp$.  We can then measure the {\bf DIS} cross section for scattering from a
nucleon in the nucleus as a function of its initial momentum.  This
will allow us to extract $F_2$ and hence the quark distributions.  In particular $F_2$ can be measured as a function of virtuality.
This experiment thus provides an opportunity to test the importance of the effects of virtuality that are discussed above.

This was initially studied with 5.7 GeV electrons incident on
deuterium, measuring the scattered electron and the recoil proton with
the CLAS spectrometer \cite{klim06}.  While this measurement did not
have the kinematic reach to unambiguously measure a change in the nucleon structure
function, they did show that protons emitted at large angles, 
$\theta_{pq} > 120^\circ$ (where $\theta_{pq}$ is the angle between
the proton and the virtual photon), were predominantly spectators. 
Later theoretical works support this observation~\cite{Palli:2009it,Cosyn:2010ux}.

In practice, experiments will measure the ratio of cross sections at
fixed recoil momentum and different values of $x'_B$ where 
\[
x'_B = \frac{Q^2}{2p_\mu q^\mu}=\frac{Q^2}{2[(M_d-E_S)\nu+{\bfp_S}\cdot \bfq]}
\]
is the value of \xB{} in the frame of the struck nucleon, $M_d$ is the
deuteron mass, and $E_S$ and $\bfp_S$  are the energy and momentum of
the spectator nucleon. This data will be used to extract \cite{Emcsrcexpt11,Emcsrcdeens}
\begin{multline}
\frac{F^{bound}_2({x'}_B^{hi},q_1^2,\bfp_S)}{F^{free}_2(x_B^{hi},Q_1^2)}
= \\
\frac{\sigma_{DIS}({x'}_B^{hi},Q_1^2,\bfp_S)}{\sigma_{DIS}({x'}_B^{low},Q_2^2,\bfp_S)}\cdot 
\frac{\sigma^{free}_{DIS}(x_B^{low},Q_2^2)}{\sigma^{free}_{DIS}(x_B^{hi},Q_1^2)}\cdot
R_{FSI} 
\end{multline}
where $\sigma_{DIS}^{free}$ is the free-nucleon DIS cross section,
$R_{FSI}$ is a correction factor for the effects of final state interactions,
$\xB^{low}\approx0.3$ where the EMC effect is very small (i.e., where
the EMC ratio is very close to 1), and ${x'}_B^{hi}>0.45$.

By measuring the ratio of the bound to free nucleon structure functions
as a function of spectator momentum (i.e., of nucleon initial
momentum), these experiments will answer the crucial question of which
nucleons are modified {\bf and to what extent}.  Little momentum dependence would imply that
the mean-field nucleons are  modified and large momentum dependence
would imply that SRC nucleons are modified. 

\begin{figure}[t]
  \centering
  \subfloat[Bound proton structure via $d(e,e'n_{recoil})X$ scattering]{\includegraphics[height=5.5cm, width=0.44\textwidth] {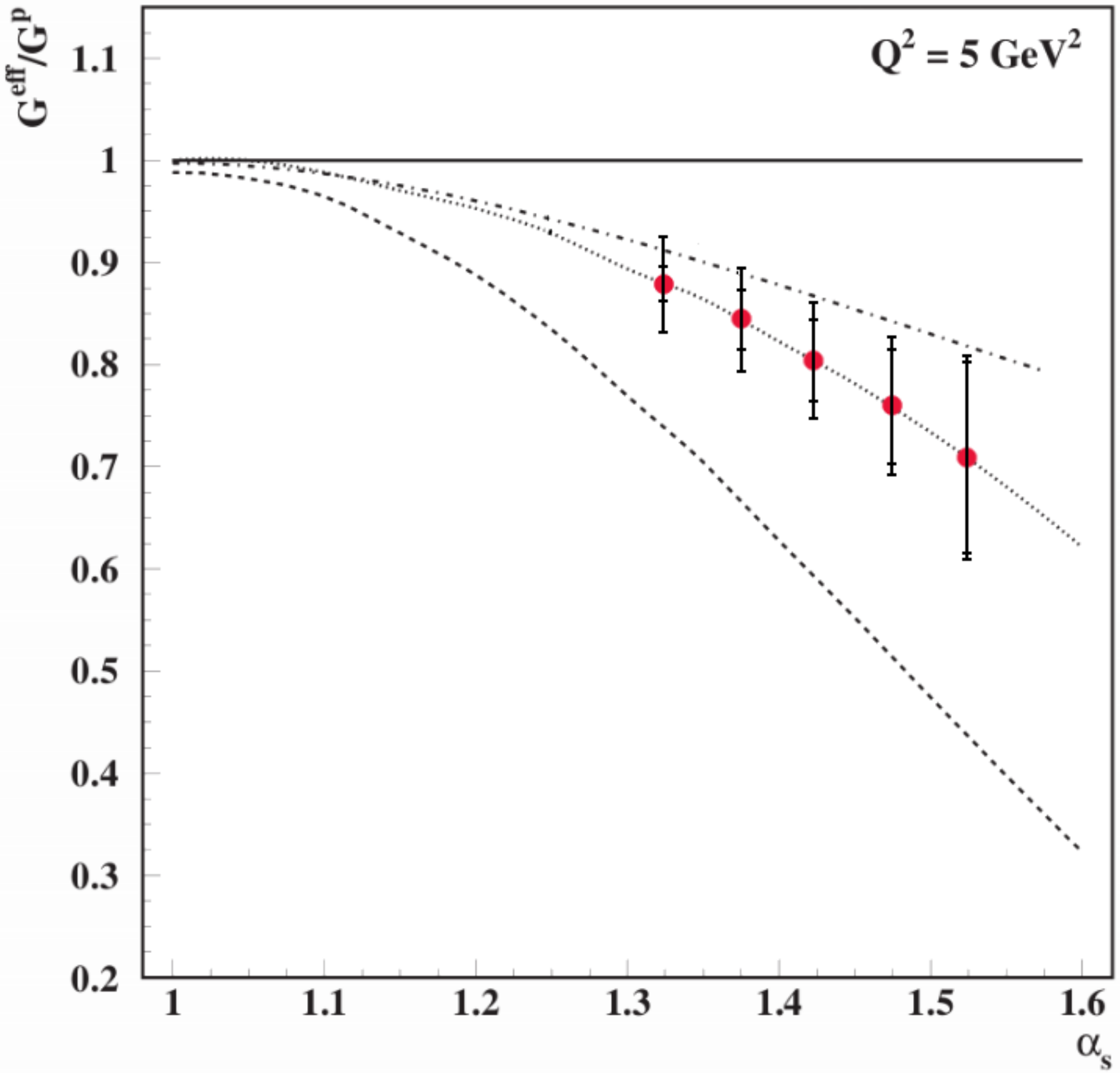} \label{fig:deens}}
  \hfill
  \subfloat[Bound neutron structure via $d(e,e'p_{recoil})X$ scattering]{\includegraphics[height=7.5cm, width=0.4\textwidth] {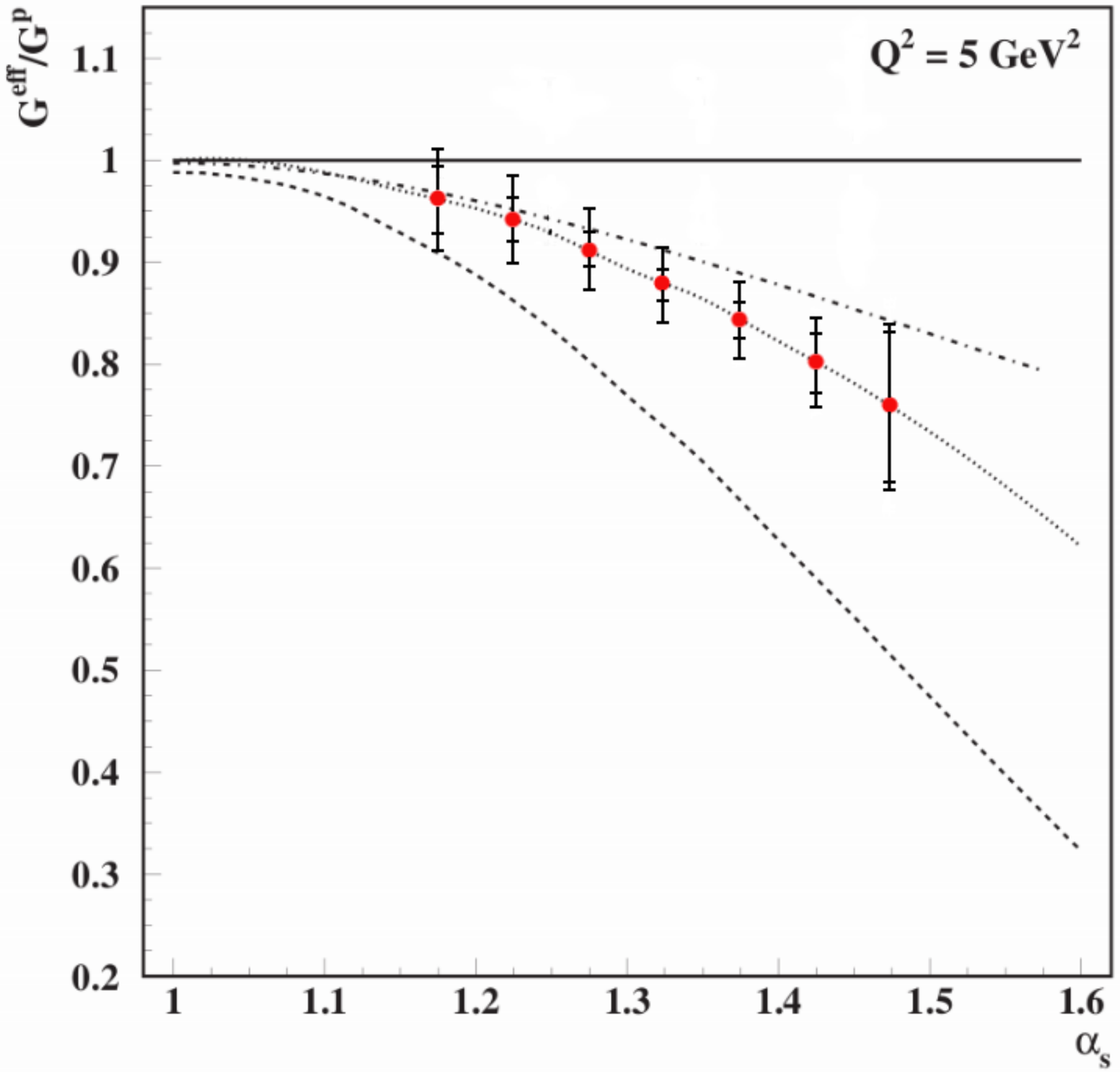} \label{fig:deeps}}
  \caption{The expected results from future Jefferson Lab tagged DIS
    measurements \cite{Emcsrcexpt11,Emcsrcdeens}.  The dashed line is obtained from the color screening model~\cite{Frankfurt:1985cv},  the dotted line is from the color delocalization model~\cite{Close:1983tn}, and dot-dashed the off-shell model \cite{Melnitchouk:1993nk}. From  \cite{Emcsrcexpt11,Emcsrcdeens}}
  \label{fig:TaggedEMC_Reasults}
\end{figure}

There are two approved Jefferson Lab experiments to measure this
reaction.  Experiment E12-11-107 \cite{Emcsrcexpt11} will measure
neutron modification by detecting the scattered electrons in the Hall
C magnetic spectrometers and the spectator protons in a set of GEM
detectors and scintillators covering scattering angles from about
$80^\circ$ to $170^\circ$.  The expected results are shown in
Fig. \ref{fig:deeps}.  Experiment E12-11-003A \cite{Emcsrcdeens} will
measure proton modification by detecting the scattered electrons in
the CLAS12 forward detector and the spectator neutrons in a large scintillator
array covering scattering angles from $160^\circ$ to $170^\circ$.  The
expected results are shown in Fig. \ref{fig:deens}.
 
A second category of experiments consists of measuring the tagged EMC
ratio.  We can ``tag'' different reaction mechanisms by detecting
either a spectator nucleon or a recoil $A-1$ nucleus.  The main idea
is that the electron scatters from a quark in one nucleon.  If that
nucleon belongs to an SRC $NN$ pair, then its partner nucleon will
leave the nucleus.  If that nucleon does not belong to an $NN$ SRC
pair, then the $A-1$ nucleus is much more likely to recoil intact.  In either case, 
\st{we will need to fully account for FSI effects} {\bf proper interpertation the reasults of such measurements requires full understanding of many-body FSI effects}
 that, to the best of our knowlege, so far were only studied for the deuteron.  Instead of the inclusive cross section ratio, the
tagged EMC ratio is
\[
R=\frac{\sigma_A(e,e'p_S)/A}{\sigma_d(e,e'p_S)/2}
\]
integrated over spectator momenta and angles.  Typically, backward angles,
$\theta_{pq}>120^\circ$, are chosen to minimize FSI.

If the spectator is a proton and
has momentum greater than 300 MeV/c, then it \st{almost certainly} {\bf is expected to} belonged
to an $np$ SRC pair.  If nucleon modification is due to nucleons belonging
to SRC pairs, then nucleon modifcation should be the same in deuterium
and in the heavier nucleus and therefore the tagged EMC ratio should be independent of \xB{}
and should be equal to $a_2(A)$, the relative probability of finding a
nucleon in an SRC pair in nucleus $A$ relative to $d$.

\st{The biggest} {\bf Large} uncertainty in interpretting these tagged EMC measurements
\st{is} {\bf stems from} the possibility that the fragments of the struck nucleon will break
up another SRC pair as they exit  the nucleus, significantly
increasing the number of backward nucleons. {\bf This effect should be smaller for light nuclei.} \st{An additional} {\bf A larger} complication arises from the nuclear spectral function that associated high-momentum nucleons with large excitation energies that {\bf will be distributed to the different fragments and} need to be taken into account.

If the measured spectator is an $A-1$ nucleus, then the
struck nucleon almost certainly did not belong to an SRC pair.
Assuming one can overcome the above complications, comparing the tagged EMC ratio for $^4$He with spectator (proton+deuteron)
and or with spectator $^3$He, can give further insight as to whether nucleon modification depends on the struck nucleon momentum or  on the struck nucleon SRC pairing.

\subsubsection{Inclusive EMC and SRC Measurements}
 The inclusive EMC and SRC measurements described in sections~\ref{EMC} and~\ref{HardScattering} were performed on a limited number of nuclei and, in the case of SRC measurements and the JLab EMC measurements, in a limited kinematic range.  Therefore, it is natural to extend both EMC and SRC measurements to additional nuclei over a wider kinematical range.

 \begin{figure}[t]
 \centering
 \includegraphics[width=8cm, height=7cm]{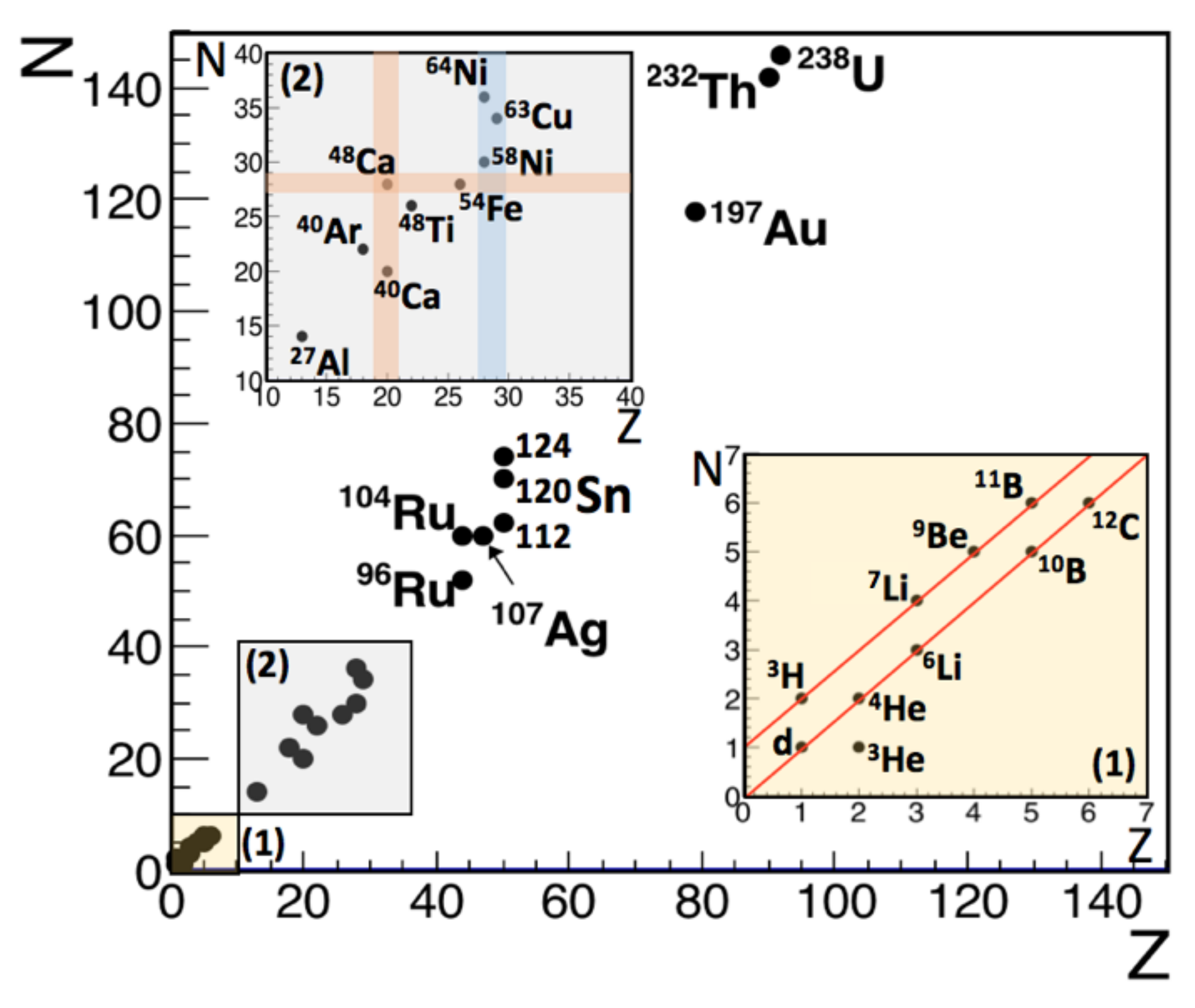} 
 \caption{ Phase-space of nuclei considered for future EMC/SRC measurements at Jefferson Lab as a function of their proton ($Z$) and neutron ($N$) numbers. The two inserts focus on the light and medium mass nuclei refims. For light nuclei one can systematicaly study a series of symmetric nucli and the detailed effect of the addition of one neutorn (proton). For medium mass nuclei the horizontal and vertical bands mark nuclei with similar numebr of protons (neutrons) and a varing number of neutrons (protons) allowing to study the effect of nuclear asymmetry.}
 \label{fig:Inclusive12GeV_Nuclei} 
 \end{figure}

 Fig.~\ref{fig:Inclusive12GeV_Nuclei} shows  nuclei that can or will be measured at Jefferson Lab as a function of their proton ($Z$) and neutron ($N$) numbers. A wise selection of nuclei allows for a systematical experimental study of SRC and the EMC effect for fixed nuclear asymmetry as a function of mass number and for fixed mass number as a function of asymmetry.  
The planned Jefferson Lab measurements \cite{solvignonexpt11,arringtonexpt06,arringtonexpt10,petratosexpt10} will systematically measure both the size of the EMC effect and the height of the SRC plateau over many nuclei from $^3$He and $^3$H to $^{208}$Pb, covering a wide range of mass numbers and nuclear asymmetries ($N/Z$).  Measurements with unstable nuclei at other laboratories could significantly extend the available range of nuclear asymmetry. 

Light and heavy nuclei can exhibit significantly different nuclear effects. Medium and heavy nuclei ($A \ge 10$) exhibit properties of nuclear saturation and can be relatively well described using effective theories for strongly interacting many-body Fermi systems. However, light nuclei  span a wide range of nuclear densities and asymmetries, with some nuclei exhibiting a rich cluster-like substructure. 

In the case of the EMC effect, special care should be given to ``standard'' nuclear structure effects that do affect the DIS cross-section ratio and can therefore potentially mimic an isospin dependent. Therefore, any measurement of the ratio of the EMC effect in isospin asymmetric nuclei (e.g. $^3$He/$^3$H, $^{48}$Ca/$^{40}$Ca etc.) should be compared to a calculation of standard nuclear effects that take into account differences in the proton and neutron momentum distributions and pairing probabilities. These can be studies indirectly by inclusive QE (e,e') reactions or directly using semi-inclusive and exclusive (e,e'N) and (e,e'pN) reactions. In the case of light nuclei ab-initio few-body calculations can also provide relevant information.

An additional advantage of light-nuclei studies is the ability to compare the experimental results with detailed ab-initio nuclear structure calculations. Assuming reaction mechanisms such as FSI, MEC and others are under control, such comparisons of experiment and theory can offer significant insight into the underlying microscopic physics. For heavy nuclei, such ab-initio calculations of short-range nuclear structure are still limited, but rapid progress is being made~\cite{wiringa14,Hagen:2015yea, Carlson:2014vla}.



In addition to extending the range of nuclei measured, it is also important to extend the measured $Q^2$ range.  This is especially important for SRC studies where the minimum initial momentum depends strongly on $Q^2$ (see Fig.~\ref{fig:pmin}).  The SRC cross section ratios of \cite{egiyan03} were measured at $1.4\le Q^2 \le 2.6$ GeV$^2$ with most of the data at $Q^2<2$ GeV$^2$ (see Fig. \ref{fig:eeratios}a).  They observed flat plateaus in the cross section ratio for $1.5\le \xB \le 1.9$, which corresponds to $250\le p_{min} \le 500$ MeV/c which is where we expect tensor correlations to dominate.  By contrast, the SRC cross section ratios of \cite{fomin12} which were measured at $Q^2=2.7$ GeV$^2$ exhibit ``plateaus'' that are not quite as flat, especially for heavier nuclei (see Fig. \ref{fig:eeratios}a).  At $Q^2=2.7$GeV$^2$, $1.5\le \xB\le1.9$ corresponds to $325\le p_{min}\le 700$ MeV/c, which extends beyond the tensor correlations region into the central correlations region.  
Measuring the $Q^2$ dependence of the SRC plateaus will help us quantitatively relate the experimental results to detailed ab-initio nuclear structure calculations - a needed comparison that was not done to date.

The $Q^2$-dependence of the EMC effect has been studied over a wide kinematical regime. However, there are still several intriguing questions about higher twist effects that should be studied systematically. The Jefferson Lab 6 GeV EMC effect measurements included data with invariant mass, $W> 1.4$ GeV, a region that is dominated by resonance production rather than DIS. The fact that the measured EMC ratios agreed with the SLAC data, measured at higher $W$, showed that resonance contributions largely cancel in the $A/d$ ratio. By covering a broader kinematic range, the future 12 GeV measurements  will help quantify this issue.
A review of the possibility of studying the large $\xB$ at the lHC is presented in~\cite{Freese:2014zda}.

\subsubsection{Semi-Inclusive and Exclusive SRC Measurements}
In the context of the EMC effect, the interpretation of planed future experiments in isospin asymmetric nuclei (e.g. $^3$He/$^3H$, $^{48}$Ca/$^{40}Ca$) requires understanding the differences in the proton and neutron momentum distributions and SRC pairing probabilities, as these enter into the baseline calculation of ``standard'' nuclear effects and EMC models calculations, that are to be compared with the experimental data.

The required information about high momentum nucleons and SRC in nuclei can be obtained by
    scattering an electron or other probe from a nucleus and detecting one
    or more of the ejected nucleons.  $A(e,e'p)$ experiments can
    measure the amounts of high momentum nucleons in different
    nuclei and how that changes with nuclear isospin asymmetry.  
    
The fact that the $A(e,e'p)$ reaction is mainly sensitive to the protons in nuclei whereas the $(e,e')$ reaction is sensitive to all nucleons in nuclei make their measurements complementary and crucial to allow for a detailed study of the dependence of SRC effects on the nuclear asymmetry.
 While more challenging to perform, A(e,e'n) measurements can complement the other reactions and yield additional information on the role of protons and neutrons in asymmetric nuclei.
    
    One Jefferson Lab experiment \cite{tritiumexpt14} will measure $^3$H
    and $^3$He$(e,e'p)$ as a function of $p_{miss}$ in kinematics where FSI are small in order
    to determine the ratio of the $^3$He and $^3$H momentum
    distributions.  In the naive SRC picture, this ratio should be two
    at low $p_{miss}$ because there are twice as many protons in
    $^3$He as in $^3$H and it should decrease to one at high
    $p_{miss}$ because there are two $pn$ pairs each in $^3$He and
    $^3$H.  
    
    Similar experiments in medium mass nuclei could measure how the number of
    high momentum protons changes as you add eight {\it neutrons} from
    $^{40}$Ca to $^{48}$Ca and by adding six more protons from $^{48}$Ca  to $^{54}$Fe  \cite{cafeexpt2016}.
    Ongoing Jefferson-Lab CLAS data-mining analysis of A(e,e'n) and A(e,e'p) scattering off $^{12}$C, $^{27}$Al,$^{56}$Fe and $^{208}$Pb
    is expected to provide new insight into isospin asymmetry effects on SRCs.

We can gain more information about SRC pairing in nuclei by knocking
    out a high-initial momentum nucleon and detecting its correlated
    partner, either with electron or proton probes, $A(e,e'pN)$ or
    $A(p,2pN)$.  By extending the range of missing momentum we can
    study the transition from tensor dominance (at $300 \le p_{miss}
    \le 500$ MeV/c) to the scalar repulsive core (at higher
    $p_{miss}$).  By focusing on lower $p_{miss}$, we can map the
    transition fro the mean-field to the SRC-dominated domain (the
    nuclear ``Migdal jump'').  By extending the $A$-dependence of SRC
    pair abundancies and properties we can learn about SRC-pair
    quantum numbers and provide data for a quantitative theory of SRCs.

\subsubsection{Polarized EMC  Measurements \label{PolEmc}}
Motivated by open questions about the EMC effect and the ``proton spin crisis'', Jefferson Lab will perform the first measurement of the spin-dependent EMC effect utilizing CLAS12 in Hall B with 11 GeV polarized electrons and polarized targets \cite{PolEmc14}.  They will determine the ratio of the double-spin asymmetries in $^7$Li (using $^7$LiD) in which a highly polarized proton is embedded in the nuclear medium, and on the proton (using $^6$LiH). The double spin asymmetry is measured as
\[
A_\para = \frac{d\sigma\downarrow\Uparrow - d\sigma\uparrow\Uparrow}{d\sigma\downarrow\Uparrow + d\sigma\uparrow\Uparrow}
\]
and is approximately equal to the ratio of polarized to unpolarized structure functions: $g_1^{^7Li}/F_1^{^7Li}$.  Many systematic uncertainties will cancel in the asymmetries and in the ratios of asymmetries.  Together with the unpolarized structure function (also to be measured at Jefferson Lab), they will also extract $g_1^{^7Li}$ and, using a sophisticated modern wave function model, extract the “in-medium” proton spin structure function function $g_1^{p\para{}^7Li}$ for a proton bound in $^7$Li. They will cover a kinematic range of $1 < Q^2 < 15$ GeV$^2$ and $0.06 < \xB < 0.8$.

\begin{figure}[t]
  \centering
  \includegraphics[width=3in] {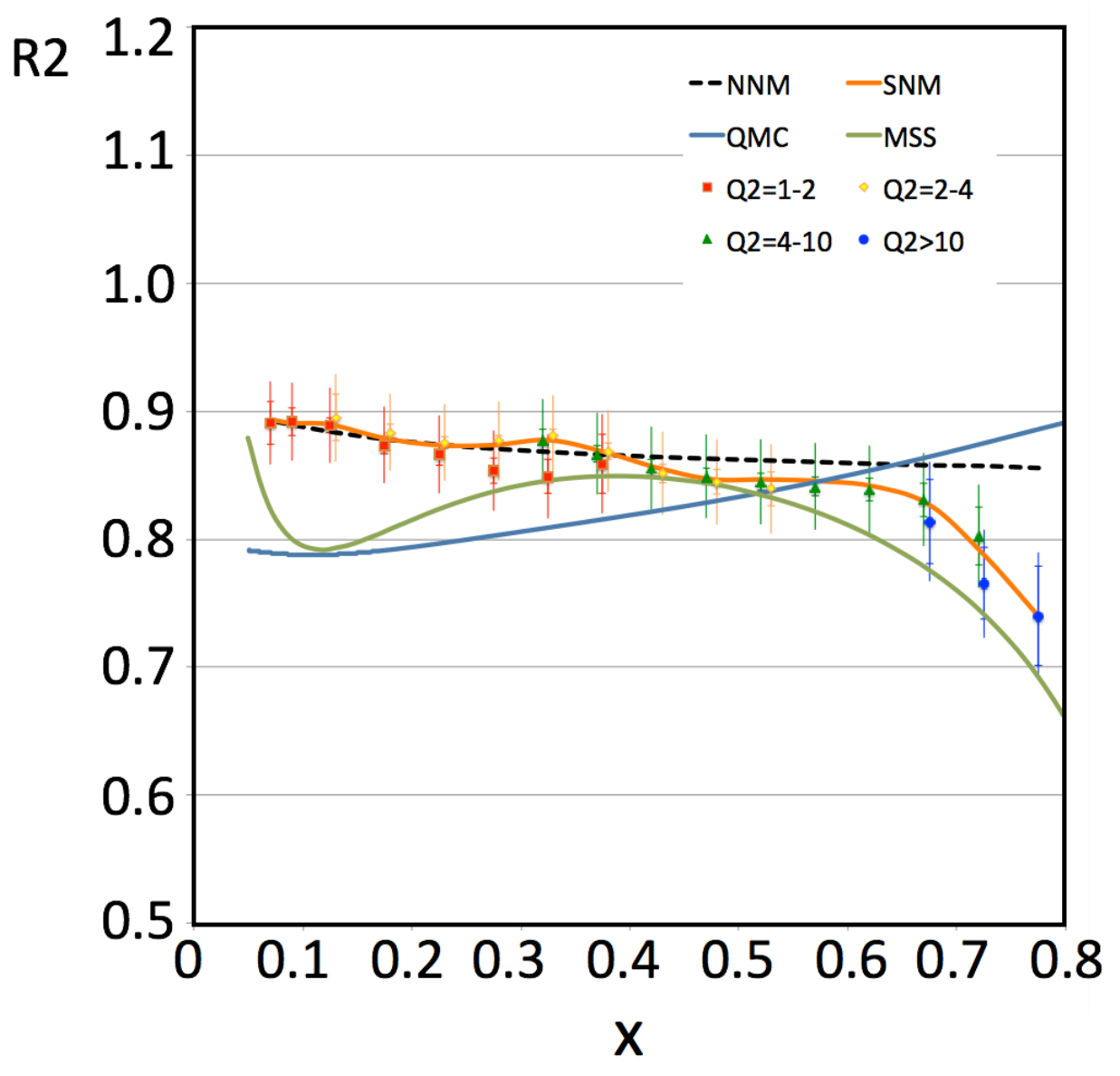} 
  \caption{The expected results of the polarized EMC effect measurement at Jefferson Lab.  The ratio of the parallel double spin asymmetry $A_\para$ for $^7\vec{\rm Li}(\vec e,e')$ to $\vec p(\vec e,e')$, normalized by multiplying it with the ``naive'' unpolarized structure function ratio for $^7$Li over hydrogen, plotted vs $\xB$.  The models are NNM (naive nuclear model with no fermi motion), SNM (standard nuclear model with fermi motion and kinematical binding energy effects), QMC (Quark-Meson Coupling model), and MSS (x-rescaling \cite{fanchiotti14}).  Figure adapted from \cite{PolEmc14}.}
  \label{fig:PolEmc}
\end{figure}

Mean field models of nucleon modification predict stronger effects than in the unpolarized structure functions.  On the other hand, since nucleons in tensor correlations tend to have opposite spin to the overall nuclear spin, the EMC effect could be minimal or even in the opposite direction.  These data will provide new constraints on models for the EMC effect, some of which predict that medium modifications of quark distributions depend strongly on the quark helicities (see Fig.~\ref{fig:PolEmc}).

\subsubsection{Parity Violating Deep Inelastic Scattering}
There is some evidence that $u$- and $d$-quark distributions are
modified differently in asymmetric nuclei.  Theoretically, since
protons move faster than neutrons in neutron-rich nuclei, if nucleon
modification depends on nucleon virtuality (as in the PLC model), then
we expect protons, with 2 $u$- and 1 $d$-quarks, to be more modified
than neutrons.

Experimentally, the NuTeV experiment compared neutrino and
anti-neutrino DIS off an Iron target and extracted a value of the
Weinberg mixing angle that differs from the standard model by about
$3\sigma$ \cite{zeller02,zeller03}.  While this led to much excitement and attempts to relate it to
physics beyond the standard model, recently it was shown that an
isospin dependent EMC effect that affects protons more than neutrons
could resolve the anomaly \cite{Cloet09}. 

A measurement of parity violation in $A(e,e')$ DIS would directly
measure the $d-u$ difference as a function of \xB \cite{pvdisexpt}.  The difference in
the 
left-right asymmetry for helicity $+1$ and $-1$ electrons is
proportional to the product of the photon and $Z$ amplitudes divided
by the square of the photon amplitude.  This asymmetry will be $10^2$
to $10^3$ parts per million for DIS scattering from a
heavy nucleus:
\[
A_{PV}\approx -\frac{G_FQ^2}{4\sqrt{2}\pi\alpha}\left[ a_1(x)+\frac{1-(1-y)^2}{1+(1-y)^2}a_3(x)\right]
\]
where $y=1-E/E'$, and $a_1$ and $a_2$ depend on the quark
distributions.  In the symmetric nucleus limit
\[
a_1\simeq
\frac95-4\sin^2\theta_W-\frac{12}{25}\frac{u_A^+-d_A^+}{u_A^+ + d_A^+}
+ \ldots
\]
where $u_A$ refers to all the up quarks in the nucleus and the
superscript $+$ refers to the sum of the quark and anti-quark
distributions.  Thus the parity violating asymmetry is sensitive to
the difference between the $u$ and $d$ quark distributions in the
nucleus.  

\label{pvdis}

\subsection{Theory}\label{FT}
The review of the theory presented here shows that there is a strong connection between the cause of the EMC effect and the short-ranged correlations that cause the high $\xB$ plateau in (e,e') scattering on nuclei. Nevertheless, there are gaps in almost every part of the theory, from the initial state wave function, to the modification of nucleon structure, to the need to include the effects of final state interactions. We therefore present an outline of the necessary improvements.

The EMC effect is a modification of nucleon structure functions. Obtaining an understanding of this effect therefore requires a working  understanding of the  valence sector of the free nucleon wave function,  so that the effects of the  medium on the relevant components can be correctly included. Lattice calculations,
 {\it e.g.,}~\cite{Ji:2013dva,Lin:2014zya},  and the Dyson-Schwinger approach~\cite{Cloet:2013jya} are making progress on computing free nucleon parton distributions. It also would be necessary to build nucleon models that are easily related to the output of these Euclidean-space theories, {\it e.g.}~\cite{Hobbs:2016xfz,Burkardt:1996gu}. A twenty-first century  calculation of medium modifications  cannot be made without inputs from such models.
 
 The calculation of deep inelastic scattering from nuclei needs to be improved in several different ways. For example, the calculations using the PLC-suppression model have been made mainly for $\xB=0.5$~\cite{Ciofi07}, where effects of Fermi motion nearly vanish. To understand the EMC ratios discussed above it is necessary to be able to make accurate calculations for a  range of values of $0.3\le \xB\le 0.7$. So far this has been done~\cite{Freese:2014zda} by assuming that no medium modification occurs for $\xB<0.45$ and linearly interpolating the region between $0.45<\xB<0.65$. 
  Calculations need to handle finite-sized nuclei without resorting to infinite nuclear matter calculations using a local density approximation. Such a program would require computation of nuclear spectral functions for finite-sized nuclei. This would involve intensive numerical work, so it would be important to 
 present such spectral functions in an easily accessible manner. 
 
We have seen that only models with medium modification arising from short-ranged effects can handle both the  EMC effect and high $\xB$ $(e,e')$ scattering.
However models in which the medium modification is driven by mean-field effects give an excellent description of the EMC effect, see {\it e.g.}~\cite{Cloet:2005rt,Cloet:2006bq,Cloet:2009qs}. It would not be realistic to think that the ultimate accurate description would make use of only one of the two possible ideas.
Therefore it is important  to build models of  medium modifications of nucleon wave functions that includes both mean-field effects and the effects of correlations. The necessary model of nuclei would need to be consistent with nuclear saturation properties, include non-nucleonic degrees of freedom and have those relativistic effects needed to compute nuclear deep inelastic scattering cross sections.  

Many treatments of final state interactions for exclusive reactions (e.g., $(e,e'p)$ and $(e,e'pN)$) use complex optical potentials, which automatically violate current conservation.  To fully understand spectroscopic factors and nucleon-nucleon correlations it is necessary to ensure that the reaction theory models conserve current.  We also need to better understand electromagnetic current operators in models of the nucleon-nucleon interactions that employ low-momentum cutoffs.

There is a need to understand higher twist effects in nuclei, so we can understand why the EMC ratios measured at JLab are nearly the same as those measured at much higher energies at SLAC and CERN.  
 
In addition to improving our understanding of the theoretical underpinnings of the causes of the EMC-SRC correlation, it is necessary to  explore the implications of the EMC-SRC correlations and of $pn$ dominance in SRC.  
The possible inversion of the kinetic energy sharing in asymmetric nuclei could significantly affect several sub-fields of physics. In astrophysics the nuclear symmetry energy is of fundamental importance. It describes the change in energy of a nuclear system when a proton is replaced by a neutron. $np$-SRC dramatically reduce the kinetic part of the symmetry energy \cite{hen15} and work is ongoing to understand other effects. Additional implications of SRCs on nuclear systems include the nuclear response to neutrino scattering \cite{fields13,Fiorentini13a}, cooling rates of neutron stars, contact interactions in Fermi systems 
\cite{hen15b,Frankfurt:2008zv} and more. While the discussion of these effects extends beyond the scope of this review, they are extensively discussed in the literature.

\section {The way we think it is and the ways to check }

This article has  focused on explaining  two seemingly unrelated phenomena: lepton-nucleus deep inelastic scattering (DIS) and  quasi-elastic (QE) electron-nucleus scattering at large values of $\xB$, and their  surprising relation. 
DIS from a nucleus is very different than DIS from a collection of free nucleons; this is the EMC effect which is parameterized in terms of the slope  of the  EMC ratio, $R$, of bound to ``free'' cross sections.  This slope cannot be explained unless the internal quark structure of a bound nucleon differs from that of a free nucleon.  

Quasi-elastic scattering, in which a nucleon is knocked out of the nucleus intact, reveals plateaus in the cross section ratios of nuclei to deuterium  at large values of $\xB$ that correspond to scattering from short range correlated (SRC) two-nucleon pairs.
Different experiments show that the slope of the EMC effect is linearly proportional to the height of the plateaus!  Further studies showed that the two-nucleon pairs consist of a neutron  and a proton. 

A review of the available  experimental and theoretical evidence shows that the relation between the EMC slope, $dR/d\xB$, and the SRC plateau height is no accident.  There is an underlying 
cause of both effects: the influence of strongly correlated neutron-proton pairs.  These correlated pairs are temporary high-density fluctuations in the nucleus in which the internal structure of the nucleons is briefly modified.  This  conclusion needs to be quantified by future experiments and improved theoretical analyses that are discussed  in this article.

The connection between the EMC effect and nucleon-nucleon correlations is very profound. Although the binding energy of a nucleon is less than a percent of its mass, the fact that the nucleon is made of quarks and gluons is manifest in two distinct sets of phenomena, via experiments that have been repeated several times. The direct influence of the quark presence in nuclei is now established.

This presence is a subtle effect as it must be, given the generally  small deviation of $R$ from unity, and does not arise via  the usual low-energy, low momentum transfer  nuclear physics observables: binding energy, spectra, radii, electroweak transition rates, {\it etc}. Nonetheless, the quark presence cannot be denied. We expect that a deeper understanding of the EMC/SRC connection will ultimately lead to an improved understanding of the nature of confinement of light quarks. 

The Jefferson-Lab 12GeV program includes a series of approved experiments targeted at improving our understanding of the EMC effect, SRCs and their connection. The forthcoming results of these experiments are expected to shad new light on the origin of the EMC effect and provide stringent constraints on current of future theoretical calculations.

    %

\section{Acknowledgments}
We thank our many colleagues for their efforts in accomplishing the research discussed in this review and also for the many insights that they have provided through discussions.
This work of was supported by the U. S. Department of Energy Office of Science, Office of Nuclear Physics under Award Numbers DE-FG02-97ER-41014, DE-FG02-94ER40818 and DE-FG02-96ER-40960 and by the Israel Science Foundation (Israel) under Grants Nos. 136/12 and 1334/16.
\section{Appendix \label{Appendix}}

\subsection{Understanding the $np$ relative wave function  \label{appendix:Understanding_np}}
The aim of this Appendix  is to  provide a qualtitative explanation  that the momentum space wave function of the  deuteron, a very weakly bound system, has a significant high momentum $k$ tail. Indeed one sees  an  approximate
 $k^{-4}$ behavior of the deuteron density for large values of $k$. This tail persists in nuclei because of short ranged correlations  between nucleons.
 
 A $1/k^4$  density comes from $1/k^2$ in the wave function which can be obtained  if the nucleon-nucleon interaction is a delta function in coordinate space,
   as occurs  in  leading order
 EFT or in the effective range expansion behavior. Such approximations are valid only at very small values of momentum $1/r_e \gg k \gg 1/a$, where there is approximate scale invariance, where
 $a$ is the scattering length of about 5 fm and $r_e$ is the effective range of about 2 fm.   The $1/k^2$ behavior of the wave function emerges
 at large values of $k$  due to the second-order effects of the 
 of the one pion
exchange (OPE) contribution to the tensor potential $V_T$.  The Schoedinger equation for the spin-one two-nucleon
system, which involves $S$ and $D$ state components, can be expressed
as an equation involving the $S$ state only by using $(-B
-H_0)|\Psi_D\rangle =V_T|\Psi_S\rangle$, where $B$ is the binding
energy of the system and $H_0$ is the Hamiltonian excluding the tensor
potential. Thus one obtains an effective $S-$state potential: $V_{\rm
  00}=V_T( -B-H_0)^{-1}V_T$, \eq{V00},  where $V_T$ connects the $S$ and $D$
states.  The intermediate Hamiltonian $H_0$ is dominated by the
effects of the centrifugal barrier and can be approximated by the
kinetic energy operator~\cite{BrownJackson}.  This second-order term is large because it
contains an isospin factor $(\mathbf\tau_1\cdot\mathbf\tau_2)^2=9$, and
because $S_{12}^2=8-2S_{12}$.  Evaluation of the $S$-state potential,
neglecting the small effects of the central potential in the intermediate $D$-state,
yields
\begin{equation}  V_{00}(k,k')\approx-M\frac{32f^4}{\mu^4 \pi^2}\int
  \frac{p^2dp}{MB+p^2}I_{02}(k,p)I_{20}(p,k'),\label{v2}\end{equation}
where $M$ is the nucleon mass, $f^2\approx 0.08$ is the  square of the $\pi N$ coupling
constant, $\mu$ is the pion mass, and $I_{LL'}$ are  partial wave projections of the OPEP extensor interaction in momentum space. These are evaluated in
\cite{Haftel:1970zz}
\bea I_{02}(p,k)=I_{20}(p,k)={k^2Q_2(z)+p^2Q_0(z)\over 2pk}-Q_1(z),\nonumber\\\eea
with $z\equiv (p^2+k^2+\mu^2)/( 2pk),$ and $Q_i$ are Legendre
functions of the second kind in the conventions of that reference.
The result, \eq{v2} corrects  errors in \cite{Hen:2014lia}.  The errors do  not affect the qualitative statements made in the cited paper, as we now demonstrate.

We  use \eq{v2} to estimate quantities of interest.
We note the asymptotic property: $\lim_{p\to\infty}
I_{02}(p,k)=1-(k^2+\mu^2)/p^2+\cdots$.  
Thus the integrand of \eq{v2} is dominated by large values of $p$ and
diverges unless there is a cutoff. This means that $V_{00}( k',k)$ is
approximately a constant, independent of $ k$ and $k'$. This is the
signature of a short ranged interaction.  We  expose this feature in more detail by assuming that for the important regions of the
integral appearing in \eq{v2} by treating the variables  $k, k'$ as small compared to  the cutoff momentum. Then 
$
 I_{02}(0,p)\approx { p^2\over p^2+\mu^2},$ 
and
\begin{equation}  V_{00}(k,k')\approx-M\frac{32f^4}{\mu^4 \pi^2}\int_0^M
  \frac{p^2dp}{MB+p^2}\left({p^2\over p^2+\mu^2}\right)^2 (1 +\cdots) 
   ,\label{v3}\end{equation}
  where we have cut off the linearly divergent integral for momenta $p>M$ and $\cdots$ represents terms of  ${\cal O}({k^2+{k'}^2\over M^2})$.  All realistic models of the $NN$ interaction employ some sort of a cutoff, and a mass scale of the nucleon mass is typical of one-boson exchange potentials~\cite{Machleidt:1987hj,Machleidt:1989tm}.
  Thus  $V_{00}(k,k')$ is approximately independent of its momentum arguments,  the hallmark of short-ranged interactions. The use of \eq{v3} provides an approximate  upper limit.

The resulting  asymptotic $1/k^4$ dependence of the square of the   wave function can be
seen by using the Lippmann-Schwinger equation in the form
\bea& \la k|\psi_S\ra\approx \la k| ( -B-H_0)^{-1} \int d^3k'V_{00}(k,k')\la k'|\psi_S\ra\,\\&
\approx {-V_{00}(0,0)\over B+{k^2\over M}}  \int d^3k' \psi_S(k')\,= {-V_{00}(0,0)\over B+{k^2\over M}}  (2\pi)^{3/2}\psi_S(r=0)\nonumber
\eea
where the subscript $S$ refers to the $S-$state and the integral over all momenta, $k'$ leads to a proportionality to the coordinate-space wave function at the origin. In terms of the usual $S$-state radial wave function  $u(r)$ we have
\bea \psi_S(r=0)= \lim_{r\to 0} {u(r)\over r} {1\over \sqrt{4\pi}}.\eea
Using known wave functions, we find $ \lim_{r\to 0} {u(r)\over r}=(0.0267,0.0584,0.0792)\,{\rm fm}^{-3/2}$ for the Nijmegen, Reid93~\cite{Stoks:1994wp}, and Argonne V18~\cite{Wiringa:1994wb} potentials respectively.  
The result \eq{v3} shows the $1/k^2$ dependence of the wave function, with overall strength determined by the detailed potential models. The density is the square of the wave function $\sim 1/k^4$ with an overall strength varying by a factor of 9, depending on the potential used.  Thus  we find a high momentum $1/k^4$ behavior far beyond the validity of the effective range approximation.  Potentials  without 
  this high-momentum density either have a very weak tensor force or a  
  cutoff at low momenta.

We may check the  rough validity of these findings by computing the $D$ state probability, $P_D$: 
\bea &P_D=\la\psi_s|V_T {1\over( B+H_0)^2} V_T|\psi_S\ra\nonumber\\
&\approx \frac{32f^4}{\mu^4 \pi^2} (2\pi)^3 \psi^2(r=0)\int_0^M {p^2dp\over (B+{p^2\over M})^2} \left({p^2\over p^2+\mu^2}\right)^2
\label{pdd}\eea
We evaluate $P_D$ using $u(r=0)$ for each of    the Nijmegen, Reid93~\cite{Stoks:1994wp}, and Argonne V18~\cite{Wiringa:1994wb} potentials respectively. Numerical evaluation  of  \eq{pdd}  equation yields $P_D=({2,10,18})$ \% for the three potentials repsectively.
The    actual value for all of these potentials 
is  about $P_D=6\% $.  These results show that qualitative treatment here is adequate only  for rough estimates that maintain   the qualitative idea that the
iterated effects of OPEP produce the $1/k^4$ behavior of the deuteron density.  
The results of this sub-section depend on the  chosen scale ($M$ here).  Choosing a sufficiently softer scale would modify the high-momentum dependence of the wave function.A detailed comparison of  the momentum dependence of known deuteron wave functions is presented in \cite{Hen:2014lia}.

Results similar that the  relevant interaction matrix element is approximately independent of its momentum arguments
 have been obtained previously. Mosel's group~\cite{Lehr:2000ua,Lehr:2001qy,Konrad:2005qm} assumed this independence and used it to 
  help to clarify the basic, fundamental origins of the nucleon spectral functions and the high-momentum tails. Using a constant interaction  matrix element, 
along with the Fermi-gas model, and solving the relevant Dyson equation gave
high-momentum tails  with a density $\sim ~1/k^4$, and 
 spectral functions essentially identical to those of more detailed computations.

\subsection{Basic terminology}
\newcommand{\bfx}{{\bf x}}\newcommand{\bfy}{{\bf y}}
We   define some basic terms.
The probability to find a nucleon at a coordinate $\bfx$ (where this  notation includes spatial position, nucleon spin and isospin) is given by
\bea 
\rho(\bfx)={1\over A}\la \Psi|\sum_{i=1}^A  \delta(\bfx-\bfx_i)|\Psi\ra,
\eea
where $|\Psi\ra$ is the relevant nuclear wave function. The quantity $\r(\bfx)$ is known as the density. The normalization is 
$ \int d\bfx \r(\bfx)=1,$
where the integral includes a sum over nucleon spin and isospin.

 The two-body density in coordinate space is given by
 \bea 
\rho^{(2)}(\bfx,\bfy)={1\over A(A-1)}\la \Psi|\sum_{i\ne j}  \delta(\bfx-\bfx_i) \delta(\bfy-\bfy_j)|\Psi\ra.\nonumber\\
\label{rho2def}\eea
The integral of the two-body density  over $\bfx$ yields the density $\r(\bfy).$
The correlation function $C(\bfx,\bfy)$ is the deviation of the two-body density from the mean field approximation:
\bea C(\bfx,\bfy)=\rho^{(2)}(\bfx,\bfy)-\rho(\bfx)\rho(\bfy).
\eea
The quantity $C(\bfx,\bfy)$ vanishes if the wave function $|\Psi\ra$ can be represented as a product of single-nucleon wave functions.
Furthermore the stated normalization conditions lead to the result
\bea \int d\bfx C(\bfx,\bfy)=0,\,\int d\bfy C(\bfx,\bfy)=0.
\eea
It is useful to also define the probability $\r_{2,1}(r)$    that if a nucleon is at a given position, another one is separated by a distance $r$.
\bea &
\r_{2,1}(r)\equiv {1\over 4\pi r^2 A}\la \Psi|\sum_{i\ne j}  \delta(r-|\bfr_i-\bfr_j|)\Psi\ra\nonumber\\&=\int d^3R \r_2(\bfR+\bfr/2,\bfR-\bfr/2).\label{r21}
\eea
where $\bfR$ is the center of mass position of the two-nucleon system.

The same kind of analysis can be done in momentum space.  Evaluation of $\r(\bfx)$ requires the square of the coordinate-space representation of $|\Psi\ra$, while that of $n(\bfk)$ requires the momentum-space representation of the same wave function.
The probability for a nucleon to have a momentum $\bfk$ is given by
\bea &
n(\bfk)=
{1\over A}\la \Psi|\sum_{i=1}^A \delta(\bfk-\bfk_i)|\Psi\ra.
\eea
It is convenient to define a two-body density $n_2(\bfK,\bfkappa)$  in momentum space, which gives the probability of two nucleons having a total momentum of 
$\bfK$ and a relative momentum $\bfkappa$:
\bea & n_2(\bfK,\bfkappa)=\nonumber\\&{1\over A(A-1)}\la \Psi|\sum_{i\ne j}  \delta(\bfK/2+\bfkappa-\bfk_i) \delta(\bfK/2-\bfkappa-\bfk_j)|\Psi\ra\nonumber\\&
\eea
Experimentalists defined a  correlation as existing if the system has $\kappa\gg K$, with  $\k> k_F$ and $K<k_F$.

 It is also useful to consider the
integrated quantity:
\bea &n_{2,1}(\bfkappa)\equiv\int d^3K n_2(\bfK,\bfkappa)\nonumber\\&={2\over A(A-1)}\la \Psi|\sum_{i\ne j}  \delta(\bfk_i-\bfk_j -2\bfkappa)|\Psi\ra ,\eea
which is the momentum space version of \eq{r21}.  

A specific model for the two-nucleon density is used in the analysis of the data relevant to this review.
 For small relative distances $r$ one writes the two-nucleon wave function $\Psi(\bfR,\bfr)$ in the following form
 \bea 
 \Psi(\bfR,\bfr)=F_A(R)\psi_D(r),\,r<<R_A,  \label{fact}
 \eea
 where $R_A$ is the radius of the nucleus, and the often used assumption is that at short distances all relative wave functions are the same as that of the deuteron $D$.
In this model 
\bea \r_{2,1}^A(r)=\int d^3R F_A^2(R) \,\psi_D^2(r)\equiv a_2(A)\,\psi_D^2(r), \,r\ll R_A\nonumber\\ \label{space}
\eea
 In momentum space 
 
  \bea 
\widetilde{ \Psi}(\bfK,\bfk)=\widetilde{ F_A}(K)\widetilde{ \psi_D}(k),  \,k\gg1/R_A
 \eea
 where the tilde denotes Fourier transform and the momentum  variables are canonically conjugate to $\bfR$ and $\bfr$.
 The one-body density $n_A(\bfk_1)$ is given by
 \bea &
 n_A(\bfk_1)=\int d^3k_2 \left|\widetilde{ \Psi}(\bfK,\bfk)\right|^2\nonumber\\&=\int d^3P \widetilde{ F_A}^2(P)\left|\widetilde{ \psi_D}(\bfk_1-\bfP/2)
\right|^2 \\
&\approx \int d^3P \widetilde{ F_A}^2(P)\,\left|\widetilde{ \psi_D}(\bfk_1)
\right|^2=a_2(A)\left|\widetilde{ \psi_D}(\bfk_1)
\right|^2,
\label{mom}\eea
where $k_1\gg 1/R_A$ is assumed and the relation in terms of $a_2(A)$ is an example of Parceval's theorem. 

The next step is to relate the quantities $n_A(k_1)$ and $\r_{2,1}(r)$. The use of \eq{space}and \eq{mom} leads immediately to the result
\bea
a_2(A)={\r_{2,1}^A(r)\over r_{2,1}^D(r)}\,={n_A(k)\over n_D(k)}, \,\,{\,( r\ll R_A,\,k>1/R_A}).
\eea 
The early workers~\cite{frankfurt93} used the ratio of momentum-space densities, and recent workers~\cite{Chen:2016bde} use the coordinate space version,
but both are the same in the leading-order approximation of each approach.

\subsection{Why center-of-mass and relative coordinates factorize}
We provide a qualitative explanation of the factorization inherent in \eq{fact}.
Start with the  non-relativistic  nuclear Hamiltonian with only two-nucleon forces, and consider infinite nuclear matter.
The basic assumption is the independent pair approximation. The idea is that the average separation between nucleons  $d=1.7$  fm,  so that when one of the nucleons of the pair makes a close encounter with a third particle  the collision occurs under conditions such that the original pair had no interactions at all~\cite{Gomes:1957zz}.
This idea was formally codified by Bethe and co-workers~\cite{Bethe:1971xm}, such that the results of the independent pair approximation appear as  the first term in the hole-line expansion.

We  explain how this works. Consider two-nucleons in nuclear matter, which interact independently of the other nucleons (except for the influence of the Pauli principle).
The two-nucleon Hamiltionian, $h$, is given by
\bea &h=h_0+h_1\\& h_0={P^2\over 4M},\,h_1={p^2\over M} +Qv,\eea
where $\bfP$ is the center-of-mass momentum operator, $\bfp$ is the relative momentum operator, $v$ is the two-nucleon potential, and $Q$ is an operator that projects both nucleon momenta to be greater than the Fermi momentum, $k_F$. Since the two-nucleon Hamiltonian is a sum of  two terms, $h=h_0+h_1$,   that commute the solution to the Schroednger equation,
$h|\psi\ra=E|\psi\ra$, is a product:
\bea \psi(\bfR,\bfr)=F_A(\bfR)\chi(\bfr)\label{Fchi}\eea
where \bea& h_0 F(\bfR)=E_{\rm cm}F_A(\bfR),\, h_1\chi(\bfr)=\epsilon \chi(\bfr), E=E_{\rm cm}+\epsilon\nonumber\\\eea
with \bea  \psi(\bfR,\bfr)=e^{i\bfK\cdot\bfR}\chi(\bfr),\label{f1}\eea
where we suppress notations regarding  spin and isospin to simplify the discussion. In general the function $\chi(\bfr)$ contains all values of angular momentum and 
has both short ranged and long-ranged aspects. The essence of \eq{fact} is that for small values of $|\bfr|$ all relative wave functions look like the deuteron wave function:
\bea \lim_{r\ll d}\chi(\bfr)= \g \psi_D(r),\label{ff1}\eea
where $\gamma$ represents the probability amplitude that the wave function $\chi$ corresponds to the deuteron quantum numbers.

It is necessary to introduce a single-particle,  mean-field  operator $U$ to extend this idea to finite-sized nuclei. In that case, \eq{f1}  is   often  replaced (see {\it e.g.} ~\cite{Haxton:1980iu} by
\bea  &   \psi(\bfR,\bfr)=\sum_{\a\b}C_{\a\b}\phi_\a(\bfr_1) \phi_\b(\bfr_2)\chi(\bfr),\label{f2}
\eea
where $\p_{\a,\b}$ are solutions of the single-particle equation, $C_{\a\b}$ are coefficients computed using the shell model. The single-particles vary over the size of the nucleus, while
the variations of $\chi(\bfr)-1$ occur over the range of the nucleon-nucleon interaction. If the size of the nucleus is much larger than this range \eq{Fchi} remains true. In these applications the Miller-Spencer correlation function~\cite{Miller:1975hu} has often been used to represent $\chi({\bfr})$.
 

\hrulefill
\hrulefill

\bibliography{eep,emc,miller}
\end{document}